%% file: BdKstarMuMuPaper.tex
\pdfoutput=1
\newcommand*{\ATLASLATEXPATH}{latex/}
\documentclass[cernpreprint, atlasdraft=false, texlive=2016, UKenglish]{\ATLASLATEXPATH atlasdoc}

\usepackage{rotating} 
\usepackage{\ATLASLATEXPATH atlaspackage}
\usepackage{\ATLASLATEXPATH atlasbiblatex}

\usepackage{\ATLASLATEXPATH atlasphysics}

\graphicspath{{logos/}{figures/}}

\makeatletter
\def\input@path{{texts/}}
\makeatother

\usepackage{xspace}
\usepackage{graphicx}
\usepackage{amsmath}
\usepackage{dcolumn}
\usepackage{epsfig}
\usepackage{psfrag}
\usepackage{lineno}
\usepackage{color}

\input{BdKstarMuMuPaper-metadata}
\hypersetup{pdftitle={ATLAS document},pdfauthor={The ATLAS Collaboration}}

\begin{document}

\maketitle


\input{texts/content.tex}

\section*{Acknowledgements}
\input{acknowledgements/Acknowledgements}

\clearpage
\appendix
\part*{Appendix}
\addcontentsline{toc}{part}{Appendix}
\input{texts/appendix.tex}

\printbibliography



\newpage \input{atlas_authlist}

\end{document}

%% file: BdKstarMuMuPaper-metadata.tex
\input{texts/TitleAbstract.tex}

\author{The ATLAS Collaboration}

\PreprintIdNumber{CERN-EP-2017-161}

\AtlasDate{\today}

\AtlasJournal{JHEP}

\AtlasJournalRef{JHEP\ 10\ (2018)\ 047}
\AtlasDOI{10.1007/JHEP10(2018)047}

%% file: texts/TitleAbstract.tex
\input{texts/sll-defs.sty}
\input{texts/paper-defs.sty}

\AtlasTitle{Angular analysis of $B^0_d \to K^{*}\mu^+\mu^-$ decays in $pp$ collisions at $\sqrt{s}= 8$ TeV with the ATLAS detector}

\AtlasAbstract{%
An angular analysis of the decay $B^0_d \to K^{*}\mu^+\mu^-$ is presented,
based on proton--proton collision data recorded by the ATLAS experiment at the LHC.
The study is using 20.3 fb$^{-1}$ of integrated luminosity collected during 2012 at centre-of-mass energy of $\sqrt{s}=8$ TeV.
Measurements of the $K^{*}$ longitudinal polarisation fraction and a set of angular parameters obtained for this decay are presented.
The results are compatible with the Standard Model predictions.
}

%% file: texts/content.tex
\section{Introduction}
\label{sec:intro}

Flavour-changing neutral currents (FCNC) have played a significant role in the construction of
the Standard Model of particle physics (SM). These processes are forbidden at 
tree level and can proceed only via loops, hence are rare. 
An important set of FCNC processes involve the transition of 
a $b$-quark to an $s\mumu$ final state mediated by electroweak box and 
penguin diagrams. If heavy new particles exist, they may contribute to FCNC decay amplitudes,
affecting the measurement of observables related to the decay under study.
Hence FCNC processes allow searches for contributions
from sources of physics beyond the SM (hereafter referred to as
new physics).   
This analysis focuses on the decay $\Bd\to K^{*0}(892)\mu^+\mu^-$, where $K^{*0}(892)\to K^+\pi^-$. Hereafter, the 
$K^{*0}(892)$ is referred to as $\kstar$
and charge conjugation is implied throughout, unless stated otherwise.
In addition to angular observables such as the forward-backward asymmetry \afb \footnote{The forward-backward 
asymmetry is given by the normalised difference between the number of positive muons going in the forward and 
backward directions with respect to the direction opposite to \Bd momentum in the dimuon rest frame.}, 
there is considerable interest in measurements of the charge
asymmetry, differential branching fraction, isospin asymmetry, and ratio of rates
of decay into dimuon and dielectron final states, all as a function of the invariant
mass squared of the dilepton system \qsqr.  All of these observable sets can be
sensitive to different types of new physics that allow for FCNCs at 
tree or loop level.
The BaBar, Belle, CDF, CMS, and LHCb collaborations have published the results of studies
of the angular distributions for 
$\Bd\to \kstar\mu^+\mu^-$~\cite{Wei:2009zv,Aaltonen:2011ja,Aaij:2013qta,Aaij:2015oid,CMS:Khachatryan:2015isa,Sirunyan:2017dhj,Lees:2015ymt,Wehle:2016yoi}.
The LHCb Collaboration has reported a potential hint, at the level of $3.4$ standard deviations, of 
a deviation from SM calculations~\cite{Aaij:2013qta,Aaij:2015oid} 
in this decay mode when using a parameterization of the angular distribution
designed to minimise uncertainties from hadronic form factors. 
Measurements using this approach were also reported by the Belle and CMS Collaborations~\cite{Wehle:2016yoi,Sirunyan:2017dhj} 
and they are consistent with the LHCb experiment's results and with the SM calculations. 
This paper presents results following the methodology outlined in Ref.~\cite{Aaij:2013qta}
and the convention adopted by the LHCb Collaboration for the definition of angular observables described in Ref.~\cite{Aaij:2013iag}.  
The results obtained here are compared with theoretical predictions that use the form factors computed
in Ref.~\cite{Straub:2015ica}.

This article presents the results of an angular analysis of the decay \bdkstmumu
with the ATLAS detector, using \lumi~\invfb of $pp$ collision data at a 
centre-of-mass energy $\sqrt{s}=8$~\tev\ delivered by the Large Hadron Collider (LHC)~\cite{Evans:2008zzb} during 2012. 
Results are presented in six different bins of \qsqr in the range $0.04$ to $6.0$ \qsqrunits, 
where three of these bins overlap. Backgrounds, including a radiative tail from $\Bd\to K^* \jpsi$ events, increase for
\qsqr above 6.0 \qsqrunits, and for this reason, data above this value are not studied.

The operator product expansion used to describe the decay $\Bd\to \kstar\mu^+\mu^-$ encodes short-distance 
contributions in terms of Wilson coefficients and long-distance contributions in terms of operators~\cite{Wilson:1972ee}. 
Global fits for Wilson coefficients have been performed using measurements of \bdkstmumu and 
other rare processes. Such studies aim to connect deviations from the SM
predictions in several processes to identify a consistent pattern
hinting at the structure of a potential underlying new-physics
Lagrangian, see Refs.~\cite{Altmannshofer:2013foa,Hurth:2014vma,Descotes-Genon:2015uva}. 
The parameters presented in this article can be used as inputs to these global fits.

\section{Analysis method}
\label{sec:method}

\begin{figure}[!ht]
\begin{center}
  \includegraphics[width=0.5\columnwidth]{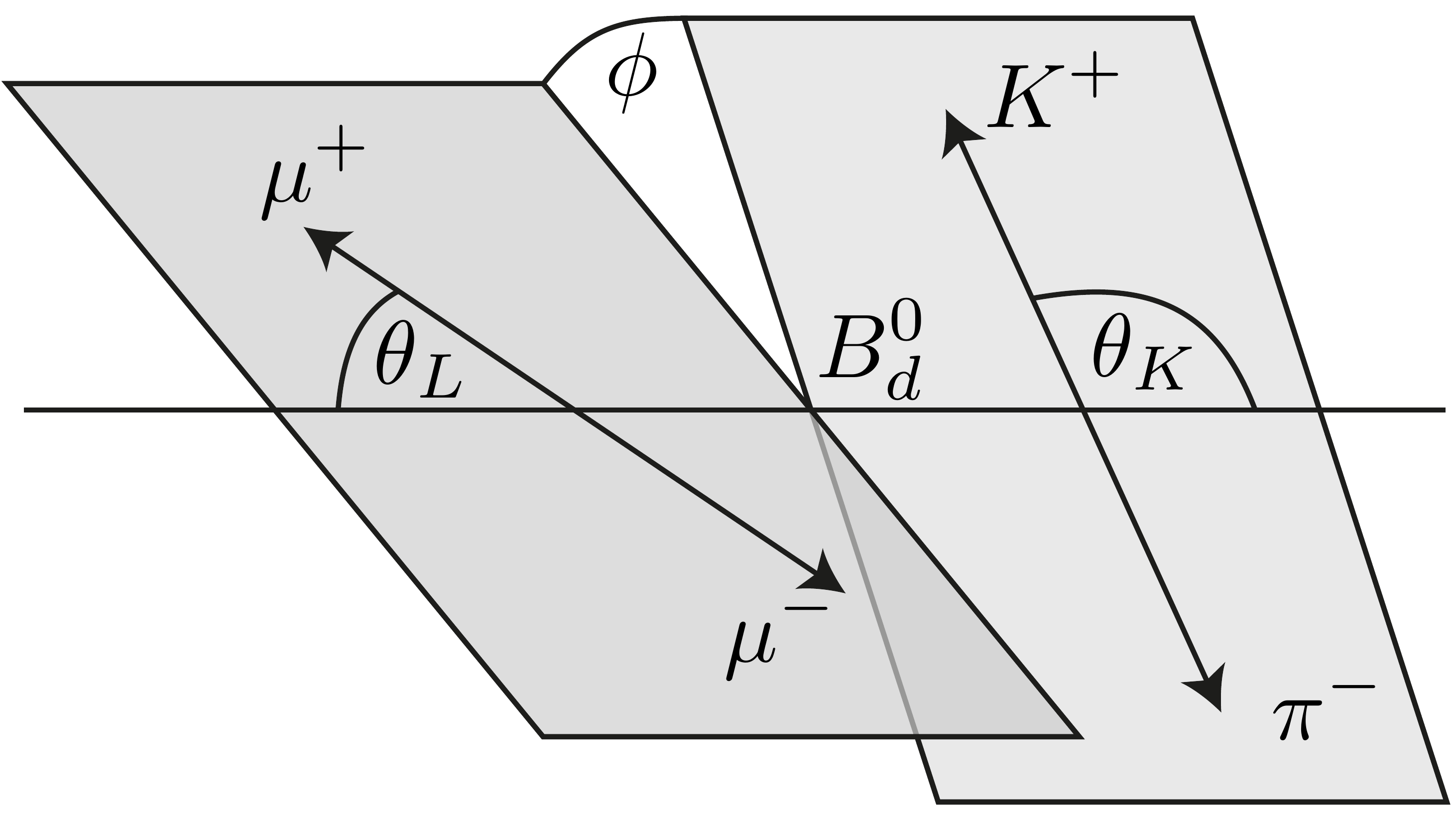}
\end{center}
\caption{An illustration of the \bdkstmumu decay showing the angles $\theta_K$, $\theta_L$ and $\phi$ defined in the text.  
Angles are computed in the rest frame of the $K^*$, dimuon system and \Bz meson, respectively.\label{fig:angles}}
\end{figure}

Three angular variables describing the decay are defined according to convention described by the LHCb Collaboration in Ref.~\cite{Aaij:2013iag}:
the angle between the $K^+$ and the direction opposite to the \Bz in the $K^*$ centre-of-mass frame (\tk);
the angle between the $\mu^+$ and the direction opposite to the \Bz in the dimuon centre-of-mass frame (\tl);
and the angle between the two decay planes formed by the $K\pi$ and the dimuon systems in the \Bz rest frame ($\phi$).  
For \Bzb mesons the definitions are given with respect to the negatively charged particles. 
Figure~\ref{fig:angles} illustrates the angles used.

The angular differential decay rate for \bdkstmumu is a function of \qsqr, \ctk, \ctl and $\phi$, and can be written in 
several ways~\cite{Dunietz:1990cj}.
The form to express the differential decay amplitude as a function of the angular parameters
uses coefficients that may be represented by the
helicity or transversity amplitudes~\cite{Descotes-Genon:2013vna} and is written as\footnote{
This equation neglects possible $K\pi$ $S$-wave contributions.  The effect of 
an $S$-wave contribution is considered following the method used by LHCb in Ref.~\cite{Aaij:2013qta}.}
\begin{eqnarray}
\frac{1}{\mathrm{d}\Gamma / \mathrm{d}\qsq}\frac{\mathrm{d}^4\Gamma}{\mathrm{d}\ctl \mathrm{d}\ctk \mathrm{d}\phi \mathrm{d}\qsq} &=& \frac{9}{32\pi}\Bigg[\frac{3(1-\fl)}{4}\stksq + \fl \ctksq + \frac{1-\fl}{4} \stksq \cos 2\theta_L \nonumber \\
&& - \fl \ctksq\cos 2\theta_L + \Sthree \stksq\stlsq\cos 2 \phi \nonumber\\
&& + \Sfour \sin 2\theta_K \sin 2\theta_L \cos\phi \nonumber + \Sfive\sin 2\theta_K\stl\cos\phi\nonumber \\
&& + \Ssix \stksq \ctl + \Sseven\sin 2\theta_K\stl\sin\phi \nonumber \\
&& + \Seight\sin 2\theta_K\sin 2\theta_L \sin\phi + \Snine \stksq\stlsq\sin 2\phi \Bigg]. \label{eq:fullangcorrelation}
\end{eqnarray} 
\noindent Here $\fl$ is the fraction of longitudinally polarised $\kstar$ mesons and the $S_i$ are 
angular coefficients.  These angular parameters are functions of the real and 
imaginary parts of the transversity amplitudes of $\Bd$ decays
into $\kstar\mu^+\mu^-$.
The forward-backward asymmetry is given by $\afb = 3\Ssix/4$. 
The predictions for the $S$ parameters depend on hadronic form factors which have significant uncertainties at leading order.  
It is possible to reduce the theoretical uncertainty in these predictions by transforming the $S_i$ using 
ratios constructed to cancel form factor uncertainties at leading order.
These ratios are given by Refs.~\cite{DescotesGenon:2012zf,Descotes-Genon:2013vna} as
\begin{eqnarray}
P_1 &=& \frac{2\Sthree}{1-\fl} \label{eq:pone}\\
P_2 &=& \frac{2}{3}\frac{\afb}{1-\fl} \label{eq:two}\\
P_3 &=& -\frac{\Snine}{1-\fl} \label{eq:three}\\
P_{j=4,5,6,8}^{\prime} &=& \frac{S_{i=4,5,7,8}}{\sqrt{\fl(1-\fl)}}.\label{eq:pprime}
\end{eqnarray}
All of the parameters introduced, \fl, $S_i$ and $P_j^{(\prime)}$, may vary with \qsqr and the data are analysed in \qsqr bins 
to obtain an average value for a given parameter in that bin.

\section{The ATLAS detector, data, and Monte Carlo samples}
\label{sec:detector}

The ATLAS experiment at the LHC is a general-purpose detector with 
a cylindrical geometry and nearly $4\pi$ coverage in solid angle \cite{PERF-2007-01}. It consists of an inner
detector (ID) for tracking, a calorimeter system and a muon spectrometer (MS). The ID consists of 
silicon pixel and strip detectors, with a straw-tube transition radiation tracker providing additional
information for tracks passing through the central region of the detector.\footnote{ATLAS uses a right-handed coordinate system with its origin at the nominal interaction point (IP) in the centre of the detector and the $z$-axis along the beam pipe. The $x$-axis points from the IP to the centre of the LHC ring, and the $y$-axis points upward. Cylindrical coordinates $(r,\Phi)$ are used in the transverse plane, $\Phi$ being the azimuthal angle around the $z$-axis. The pseudorapidity is defined in terms of the polar angle $\theta$ as $\eta=-\ln\tan(\theta/2)$.} 
The ID has a coverage of $|\eta| < 2.5$, and is immersed in a 2T axial magnetic field generated by a superconducting solenoid.
The calorimeter system, consisting of liquid argon and scintillator-tile sampling
calorimeter subsystems, surrounds the ID.  The outermost part of the detector is
the MS, which employs several detector technologies in order to provide
muon identification and a muon trigger.  A toroidal magnet system is embedded in the MS.
The ID, calorimeter system and MS have full azimuthal coverage.

The data analysed here were recorded in 2012 during Run 1 of the LHC.  
The centre-of-mass energy of the $pp$ system was $\sqrt{s}=8$~TeV. 
After applying data-quality criteria, the data sample analysed corresponds
to an integrated luminosity of \lumi~\invfb.
A number of Monte Carlo (MC) signal and background event samples were generated,
with $b$-hadron production in $pp$ collisions simulated with \PYTHIAV{8.186}~\cite{Sjostrand:2006za,Sjostrand:2007gs}.
The AU2 set of tuned parameters~\cite{ATL-PHYS-PUB-2012-003} is used together with the CTEQ6L1 PDF set~\cite{Pumplin:2002vw}.
The EvtGen 1.2.0 program~\cite{Lange:2001uf} is used for the properties of b- and c-hadron decays.
The simulation
included modelling of multiple interactions per $pp$ bunch crossing in the LHC with \PYTHIA soft QCD processes.
The simulated events were then passed through the full ATLAS detector simulation program based on 
\GEANT4~\cite{Agostinelli:2002hh,SOFT-2010-01} and reconstructed in the same way as data.  
The samples of MC generated events are described further in Section~\ref{sec:mlfit}.

\section{Event selection}
\label{sec:selection}
Several trigger signatures constructed from the MS and ID inputs are selected based on availability 
during the data-taking period, prescale factor and efficiency for signal identification.
Data are combined from 19 trigger chains where 21\%, 89\% or 5\% of selected events pass
one or more triggers with one, two, or at least three muons identified online in the MS, respectively.
Of the events passing the requirement of at least two muons, 
the largest contribution comes from the~chain requiring one muon with a 
transverse momentum $\pt>4$~\gev and the other muon with $\pt>6$~\gev.  
This combination of triggers ensures that the analysis remains sensitive to events down to the 
kinematic threshold of $\qsqr = 4m_\mu^2$, where $m_\mu$ is the muon mass.
The effective average trigger efficiency for selected signal events is about 29\%, determined from signal MC simulation.

Muon track candidates are formed offline by combining information from both the ID and MS~\cite{Aad:2014rra}.  
Tracks are required to satisfy $|\eta|<2.5$.
Candidate muon (kaon and pion) tracks in the ID are required to satisfy $\pt > 3.5$ ($0.5$)~\gev.
Pairs of oppositely charged muons are required to originate from a common vertex 
with a fit quality $\chi^2/\mathrm{NDF}<10$.

Candidate $K^*$ mesons are formed using pairs of oppositely charged kaon and pion candidates
reconstructed from hits in the ID.
Candidates are required to satisfy $\pt(\kstar) > 3.0$~\gev.
As the ATLAS detector does not have a dedicated charged-particle 
identification system, candidates are reconstructed with both 
possible $K\pi$ mass hypotheses. The selection implicitly relies on the kinematics of the 
reconstructed $K^*$ meson to determine which of the two tracks corresponds to the kaon. If both candidates
in an event satisfy selection criteria, they are retained and one of them is selected in the
next step following a procedure described below. 
The $K\pi$ invariant mass is required to lie in a window of twice the natural width around the nominal
mass of 896~\mev, i.e.\ in the range $[846, 946]$~\mev.
The charge of the kaon candidate is used to assign the flavour of the reconstructed \Bd candidate.  

The \Bd candidates are reconstructed from a $K^*$ candidate and a pair of oppositely charged
muons.    The four-track vertex is fitted and required to satisfy $\chi^2/\mathrm{NDF}<2$ to
suppress background.
A significant amount of combinatorial, \Bd, \Bp, \Bs and $\Lambda_b$ background contamination remains
at this stage. Combinatorial background is suppressed by requiring a \Bd candidate lifetime significance
$\tau/\sigma_\tau > 12.5$, where the decay time uncertainty $\sigma_\tau$ is calculated
from the covariance matrices associated with the four-track vertex fit and with the primary
vertex fit. 
Background from final states partially reconstructed as $B\to \mu^+\mu^- X$ accumulates at invariant mass
below the \Bd mass and contributes to the signal region.  
It is suppressed by imposing an asymmetric mass cut around
the nominal \Bd mass, $5150$~\mev$ < m_{K\pi\mu\mu} < 5700$~\mev.  The high-mass sideband is retained, as the 
parameter values for the combinatorial background shapes are extracted from the fit to data
described in Section~\ref{sec:mlfit}.
To further suppress background, it is required that the angle $\Theta$, defined between the 
vector from the primary vertex to the \Bd candidate decay vertex and the \Bd 
candidate momentum, satisfies $\cos\Theta > 0.999$. Resolution effects on \ctk, \ctl and $\phi$ were found to
have a negligible effect on the ATLAS $\Bs\to J/\psi\phi$ analysis~\cite{BPHY-2014-03}.
It is assumed to also be the case for \bdkstmumu.

On average 12\% of selected events in the data have more than one reconstructed \Bd candidate. The fraction
is 17\% for signal MC samples and 2--10\% for exclusive background MC samples. 
A two-step selection process is used for such events. 
For 4\% of these events it is possible to select a candidate with the smallest value of the 
\Bd vertex $\chi^2/\mathrm{NDF}$.
However, the majority, about 96\%, of multiple candidates arise from four-track combinations where the 
kaon and pion assignments are ambiguous. 
As these candidates have degenerate values for the \Bd candidate vertex $\chi^2/\mathrm{NDF}$, a second 
selection step is required. 
The \Bd candidate reconstructed with the smallest value of $|m_{K\pi} - m_{\kstar}|/\sigma(m_{K\pi})$ is retained for analysis, 
where $m_{K\pi}$ is the \kstar candidate mass, $\sigma(m_{K\pi})$ is the per-event uncertainty in this quantity, and $m_{\kstar}$ is the world 
average value of the \kstar mass.

The selection procedure results in an incorrect flavour tag (mistag) for some signal events. 
The mistag probability of a \Bz (\Bzb) meson is denoted by $\omega$ ($\overline \omega$) 
and is determined from MC simulated events to be $\mistag \, (\mistagbar)$. 
The mistag probability varies slightly with \qsqr such that
the difference $\omega -\overline \omega$ remains consistent with zero.  Hence 
the average mistag rate $\langle \omega \rangle$ in a given \qsqr bin is used
to account for this effect.
If a candidate is mistagged, the values of \ctl, \ctk and $\phi$ change sign, 
while the latter two are also slightly shaped by the swapped hadron track mass hypothesis.
Sign changes in these angles affect the overall sign of the terms
multiplied by the coefficients \Sfive, \Ssix, \Seight and \Snine (similarly for the corresponding $P^{(\prime)}$ parameters)
in Equation~(\ref{eq:fullangcorrelation}).  
The corollary is that mistagged events result in a dilution factor of $(1 - 2\langle \omega \rangle )$ for
the affected coefficients.

The region $\qsqr\in[0.98, 1.1]$ \qsqrunits is vetoed to remove any potential contamination from the $\phi(1020)$ resonance.
The remaining data with $\qsqr \in [0.04, 6.0]$ \qsqrunits are analysed in order to extract the signal parameters of interest.
Two $K^*c\overline c$ control regions are defined for \Bd decays into $K^*\jpsi$ and $K^*\psitwos$, respectively as
$\qsqr\in[8, 11]$ and $[12, 15]$ \qsqrunits.  The control samples are used to extract values for nuisance parameters 
describing the signal probability density function (pdf) from data as discussed in Section~\ref{sec:fit:control}.

For \qsqr $<6$ \qsqrunits the selected data sample consists of \nselectedtotallow events and is composed of signal $\Bd \to K^*\mumu$ decay 
events as well as background that is dominated by a combinatorial component that does 
not peak in $m_{K\pi\mu\mu}$ and does not exhibit a resonant structure in \qsqr.
Other background contributions are considered when estimating systematic uncertainties.
Above $6$ \qsqrunits the background contribution increases significantly, including
events coming from $\Bd\to K^*\jpsi$ with a radiative $\jpsi\to \mu^+\mu^- \gamma$ decay.
Scalar $K\pi$ contributions are neglected in the nominal fit and considered
only when addressing systematic uncertainties.
The data are analysed in the \qsqr bins \qsqrbinA, \qsqrbinB and \qsqrbinC \qsqrunits,
where the bin width is chosen to provide a sample of signal events sufficient to perform an 
angular analysis. The width is much larger than the $q^2$ resolution obtained from MC simulated 
signal events and observed in data for \Bd decays into $K^*\jpsi$ and $K^*\psitwos$.  
Additional overlapping bins \qsqrbinD, \qsqrbinE and \qsqrbinF \qsqrunits are analysed in order to 
facilitate comparison with results of other experiments and with theoretical predictions.

\section{Maximum-likelihood fit}
\label{sec:mlfit}

Extended unbinned maximum-likelihood fits of the angular distributions of the signal decay are
performed on the data for each \qsqr bin.
The discriminating variables used in the fit are $m_{K\pi\mu\mu}$, the cosines 
of the helicity angles ($\ctk$ and $\ctl$), and $\phi$.  The likelihood $\like$ for a given \qsqr bin
is 
\begin{eqnarray}
\like = \frac{{\mathrm{e}}^{-n}}{N!} \prod\limits_{k=1}^N \sum\limits_l n_l P_{kl} (m_{K\pi\mu\mu}, \ctk,\ctl,\phi; \widehat p, \widehat \theta),\label{eq:like}
\end{eqnarray}
where $N$ is the total number of events, the sum runs over signal and background
components, $n_l$ is the fitted yield for the $l\textsuperscript{th}$ component, $n$ is the sum over $n_l$, and $P_{kl}$ is the
pdf evaluated for event $k$ and component $l$. In the nominal fit, $l$ iterates only over one signal and one background component.
The $\widehat p$ are parameters of interest ($F_L$, $S_i$) and $\widehat \theta$ are
nuisance parameters.  The remainder of this section discusses the signal model (Section~\ref{sec:fit:signal}), treatment of background (Section~\ref{sec:fit:background}), use of $\kstar c\overline{c}$ decay control samples (Section~\ref{sec:fit:control}), fitting procedure and validation (Section~\ref{sec:fit:process}).

\subsection{Signal model}
\label{sec:fit:signal}

The signal mass distribution is modelled by a Gaussian distribution with the width given by the per-event uncertainty
in the $K\pi\mu\mu$ mass, $\sigma(m_{K\pi\mu\mu})$, as estimated from the track fit, multiplied by a unit-less scale factor $\xi$, i.e. the width given by $\xi \cdot \sigma(m_{K\pi\mu\mu})$.
The mean values of the \Bd candidate mass ($m_0$) and $\xi$ of the signal Gaussian pdf are determined from fits to data in the control regions 
as described in Section~\ref{sec:fit:control}.
The simultaneous extraction of all coefficients
using the full angular distribution of Equation~(\ref{eq:fullangcorrelation}) requires a certain
minimum signal yield and signal purity to avoid a pathological fit behaviour.
A significant fraction of
fits to ensembles of simulated pseudo-experiments do not converge using the full distribution.
This is mitigated using trigonometric transformations to  fold  certain angular distributions
and thereby simplify Equation~(\ref{eq:fullangcorrelation}) such that only
three parameters are extracted in one fit: \fl, \Sthree and one of the other
$S$ parameters.
For these folding schemes the angular parameters of interest, denoted by $\widehat p$ in Equation~(\ref{eq:like}), are $(\fl, \Sthree, S_i)$ where $i=4, 5, 7, 8$.
These translate into $(\fl, P_1, P^\prime_j)$, where $j=4, 5, 6, 8$, using Equation~(\ref{eq:pprime}).
Following Ref.~\cite{Aaij:2013qta}, the transformations listed
below are used:
\begin{eqnarray}
\fl, \Sthree, \Sfour, P^{'}_{4}: \begin{cases}
\phi \rightarrow -\phi  & \text{for $\phi < 0$}  \\
\phi \rightarrow \pi - \phi  & \text{for $\theta_{L} > \frac{\pi}{2}$} \\
\theta_{L} \rightarrow \pi - \theta_{L}  & \text{for $\theta_{L} > \frac{\pi}{2}$}, \\
\end{cases}
\label{eq:s4transformation}
\end{eqnarray}
\begin{eqnarray}
\fl, \Sthree, \Sfive, P^{'}_{5}: \begin{cases}
\phi \rightarrow -\phi  & \text{for $\phi < 0$}  \\
\theta_{L} \rightarrow \pi - \theta_{L}  & \text{for $\theta_{L} > \frac{\pi}{2}$}, \\
\end{cases}
\label{eq:s5transformation}
\end{eqnarray}
\begin{eqnarray}
\fl, \Sthree, \Sseven, P^{'}_{6}: \begin{cases}
\phi \rightarrow \pi - \phi  & \text{for $\phi > \frac{\pi}{2}$}  \\
\phi \rightarrow - \pi - \phi  & \text{for $\phi < -\frac{\pi}{2}$} \\
\theta_{L} \rightarrow \pi - \theta_{L}  & \text{for $\theta_{L} > \frac{\pi}{2}$}, 
\end{cases}
\label{eq:s7transformation}
\end{eqnarray}
\begin{eqnarray}
\fl, \Sthree, \Seight, P^{'}_{8}: \begin{cases}
\phi \rightarrow \pi - \phi  & \text{for $\phi > \frac{\pi}{2}$} \\
\phi \rightarrow -\pi -\phi  & \text{for $\phi < -\frac{\pi}{2}$}  \\
\theta_{L} \rightarrow \pi - \theta_{L}  & \text{for $\theta_{L} > \frac{\pi}{2}$} \\
\theta_{K} \rightarrow \pi - \theta_{K}  & \text{for $\theta_{L} > \frac{\pi}{2}$}. \\
\end{cases}
\label{eq:s8transformation}
\end{eqnarray}

On applying transformation~(\ref{eq:s4transformation}), (\ref{eq:s5transformation}), (\ref{eq:s7transformation}), and (\ref{eq:s8transformation}),
 the angular variable ranges become 
\begin{eqnarray*}
\ctl\in[0, 1], \ctk\in[-1,1] \text{ and } \phi\in[0, \pi],\nonumber\\
\ctl\in[0, 1], \ctk\in[-1,1] \text{ and } \phi\in[0, \pi],\nonumber\\
\ctl\in[0, 1], \ctk\in[-1,1] \text{ and } \phi\in[-\pi/2, \pi/2],\nonumber\\
\ctl\in[0, 1], \ctk\in[-1,1] \text{ and } \phi\in[-\pi/2, \pi/2],\nonumber
\end{eqnarray*}
respectively.
A consequence of using the folding schemes is that \Ssix (\afb) and \Snine cannot be extracted from the data.
The values and uncertainties of \fl and \Sthree obtained from the four fits are consistent 
with each other and the results reported are those found to have the smallest systematic
uncertainty.

\begin{figure}[!ht]
\begin{center}
 \includegraphics[width=0.49\columnwidth]{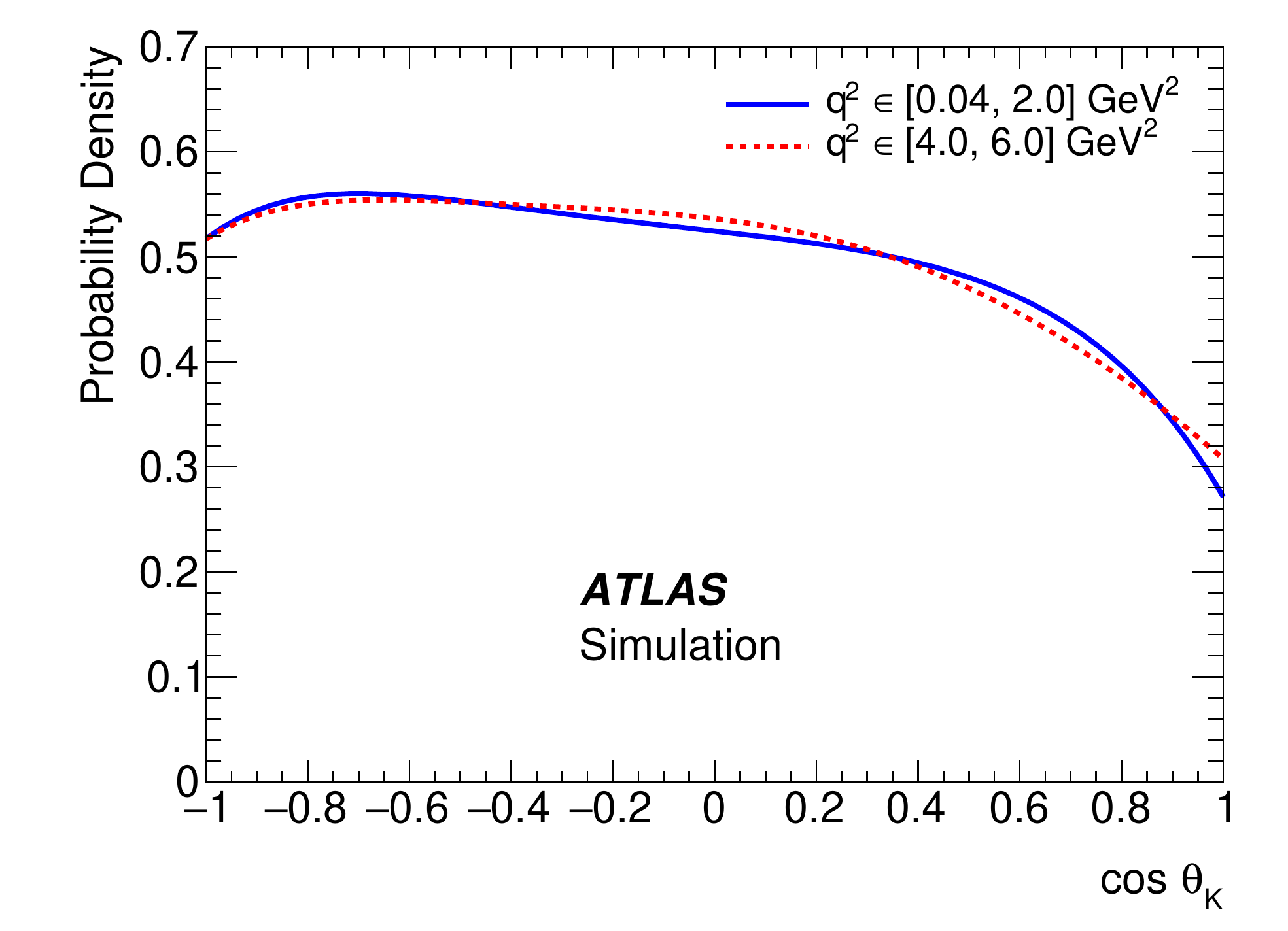}
 \includegraphics[width=0.49\columnwidth]{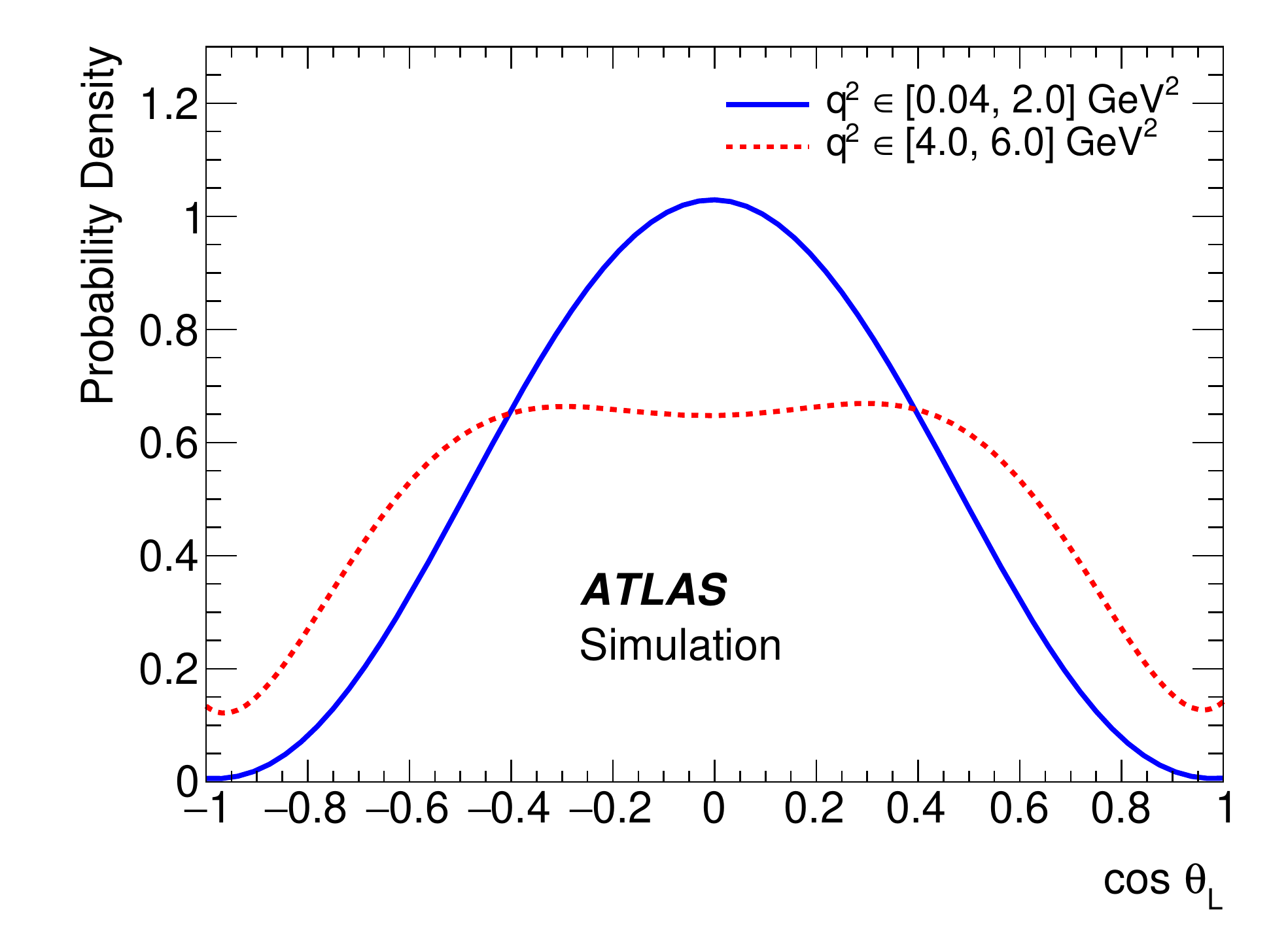}
\end{center}
\caption{The acceptance functions for (left) \ctk and (right) \ctl for (solid) $\qsqr \in \qsqrbinA$ \qsqrunits and (dashed) $\qsqr \in \qsqrbinC$ \qsqrunits,
that shape the angular decay rate of Equation~(\ref{eq:fullangcorrelation}).
\label{fig:acceptance}}
\end{figure}

Three MC samples are used to study the signal reconstruction and acceptance.
Two of them follow the SM prediction for the decay angle distributions taken
from Ref.~\cite{Ali:1999mm}, with separate samples generated for \Bd
and \Bdbar decays.  The third MC sample has $\fl=1/3$ and the angular
distributions are generated uniformly in \ctk, \ctl and $\phi$.  The samples are
used to study the effect of potential mistagging and reconstruction differences
between particle and antiparticle decays and for determination of the
acceptance.  The acceptance function is defined as the ratio of
reconstructed and generated distributions of \ctk, \ctl, $\phi$, i.e.\ it is compensating for the bias
in the angular distributions resulting from triggering, reconstruction and selection of events.
It is described by sixth-order (second-order) polynomial
distributions for \ctk and \ctl 
($\phi$) and is assumed to factorise for each angular distribution, i.e.\
using $\varepsilon(\ctk, \ctl, \phi)=\varepsilon(\ctk)\varepsilon(\ctl)\varepsilon(\phi)$.
A systematic uncertainty is assessed in order to account for this assumption.  The acceptance function multiplies the 
angular distribution in the fit, i.e.\ the signal pdf is 
\begin{eqnarray*}
P_{kl} = \varepsilon(\ctk)\varepsilon(\ctl)\varepsilon(\phi) g(\ctk, \ctl, \phi) \cdot G(m_{K\pi\mu\mu}),
\end{eqnarray*}
where $g(\ctk, \ctl, \phi)$ is an angular differential decay rate resulting from one of the four folding schemes
applied to Equation~(\ref{eq:fullangcorrelation}) and $G(m_{K\pi\mu\mu})$ is the signal mass distribution.
The MC sample generated with uniform \ctk, \ctl and $\phi$ distributions is used to determine the nominal acceptance 
functions for each of the transformed variables defined in Equations~(\ref{eq:s4transformation})--(\ref{eq:s8transformation}).
The other samples are used to estimate the related systematic uncertainty.
Among the angular variables the \ctl distribution is the most affected by the acceptance.
This is a result of the minimum transverse momentum requirements on
the muons in the trigger and the larger inefficiency to reconstruct
low-momentum muons, such that large values of $|\ctl|$ are inaccessible at low \qsqr.
As \qsqr increases, the acceptance effects become less severe.  The \ctk distribution 
is affected by the ability to reconstruct the $K\pi$ system, but that effect shows no
significant variation with \qsqr.  There is no significant acceptance effect for $\phi$.
Figure~\ref{fig:acceptance} shows the acceptance functions used for \ctk and \ctl for two different 
\qsqr ranges for the nominal angular distribution given in Equation~(\ref{eq:fullangcorrelation}).

\subsection{Background modes}
\label{sec:fit:background}

The fit to data includes a combinatorial background component that does not peak in
the $m_{K\pi\mu\mu}$ distribution.  It is assumed
that the background pdf factorises into a product of one-dimensional 
terms.
The mass distribution of this component is described by an exponential function
and second-order Chebychev polynomials are used to model the \ctk, \ctl and $\phi$ distributions. 
The values of the nuisance
parameters describing these shapes are obtained from fits to the data independently for each \qsqr bin.

Inclusive samples of $b\overline{b}\to \mu^+\mu^- X$ and $c\overline{c}\to \mu^+\mu^- X$ decays
and eleven exclusive \Bd, \Bs, \Bu and \LambdaB background samples are studied in order to 
identify contributions of interest to be included in the fit model, or to be 
considered when estimating systematic uncertainties. 
The relevant exclusive modes found to be of interest are 
discussed below.
Events with \Bc decays are suppressed by excluding the \qsqr range containing the \jpsi and \psitwos,
and by charm meson vetoes discussed in Section~\ref{sec:systematic}. 
The
exclusive background decays considered for the signal mode are 
$\Lambda_b\to \Lambda(1520) \mu^+\mu^-$, $\Lambda_b\to p K^- \mu^+\mu^-$,
$B^+\to K^{(*)+} \mu^+\mu^-$ and $B_s^0 \to \phi \mu^+\mu^-$.
These background contributions are accounted for as systematic uncertainties estimated
as described in Section~\ref{sec:systematic}.

Two distinct background contributions not considered above are 
observed in the \ctk and \ctl distributions.  
They are not accounted for in the
nominal fit to data, and are treated as systematic effects.
A peak is found in the \ctk distribution near 1.0 and appears to have contributions from at least two distinct sources. One of these
arises from misreconstructed \Bp decays, such as $\Bp\to K^+\mu\mu$ and $\Bp\to \pi^+\mu\mu$.
These decays can be reconstructed as signal if another track is combined with the hadron to form
a \kstar candidate in such a way that the event passes the reconstruction and selection.
The second contribution comes from combinations of two charged tracks that pass the selection
and are reconstructed as a \kstar candidate.  These fake \kstar candidates accumulate
around \ctk of 1.0 and are observed in the $K\pi$ mass sidebands away from 
the \kstar meson.  They are distinct from the structure of expected $S$-, $P$- and 
$D$-wave $K\pi$ decays resulting from a signal $\Bd\to K\pi\mu\mu$ transition.
The origin of this source of background is not fully understood.  The observed excess may arise
from a statistical fluctuation, an unknown background process, or a combination of both.
Systematic uncertainties are assigned to evaluate the effect of these two background contributions, as described in Section~\ref{sec:systematic}.

Another peak is found in the \ctl distribution near values of $\pm$0.7. It is associated with partially reconstructed $B$ decays into final 
states with a charm meson. This is studied using Monte Carlo simulated events
for the decays $D^0\to K^-\pi^+$, $D^+\to K^-\pi^+\pi^+$ and $D^+_s\to K^+K^-\pi^+$.  Events with a $B$ meson decaying 
via an intermediate charm meson $D^0$, $D^+$ or $D^+_s$ are found to pass the selection 
and are reconstructed in such a way that they accumulate around 0.7 in $|\ctl|$.  
These are removed from the data sample when estimating systematic uncertainties, as described in Section~\ref{sec:systematic}.

\subsection{\boldmath $\kstar c\overline{c}$ control sample fits}
\label{sec:fit:control}

\begin{figure}[!ht]
\begin{center}
 \includegraphics[width=0.49\columnwidth]{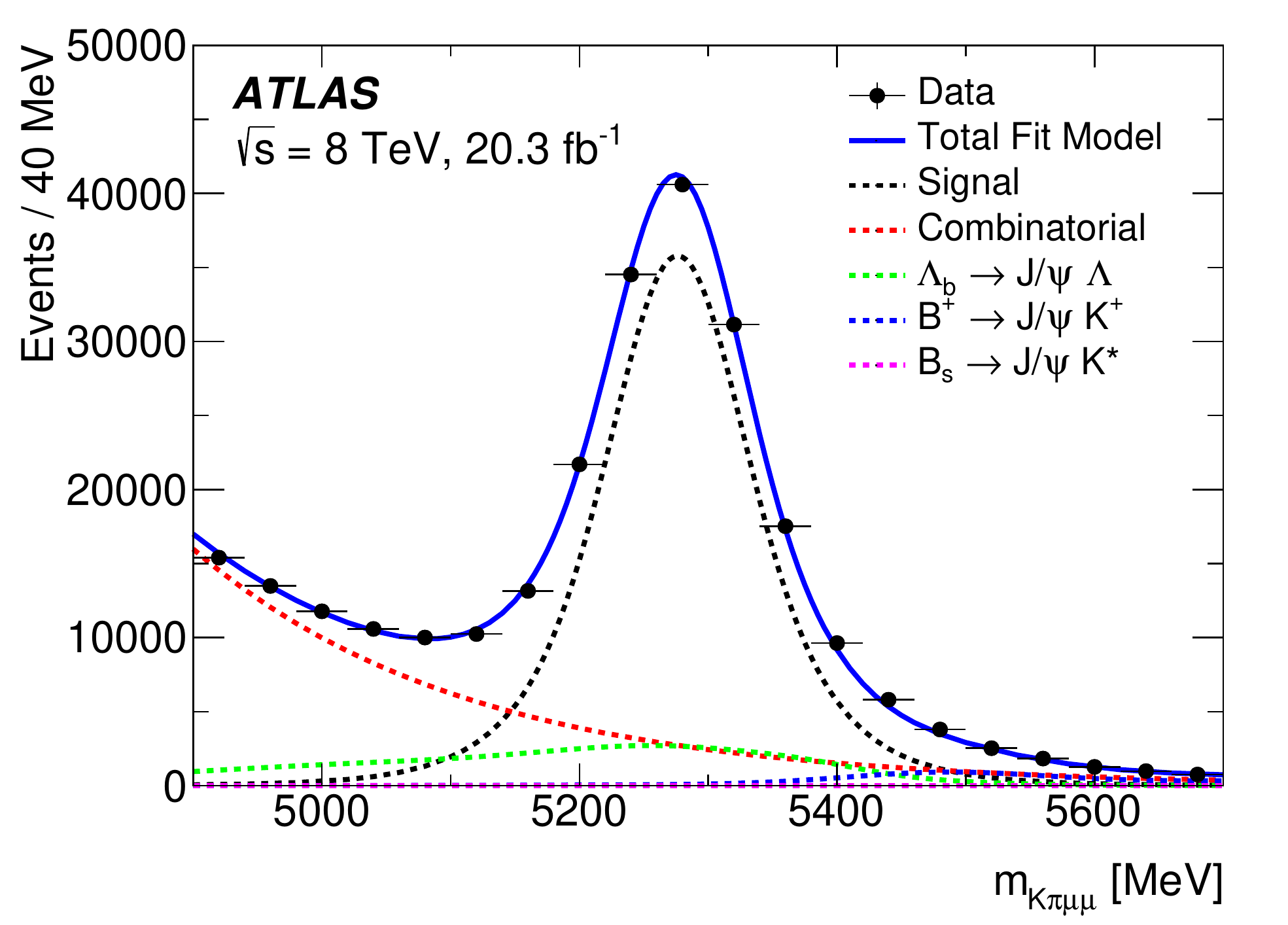}
 \includegraphics[width=0.49\columnwidth]{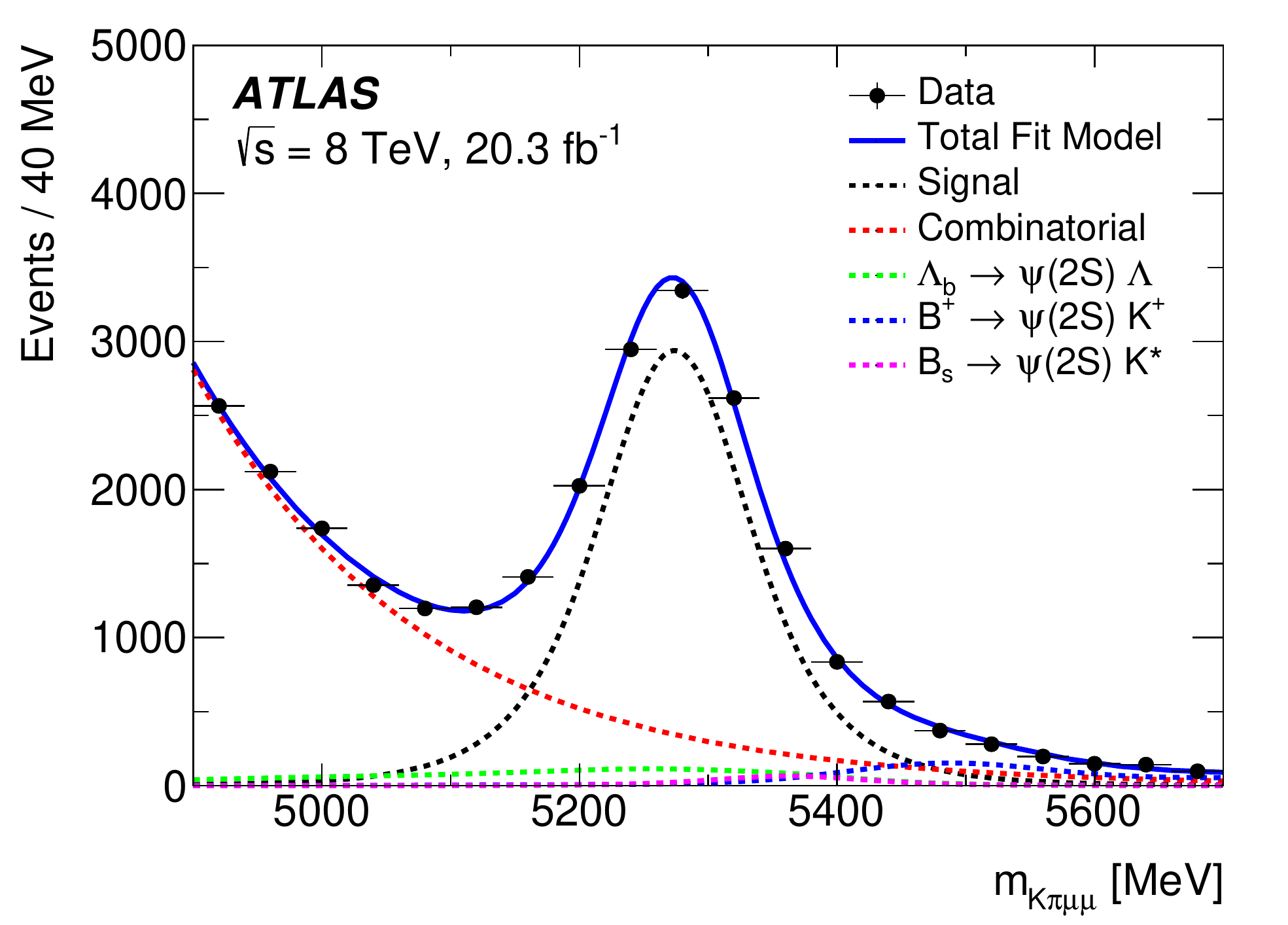}
\end{center}
\caption{Fits to the $K\pi\mu\mu$ invariant mass distributions for the
(left) $K^* \jpsi$ and (right) $K^*\psitwos$ control region samples.  The data are shown as points and 
the total fit model as the solid lines.  The dashed lines represent (black) signal, 
(red) combinatorial background, (green) $\Lambda_b$ background, (blue) $B^+$ background
and (magenta) $B_s^0$ background components. 
\label{fig:control}}
\end{figure}

The mass distribution obtained from the simulated samples for $\kstar c\overline{c}$
decays, respectively as $\qsqr\in[8, 11]$ and $[12, 15]$ \qsqrunits, and the signal mode, in different bins of \qsqr, are found to be
consistent with each other. 
Values of
$m_0$ and $\xi$ for $\Bd\to\kstar\jpsi$ and $\Bd\to\kstar\psitwos$ 
events are used for the signal pdf and extracted from fits to the data.
An extended unbinned maximum-likelihood fit is performed in
the two $\kstar c\overline{c}$ control region samples.
There are three exclusive backgrounds included: $\Lambda_b\to \Lambda c\overline{c}$, $B^+\to K^+ c\overline{c}$ and $B_s^0\to \kstar c\overline{c}$.
The $\kstar c\overline{c}$ pdf\ has the same form as the signal model, combinatorial background is
described by an exponential distribution, and double and triple Gaussian pdfs
determined from MC simulated events are used to describe the exclusive background contributions.
A systematic uncertainty is evaluated by allowing for 0, 1, 2 and 3 exclusive background components.
The control sample fit projections for the variant of the fit including all three exclusive
backgrounds can be found in Figure~\ref{fig:control}.  
The impact of the used exclusive background model on the peak position and scale factor of the signal pdf is negligible.
From these fits the statistical and 
systematic uncertainties in the values of $m_0$ and $\xi$ are extracted for the \Bd
component in order to be used in the $\Bd\to \kstar\mu^+\mu^-$ fits.
From the \jpsi control data it is determined that
the values for the nuisance parameters describing the signal model pdf in the $K\pi\mu\mu$ mass are
$m_0=5276.6\pm 0.3\pm 0.4$~\mev and $\xi=1.210 \pm 0.004 \pm 0.002$, 
where the uncertainties are statistical and systematic, respectively.  
The \psitwos sample yields compatible results albeit with larger uncertainties.  
These results are similar to those obtained from the MC simulated samples, and 
the numbers derived from the $K^* \jpsi$ data are used for the signal region fits.

\subsection{Fitting procedure and validation}
\label{sec:fit:process}
A two-step fit process is performed for the different signal bins in \qsqr. 
The first step is a fit to the $K\pi\mu^+\mu^-$ invariant mass
distribution, using the event-by-event uncertainty in the reconstructed mass as a conditional variable.  For this fit, the parameters
$m_0$ and $\xi$ are fixed to the values obtained from fits to data control samples as described in Section~\ref{sec:fit:control}.
A second step adds the (transformed) \ctk, \ctl and $\phi$ variables to the likelihood in order to
extract \fl and the $S$ parameters along with the values for the nuisance parameters related to the combinatorial background shapes.
Some nuisance parameters, namely $m_0$, $\xi$, signal and background yields, and the exponential shape parameter for the 
background mass pdf, are fixed to the results obtained from the first step.

The fit procedure is validated using ensembles of simulated pseudo-experiments generated with the 
\fl and $S$ parameters corresponding to those obtained from the data.  The purpose of these 
experiments is to measure the intrinsic fit bias resulting from the likelihood estimator used to 
extract signal parameters.  These ensembles are also used to check that the uncertainties extracted
from the fit are consistent with expectations.
Ensembles of simulated pseudo-experiments are performed in which signal
MC events are injected into samples of background events generated from the likelihood.  
The signal yield determined from the first
step in the fit process is found to be unbiased.
The angular parameters extracted from the nominal fits have biases with magnitudes ranging between 0.01 and 0.04, 
depending on the fit variation and \qsqr bin.  
A similar procedure is used to estimate the effect of neglecting $S$-wave contamination in the data sample.
Neglecting the $S$-wave component in the fit model results in a bias between 0.00 and 0.02 in the angular parameters. 
Similarly, neglecting exclusive background contributions from $\Lambda_b$, $B^+$ and $B_s^0$ decays
that peak in $m_{K\pi\mu\mu}$ near the \Bd mass results in a bias of less than 0.01 on the angular parameters.
All these effects are included in the systematic uncertainties described in section~\ref{sec:systematic}.
The $P^{(\prime)}$ parameters are obtained using the fit results and covariance matrices from the 
second fit along with Equations~(\ref{eq:pone})--(\ref{eq:pprime}).

\section{Results}
\label{sec:result}

\begin{figure}[!ht]
\begin{center}
  \begin{picture}(550,320)
\put(0,160){
\resizebox{\columnwidth}{!}{
  \includegraphics[width=0.45\columnwidth]{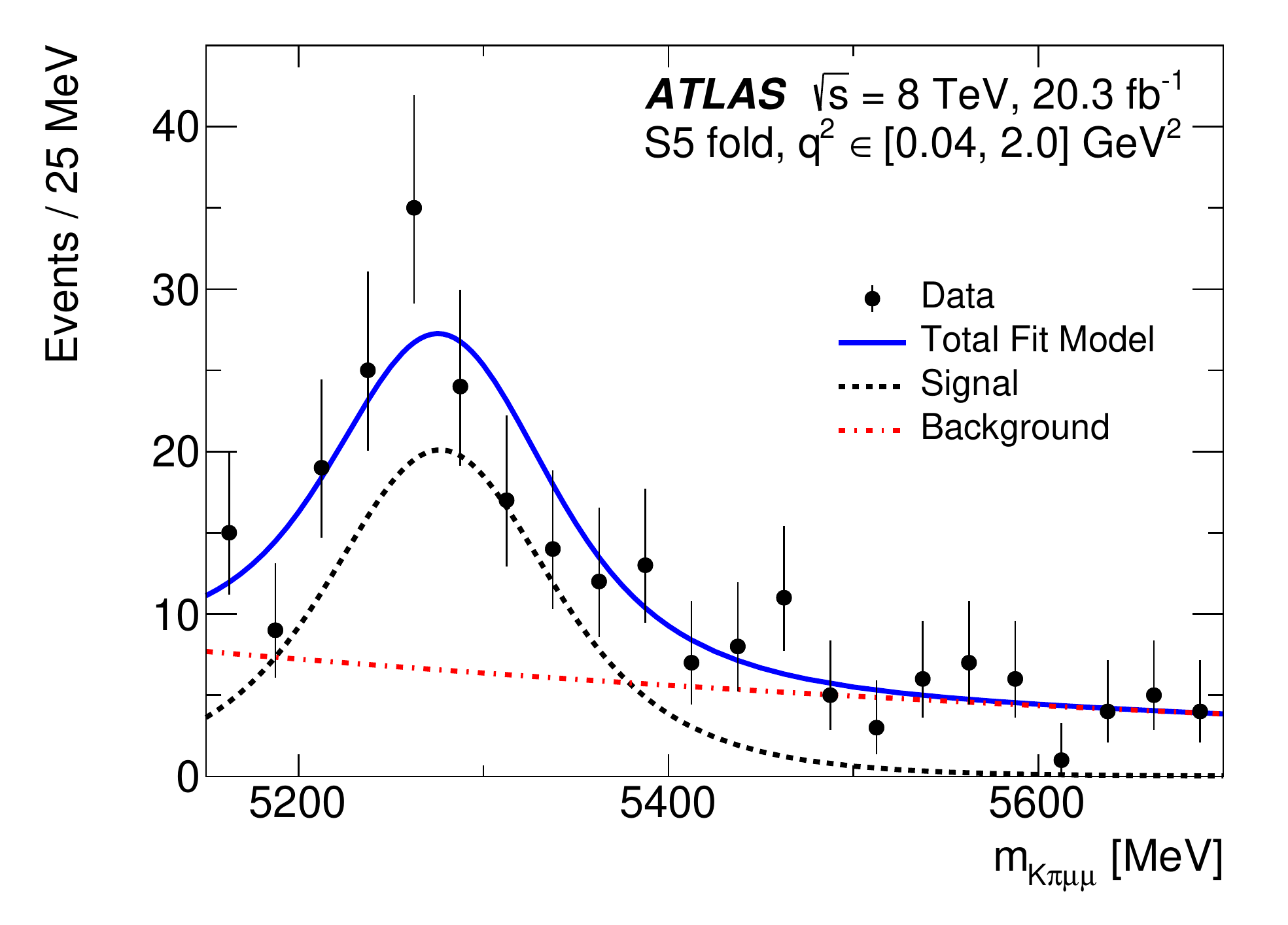}
  \includegraphics[width=0.45\columnwidth]{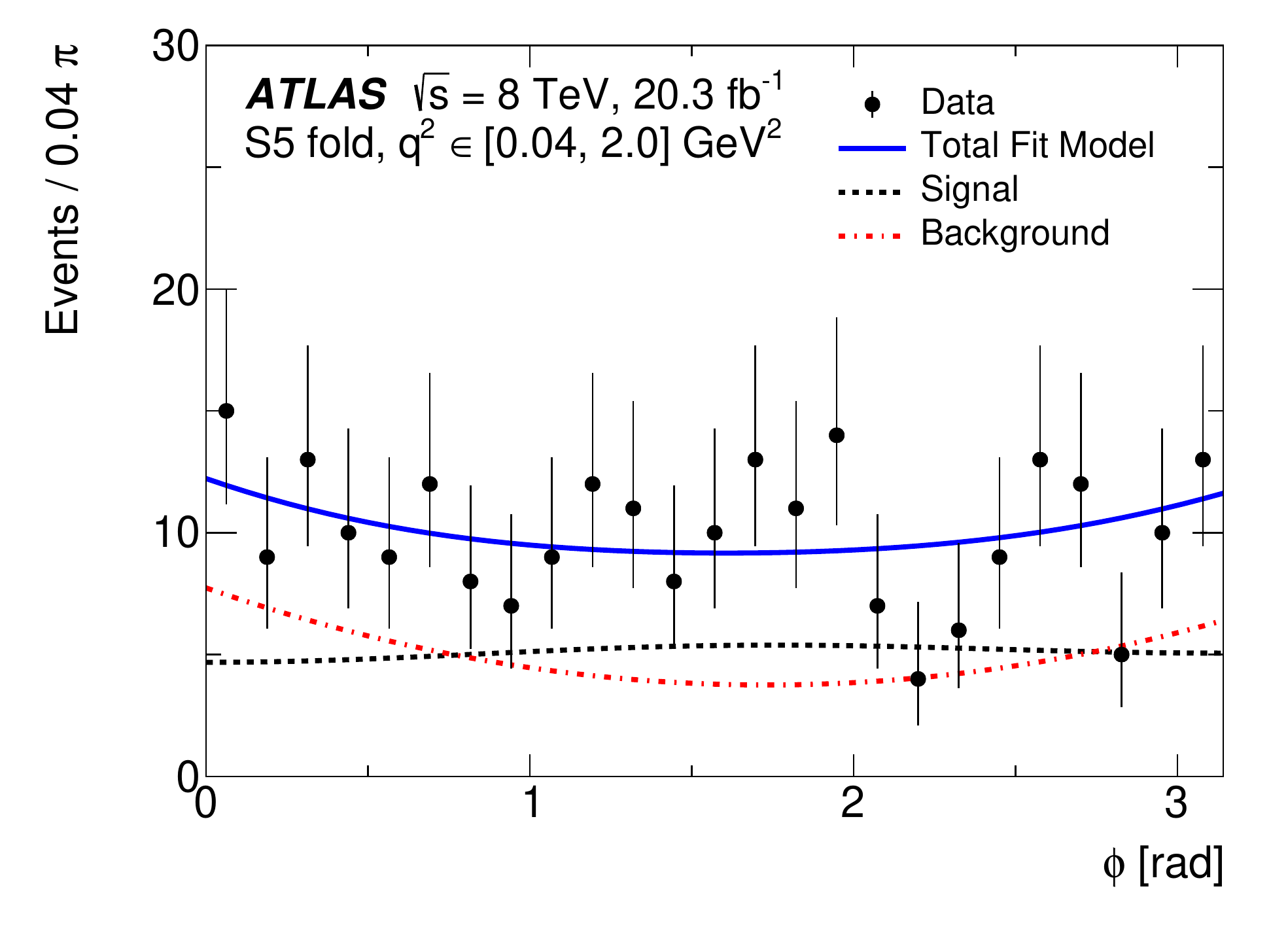}
}
}
\put(0,0){
\resizebox{\columnwidth}{!}{
  \includegraphics[width=0.45\columnwidth]{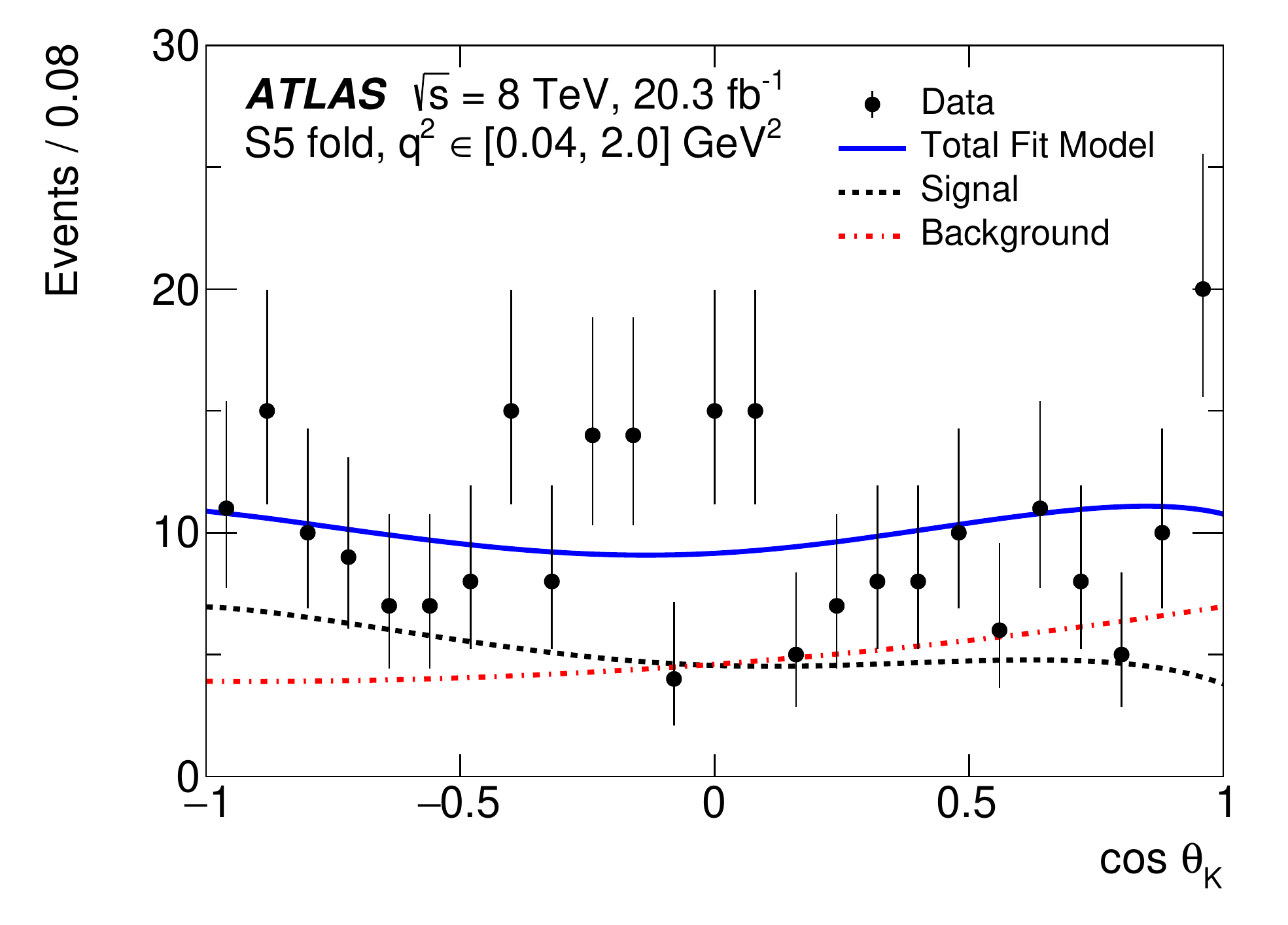}
  \includegraphics[width=0.45\columnwidth]{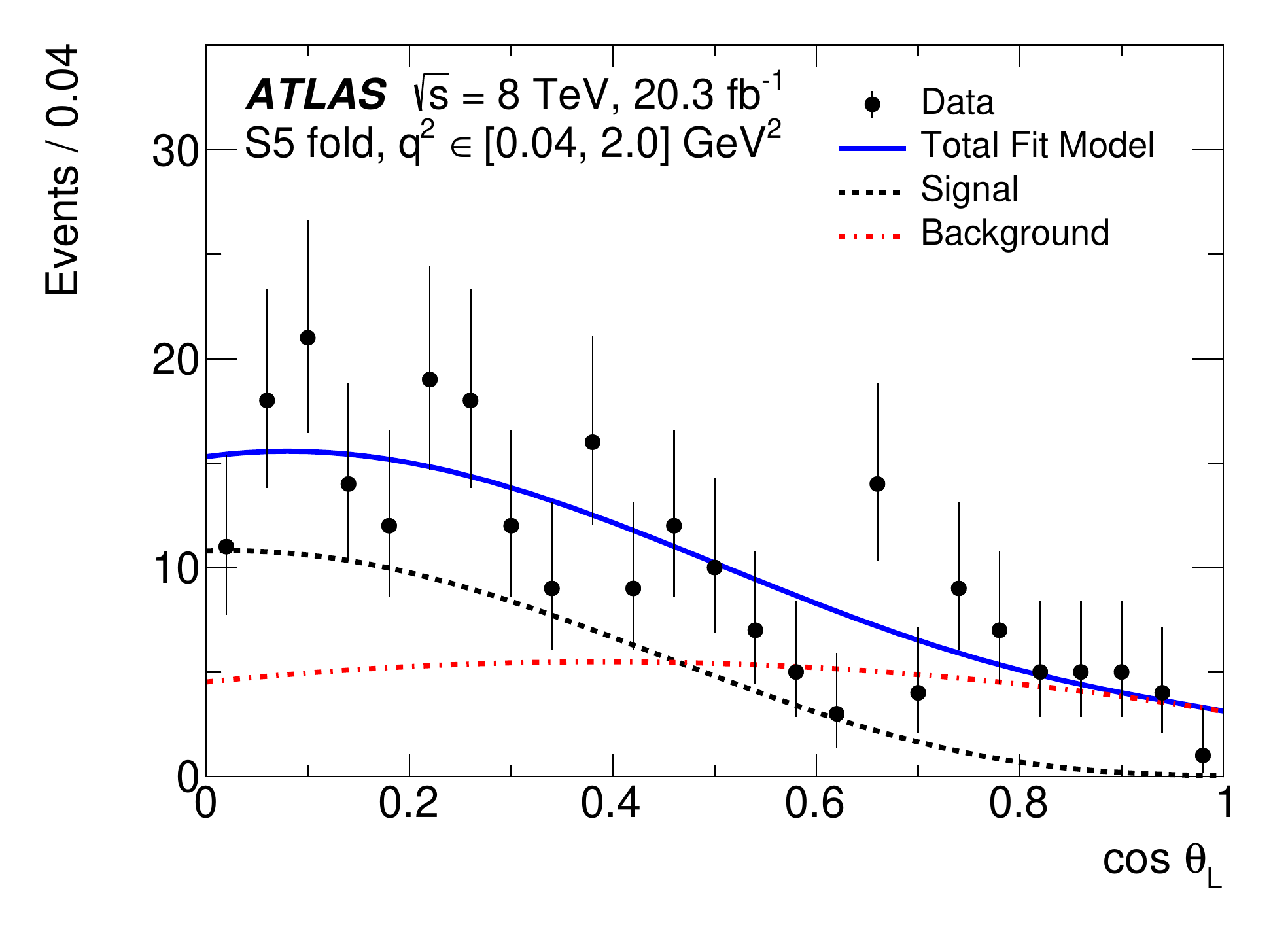}
}
}
\end{picture}
\end{center}
\caption{The distributions of (top left) $m_{K\pi\mu\mu}$,  (top right) $\phi$, (bottom left) \ctk, and (bottom right) \ctl obtained for $\qsqr\in\qsqrbinA$ \qsqrunits.
The (blue) solid line is a projection of the total pdf, the (red) dot-dashed line represents the background, and the (black) dashed line represents the signal component.  These plots are obtained from a fit using the \Sfive folding scheme. \label{fig:resultqsqrA}}
\end{figure}

\begin{figure}[!ht]
\begin{center}
 \begin{picture}(550,320)
\put(0,160){
\resizebox{\columnwidth}{!}{
  \includegraphics[width=0.45\columnwidth]{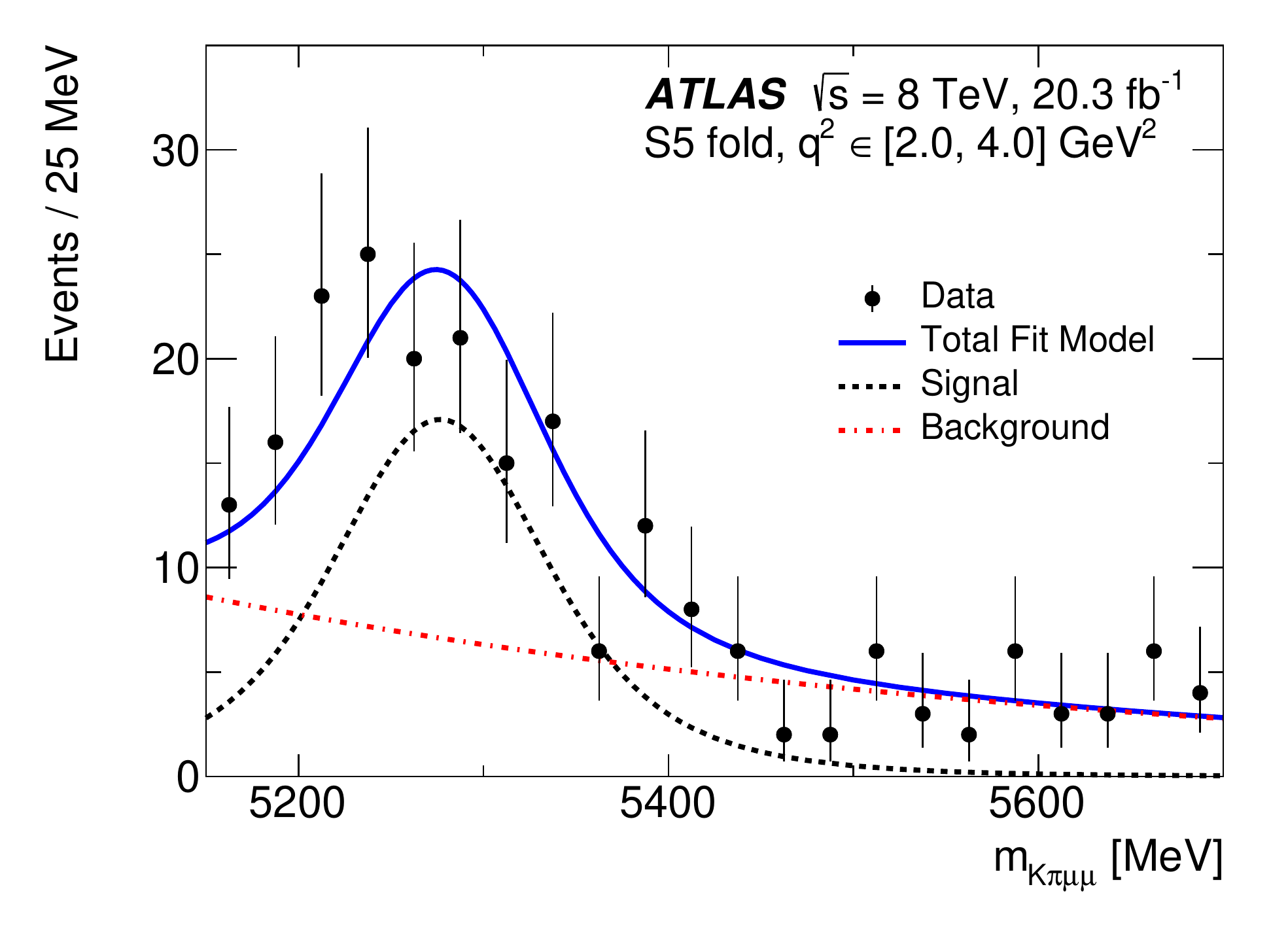}
  \includegraphics[width=0.45\columnwidth]{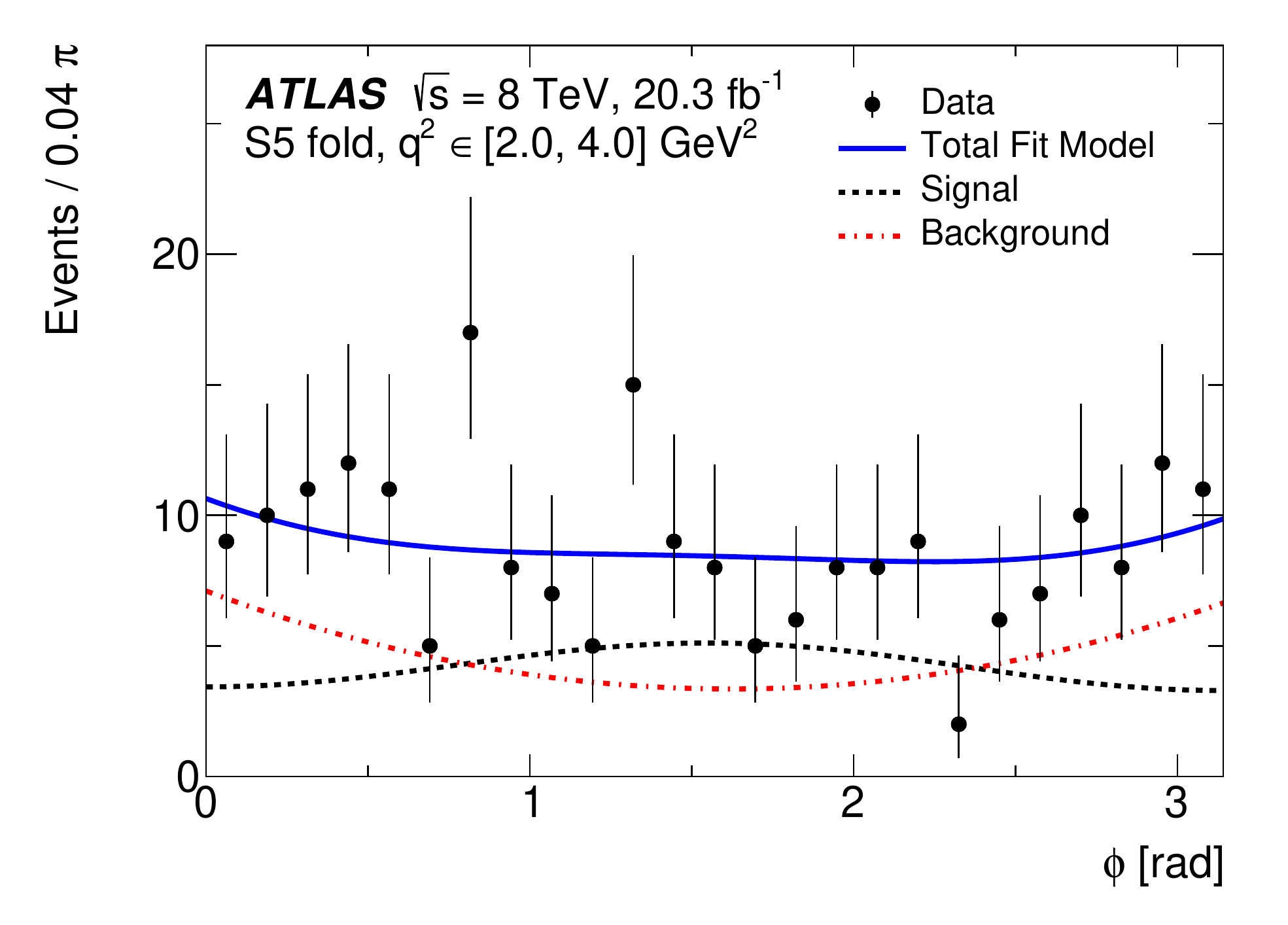}
 }
}
\put(0,0){
 \resizebox{\columnwidth}{!}{
  \includegraphics[width=0.45\columnwidth]{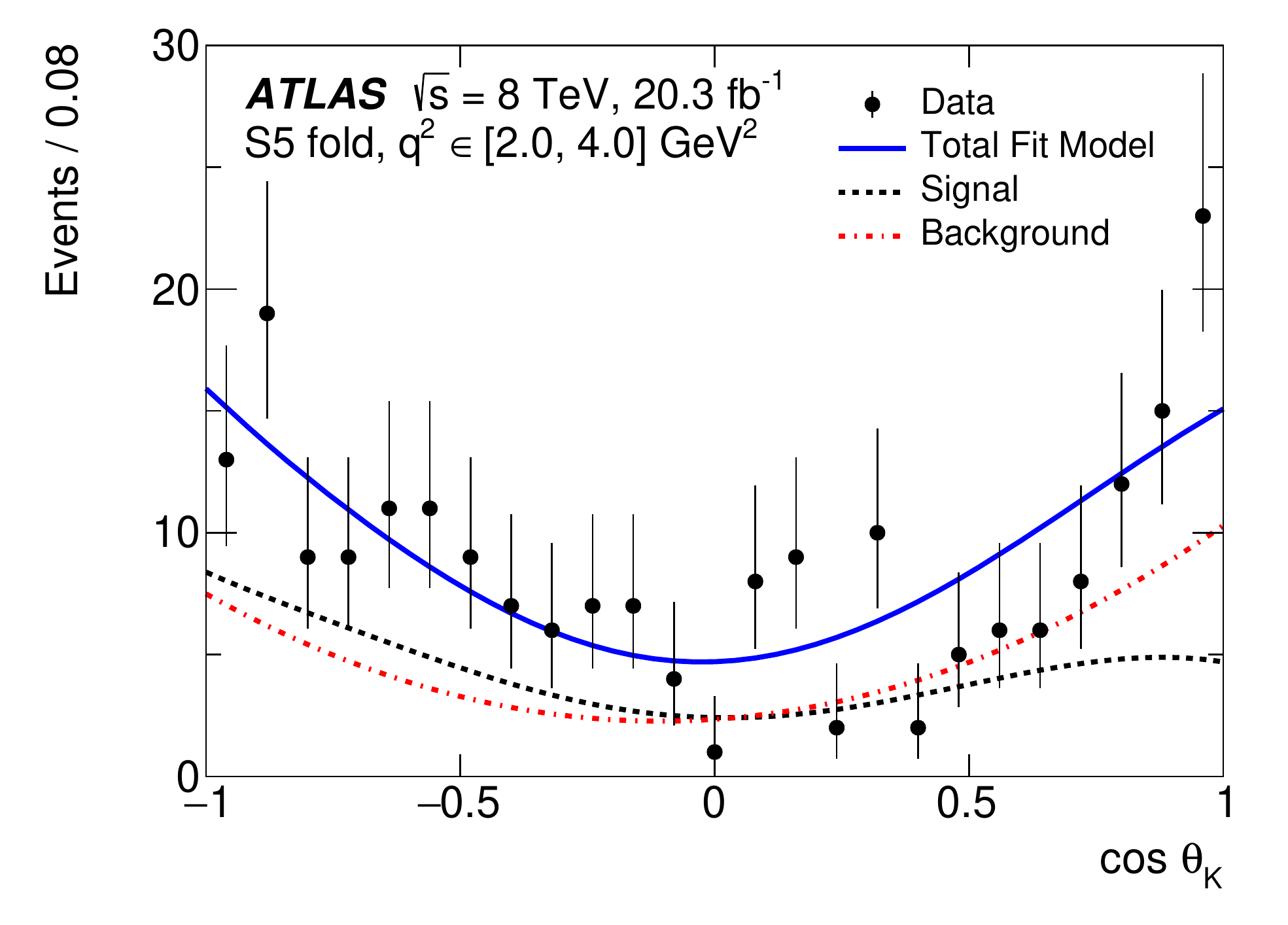}
  \includegraphics[width=0.45\columnwidth]{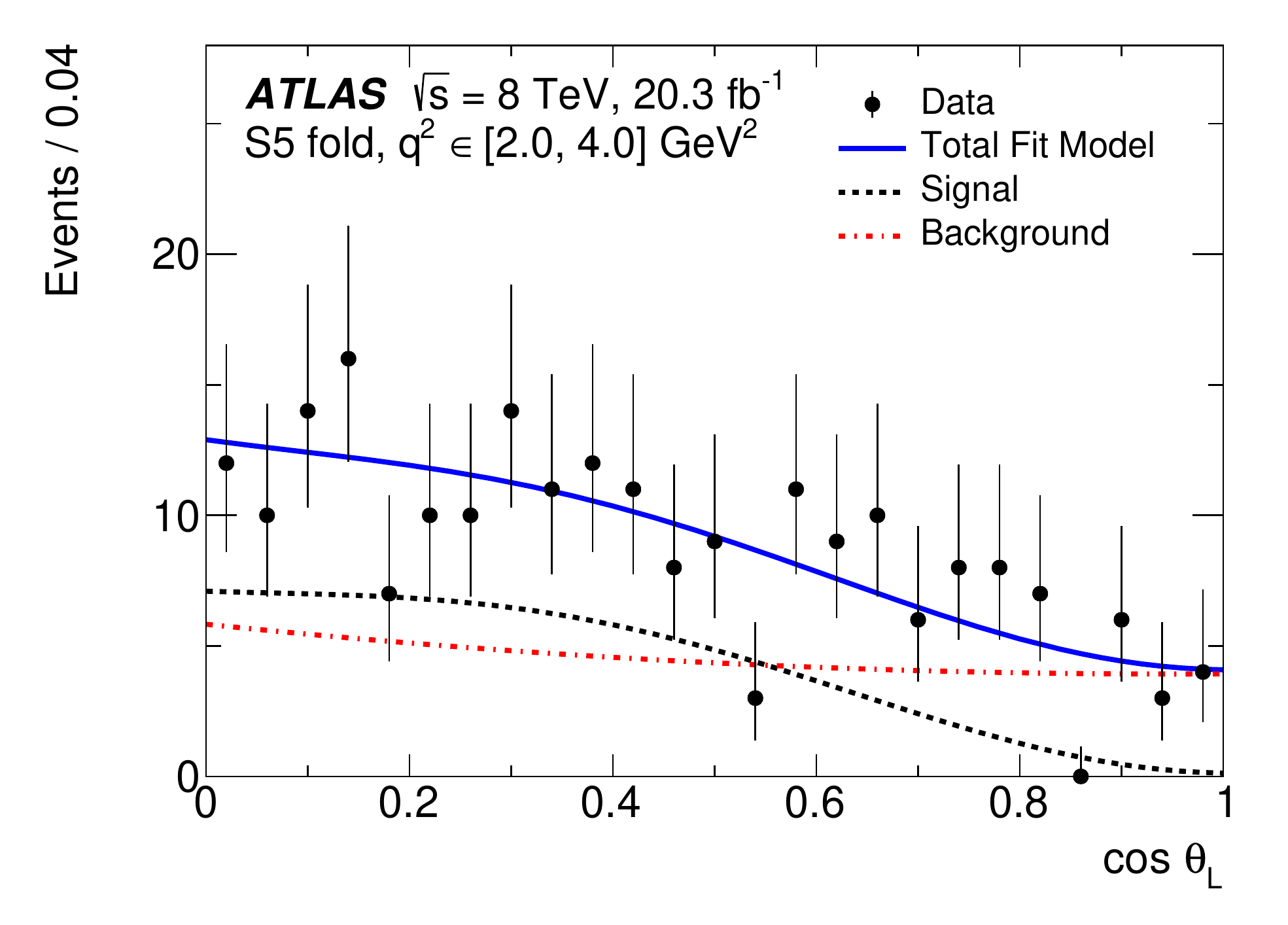}
 }
}
\end{picture}
\end{center}
\caption{The distributions of (top left) $m_{K\pi\mu\mu}$,  (top right) $\phi$, (bottom left) \ctk, and (bottom right) \ctl obtained for $\qsqr\in\qsqrbinB$ \qsqrunits.
The (blue) solid line is a projection of the total pdf, the (red) dot-dashed line represents the background, and the (black) dashed line represents the signal component.  These plots are obtained from a fit using the \Sfive folding scheme.
\label{fig:resultqsqrB}}
\end{figure}

\begin{figure}[!ht]
\begin{center}
 \begin{picture}(550,320)
\put(0,160){
  \resizebox{\columnwidth}{!}{
  \includegraphics[width=0.45\columnwidth]{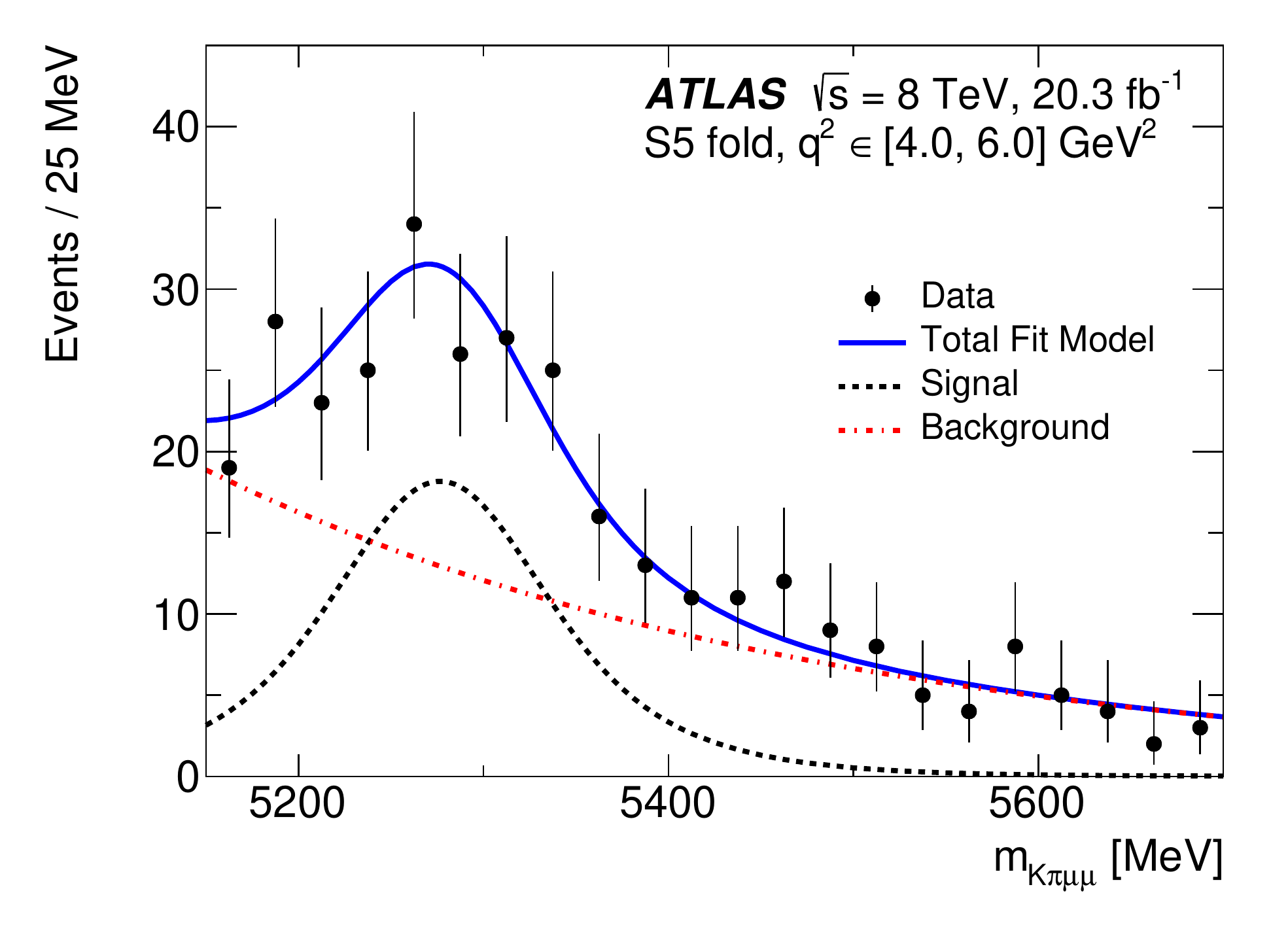}
  \includegraphics[width=0.45\columnwidth]{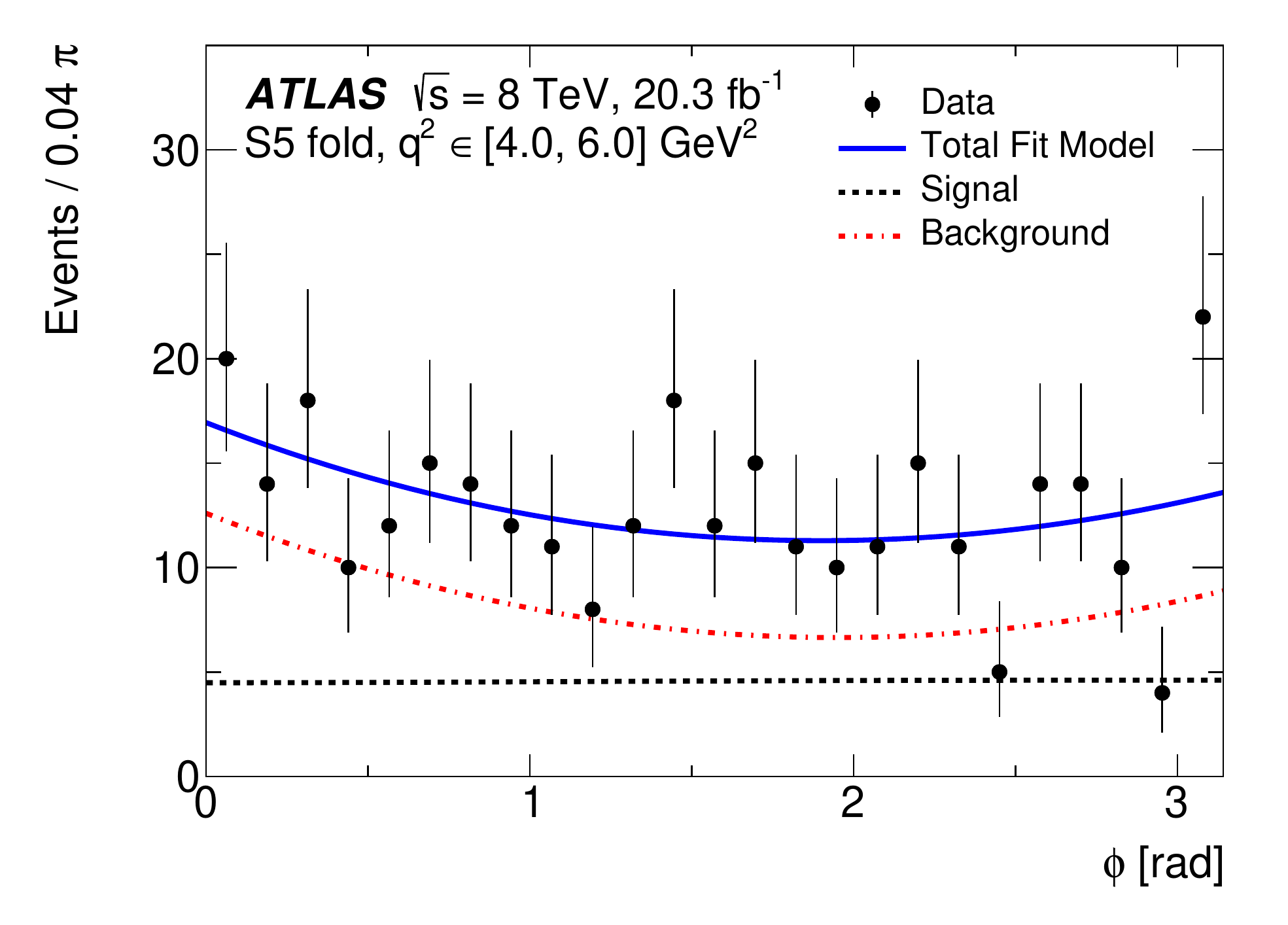}
 }
}
\put(0,0){
 \resizebox{\columnwidth}{!}{
  \includegraphics[width=0.45\columnwidth]{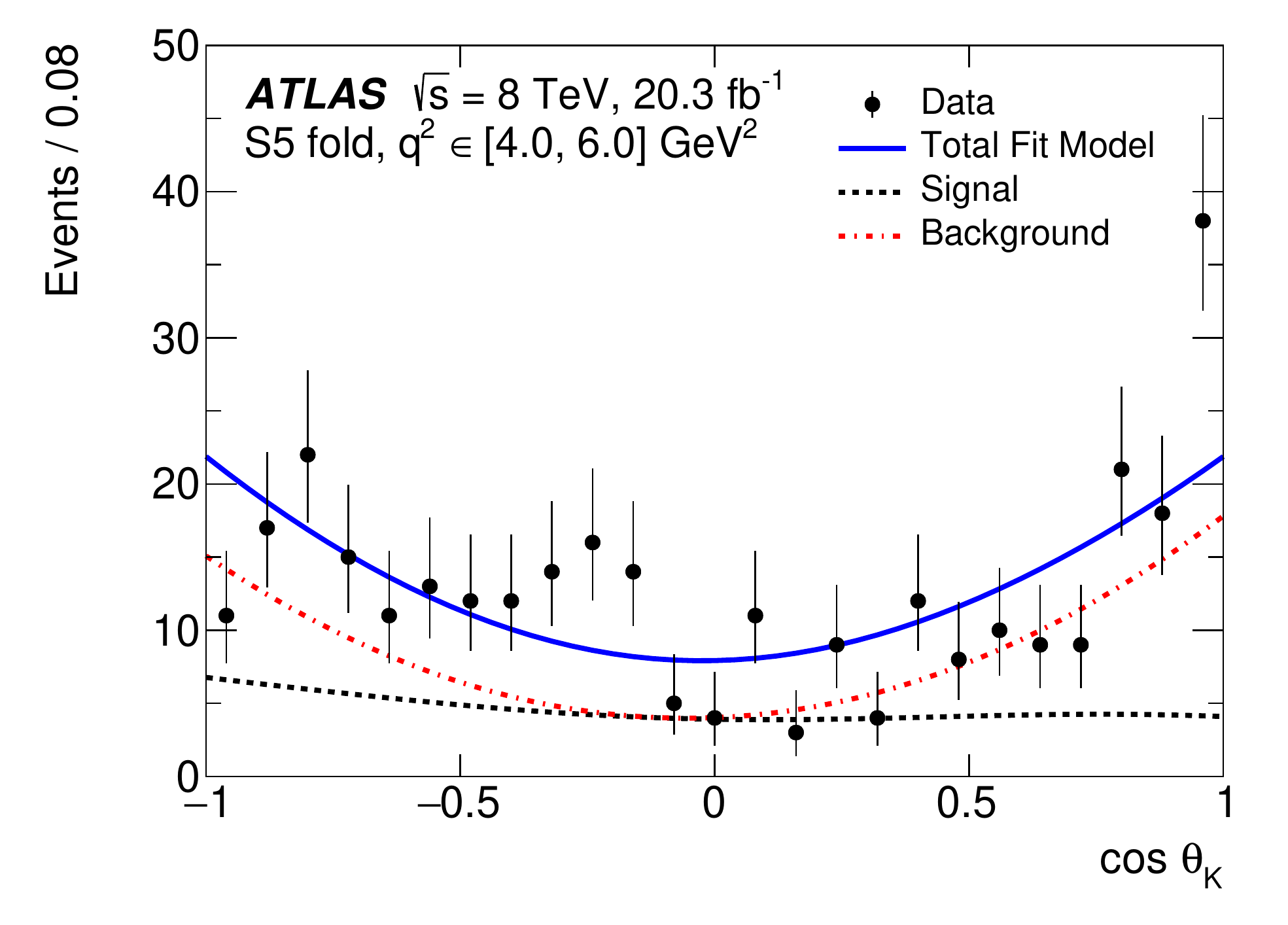}
  \includegraphics[width=0.45\columnwidth]{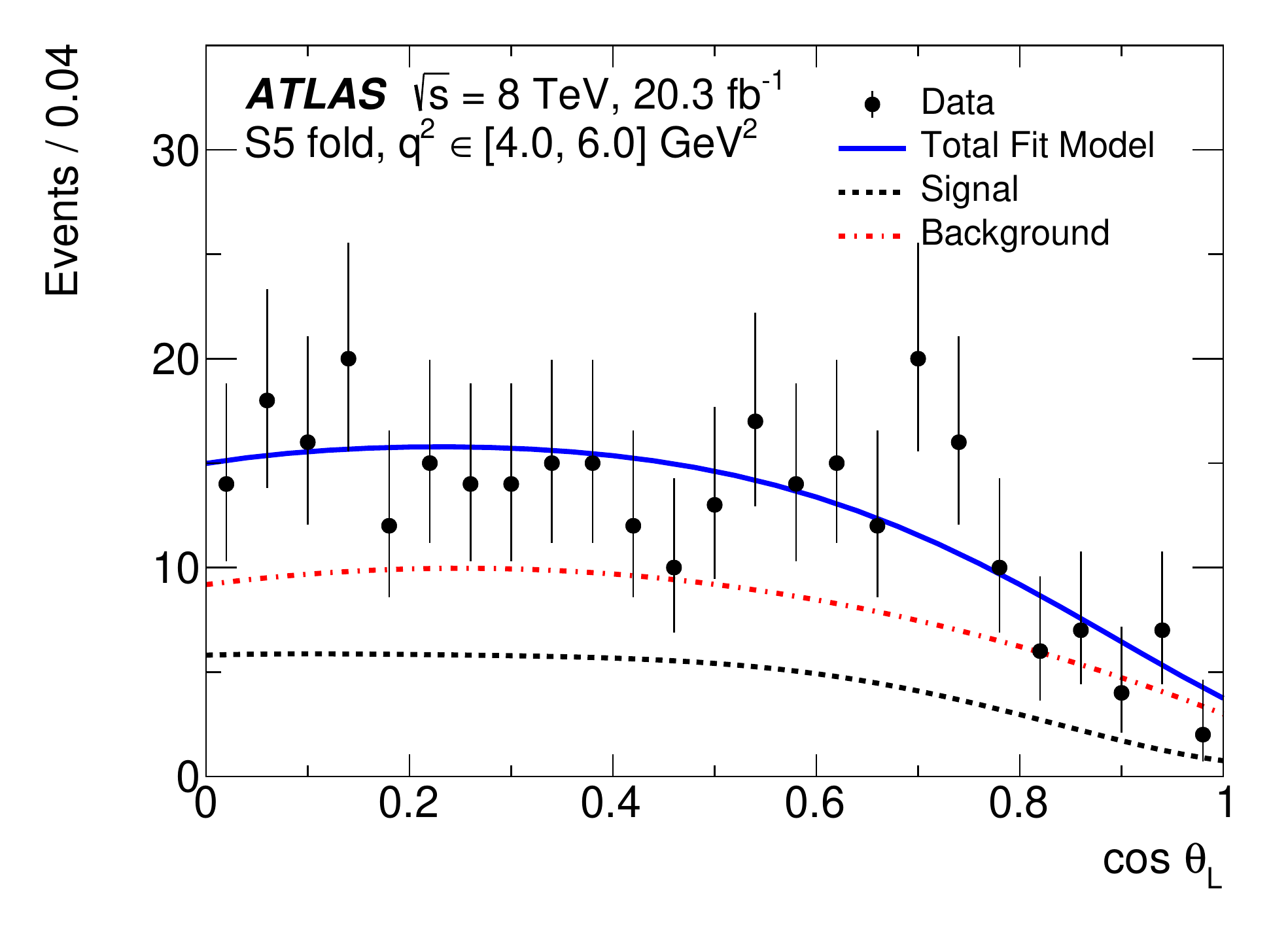}
 }
}
\end{picture}
\end{center}
\caption{
The distributions of (top left) $m_{K\pi\mu\mu}$,  (top right) $\phi$, (bottom left) \ctk, and (bottom right) \ctl obtained for $\qsqr\in\qsqrbinC$ \qsqrunits.
The (blue) solid line is a projection of the total pdf, the (red) dot-dashed line represents the background, and the (black) dashed line represents the signal component.  These plots are obtained from a fit using the \Sfive folding scheme.
\label{fig:resultqsqrC}}
\end{figure}

\begin{figure}[!ht]
\begin{center}
 \begin{picture}(550,320)
\put(0,160){
 \resizebox{\columnwidth}{!}{
  \includegraphics[width=0.45\columnwidth]{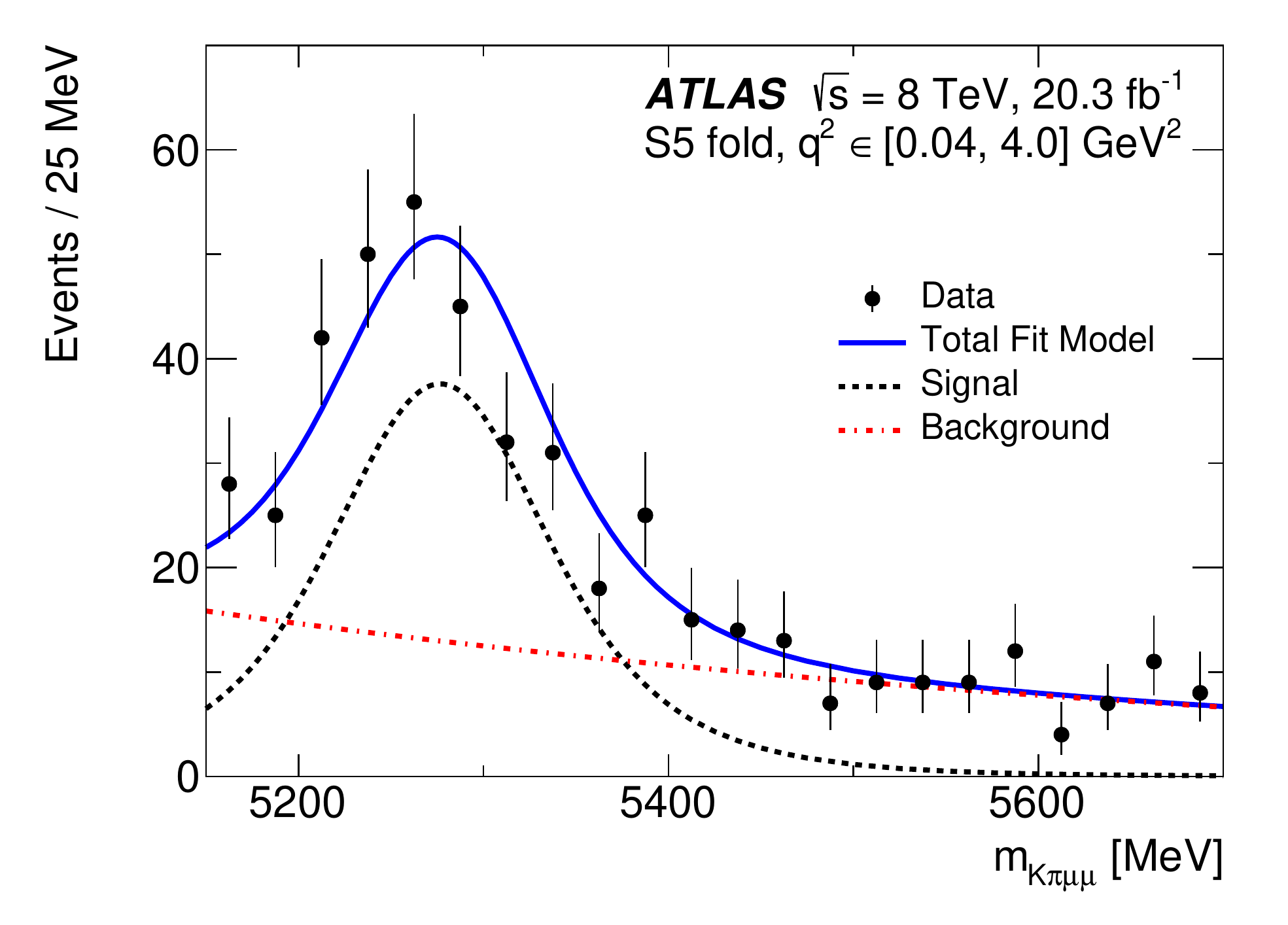}
  \includegraphics[width=0.45\columnwidth]{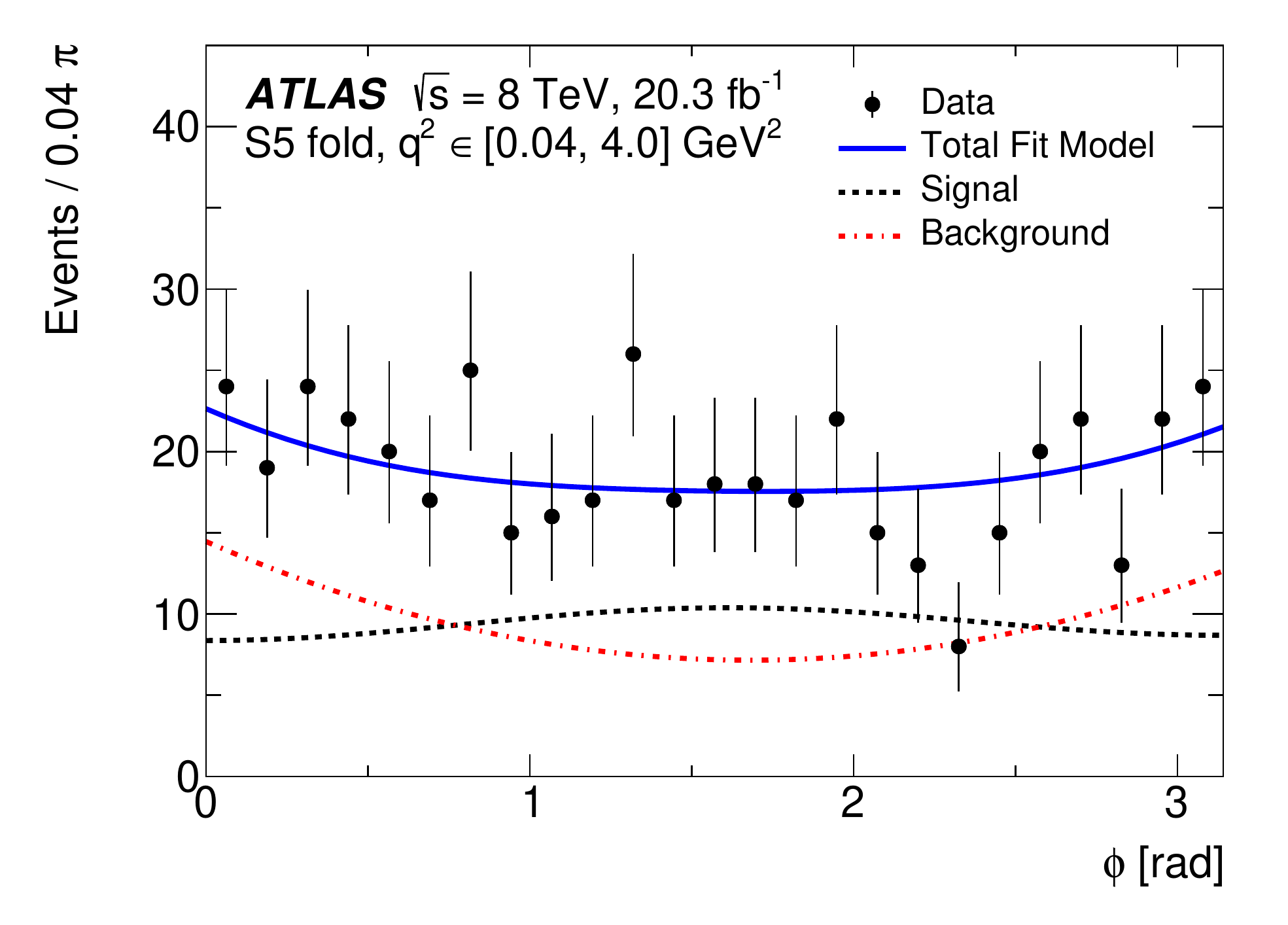}
 } 
}
\put(0,0){
 \resizebox{\columnwidth}{!}{
  \includegraphics[width=0.45\columnwidth]{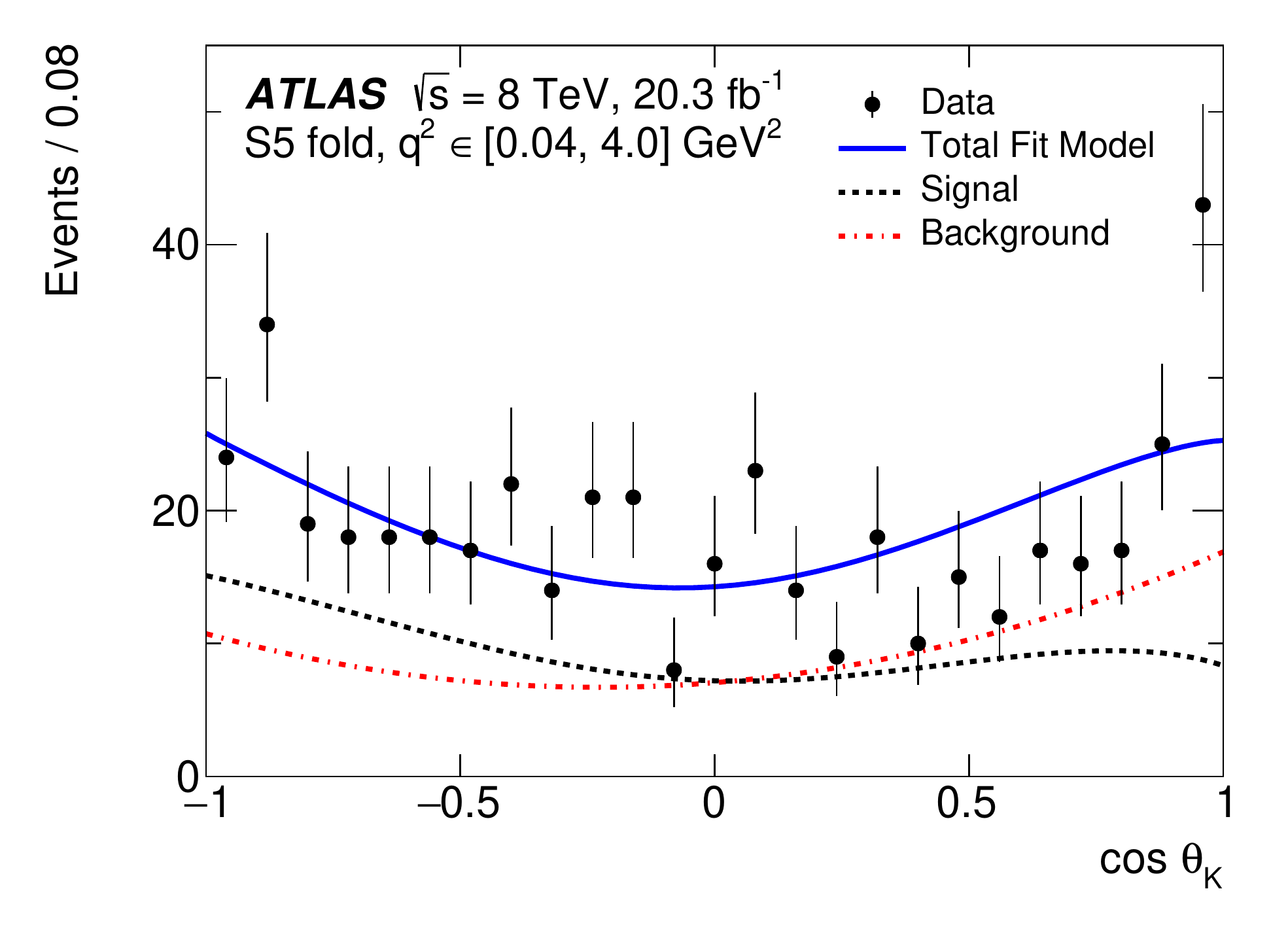}
  \includegraphics[width=0.45\columnwidth]{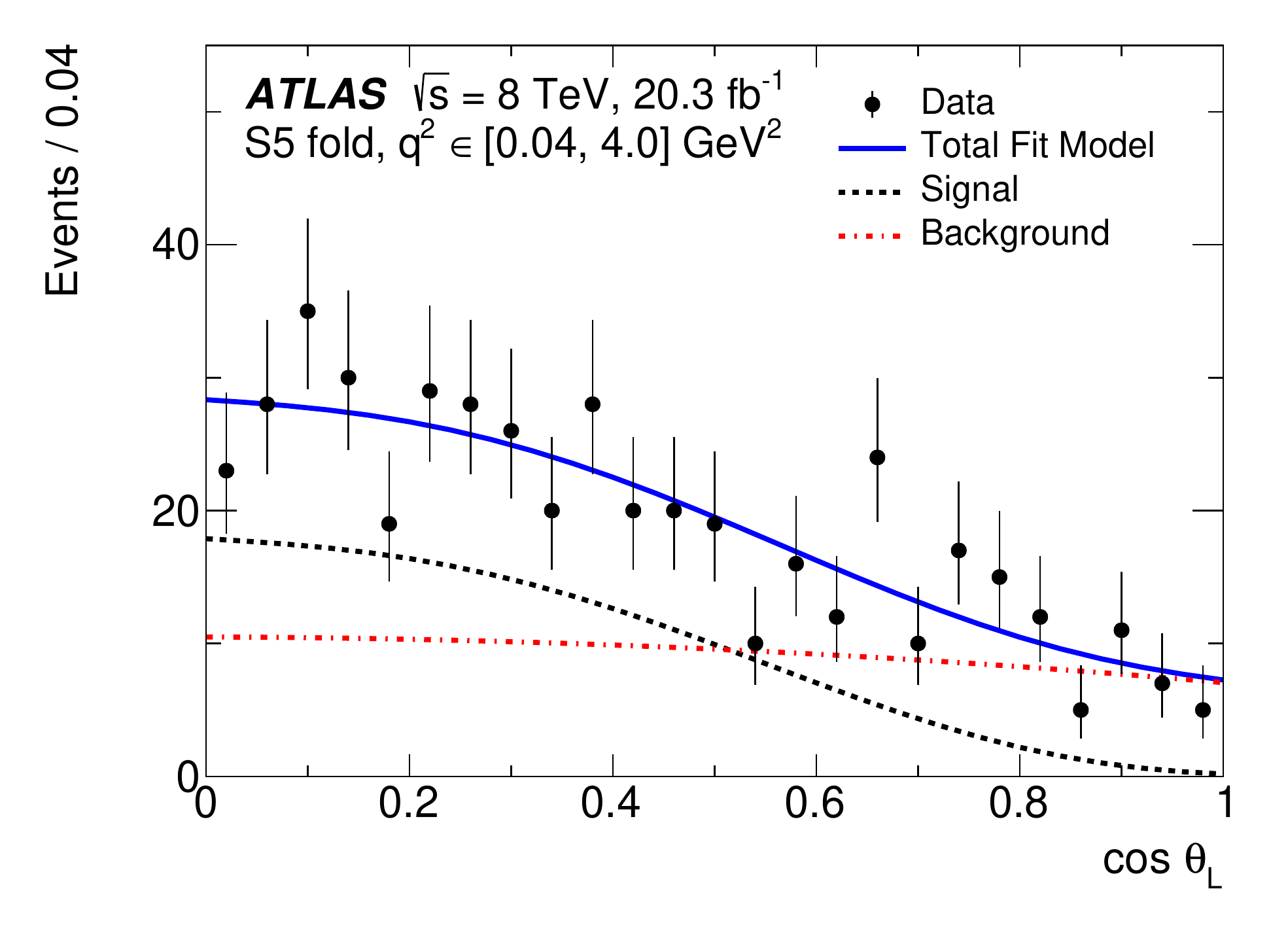}
 }
}
\end{picture}
\end{center}
\caption{
The distributions of (top left) $m_{K\pi\mu\mu}$,  (top right) $\phi$, (bottom left) \ctk, and (bottom right) \ctl obtained for $\qsqr\in\qsqrbinD$ \qsqrunits.
The (blue) solid line is a projection of the total pdf, the (red) dot-dashed line represents the background, and the (black) dashed line represents the signal component. These plots are obtained from a fit using the \Sfive folding scheme.
\label{fig:resultqsqrD}}
\end{figure}

\begin{figure}[!ht]
\begin{center}
 \begin{picture}(550,320)
\put(0,160){
 \resizebox{\columnwidth}{!}{
  \includegraphics[width=0.45\columnwidth]{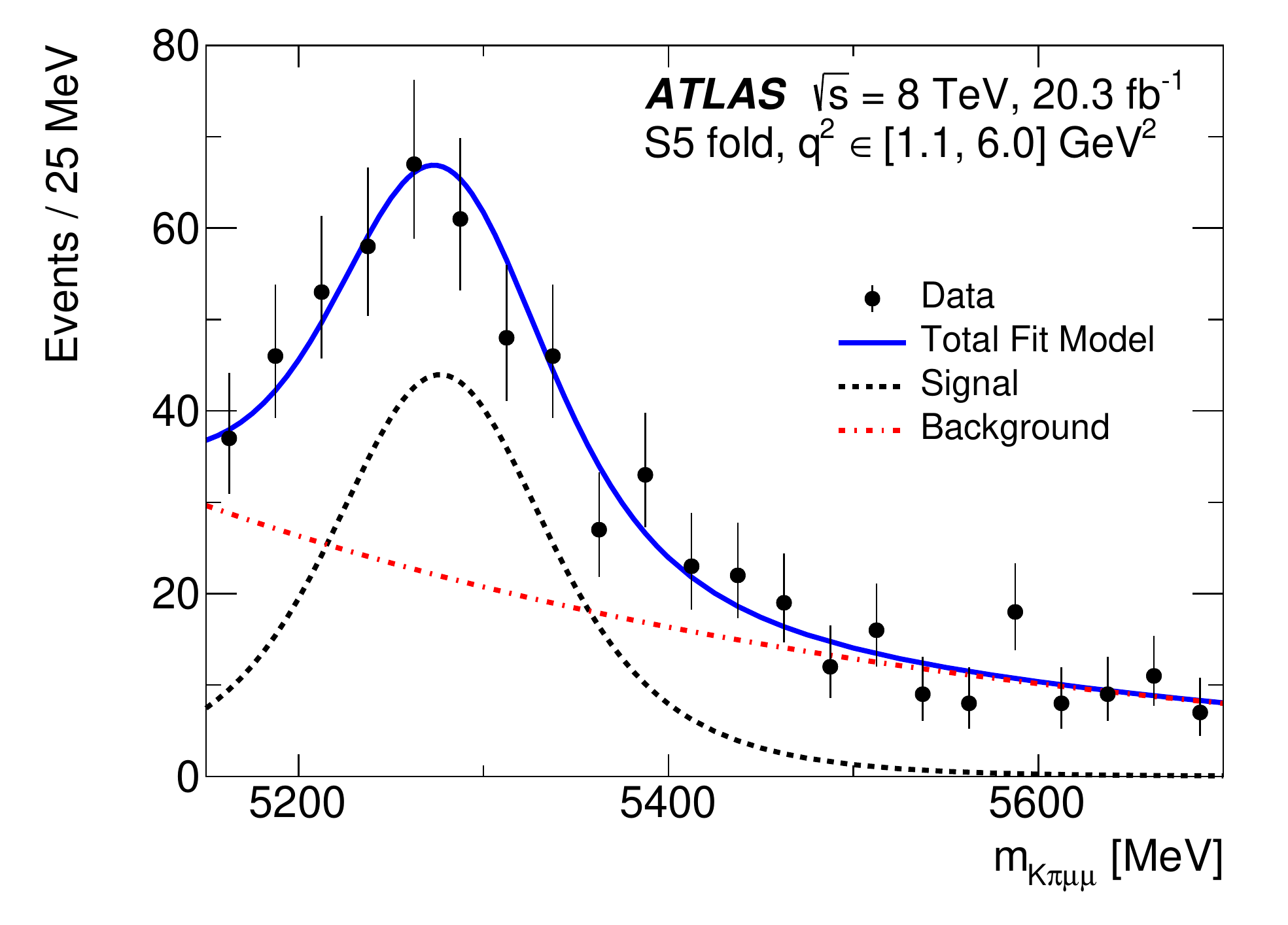}
  \includegraphics[width=0.45\columnwidth]{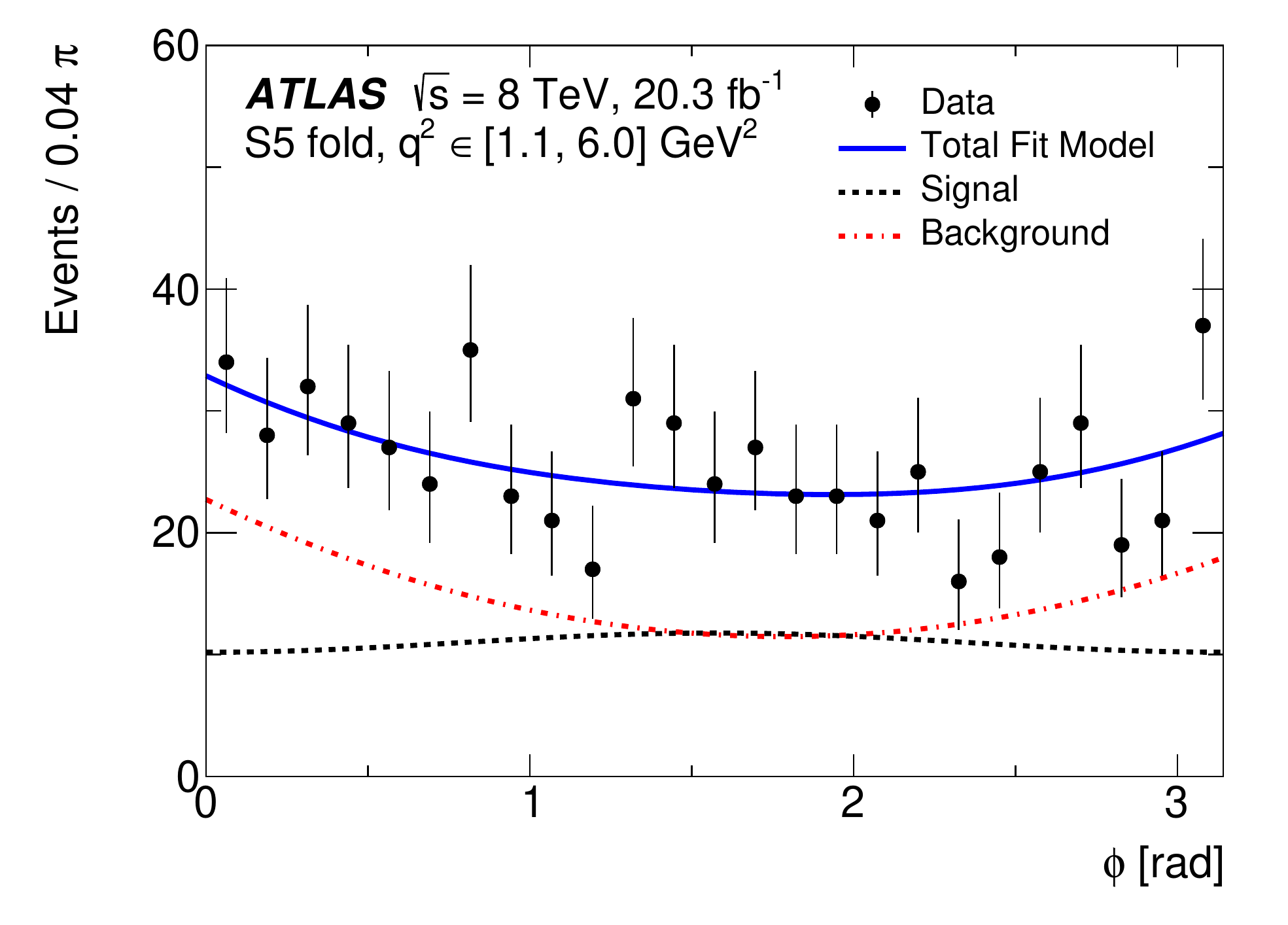}
 }
}
\put(0,0){
 \resizebox{\columnwidth}{!}{
  \includegraphics[width=0.45\columnwidth]{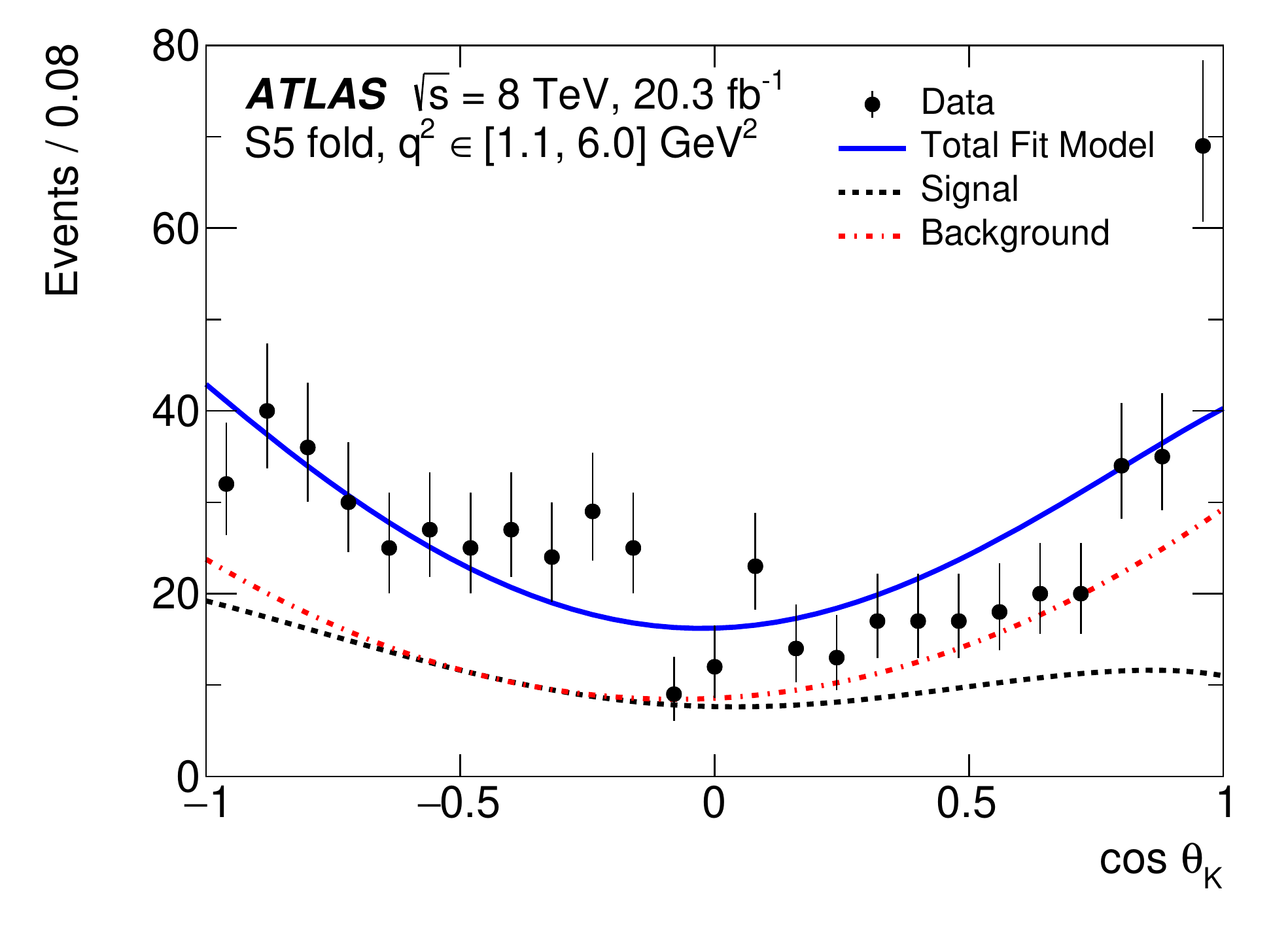}
  \includegraphics[width=0.45\columnwidth]{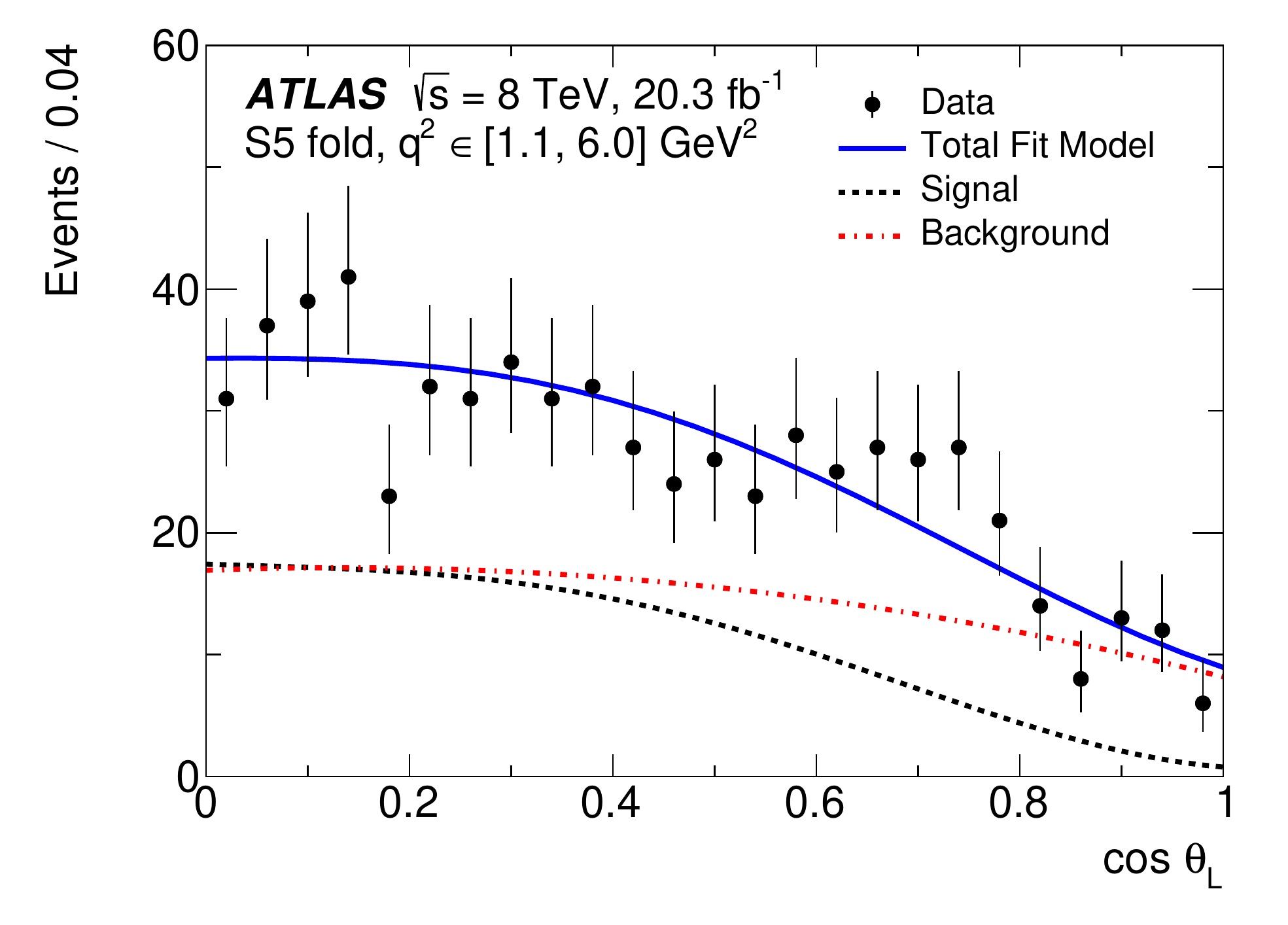}
 }
}
\end{picture}
\end{center}
\caption{
The distributions of (top left) $m_{K\pi\mu\mu}$,  (top right) $\phi$, (bottom left) \ctk, and (bottom right) \ctl obtained for $\qsqr\in\qsqrbinE$ \qsqrunits.
The (blue) solid line is a projection of the total pdf, the (red) dot-dashed line represents the background, and the (black) dashed line represents the signal component. These plots are obtained from a fit using the \Sfive folding scheme.
\label{fig:resultqsqrE}}
\end{figure}

\begin{figure}[!ht]
\begin{center}
 \begin{picture}(550,320)
\put(0,160){
 \resizebox{\columnwidth}{!}{
  \includegraphics[width=0.45\columnwidth]{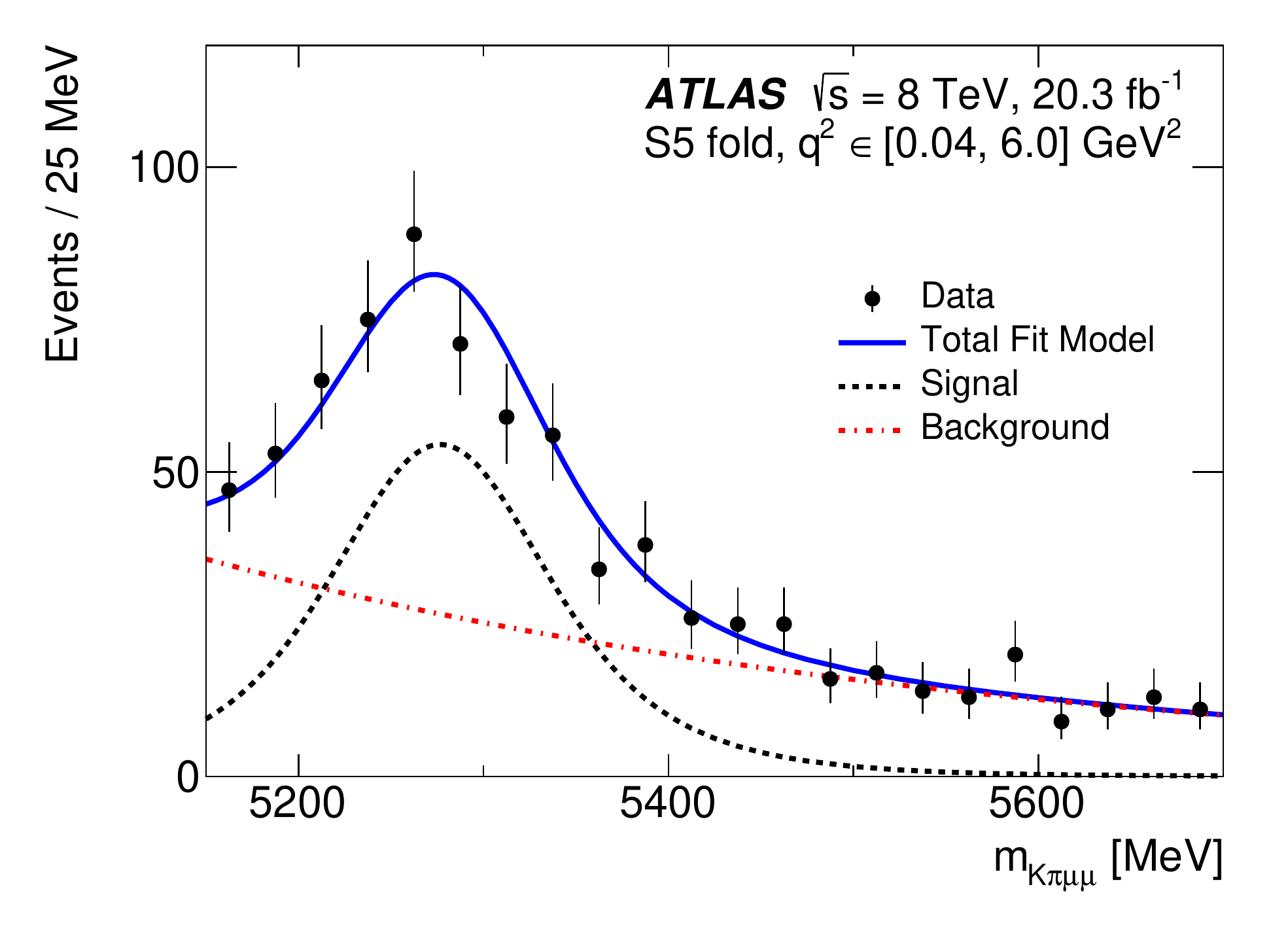}
  \includegraphics[width=0.45\columnwidth]{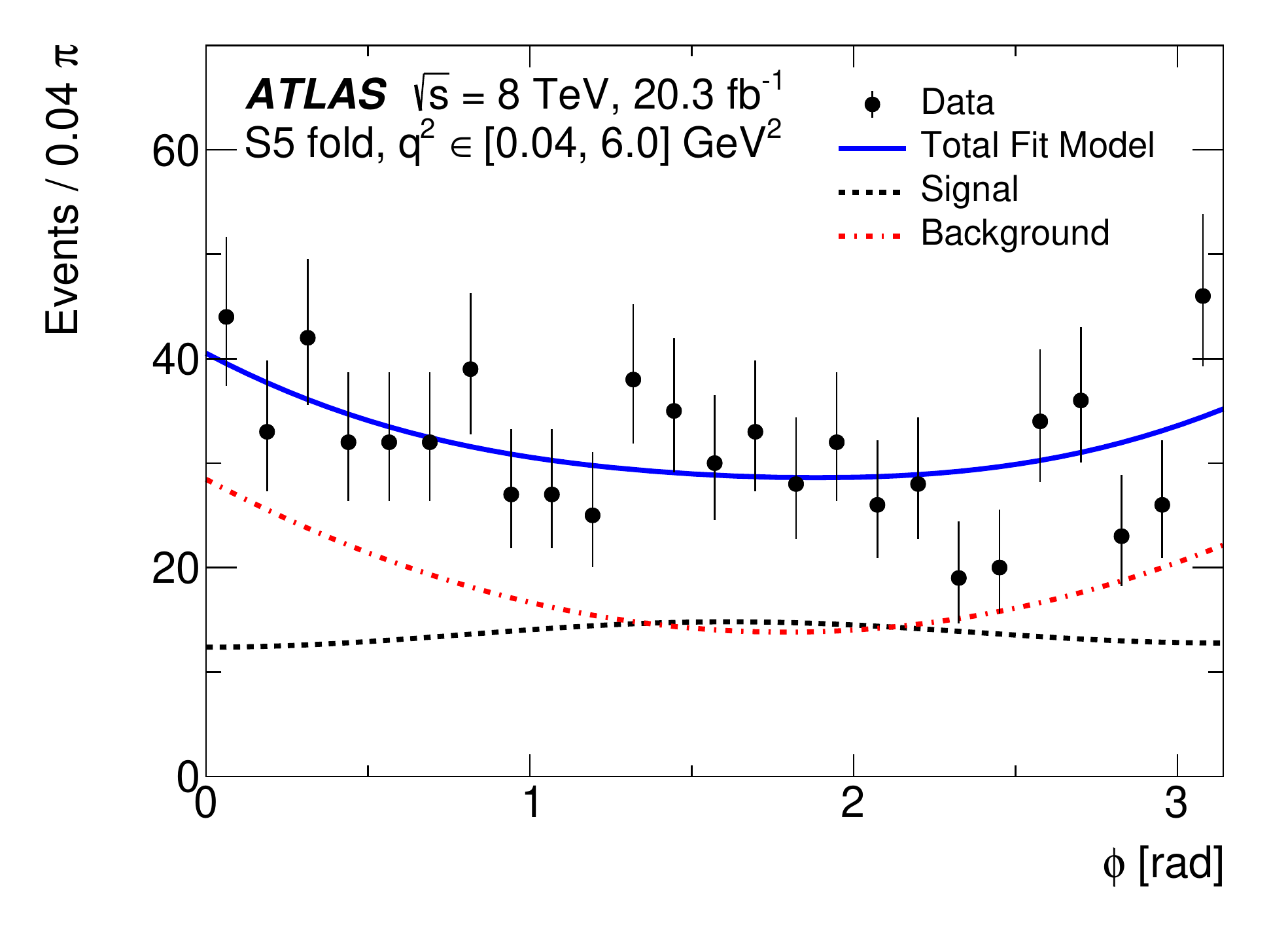}
 }
}
\put(0,0){
 \resizebox{\columnwidth}{!}{
  \includegraphics[width=0.45\columnwidth]{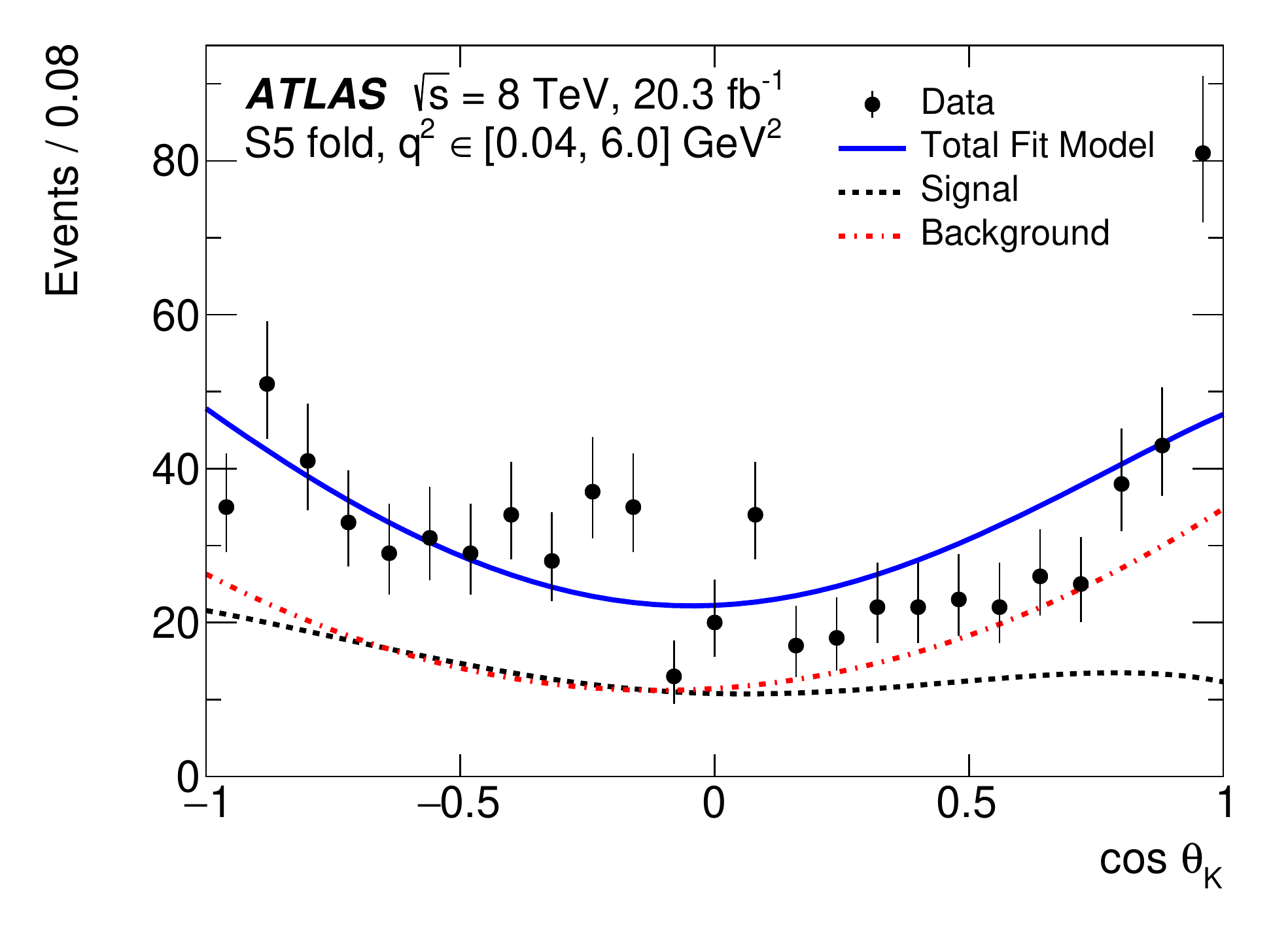}
  \includegraphics[width=0.45\columnwidth]{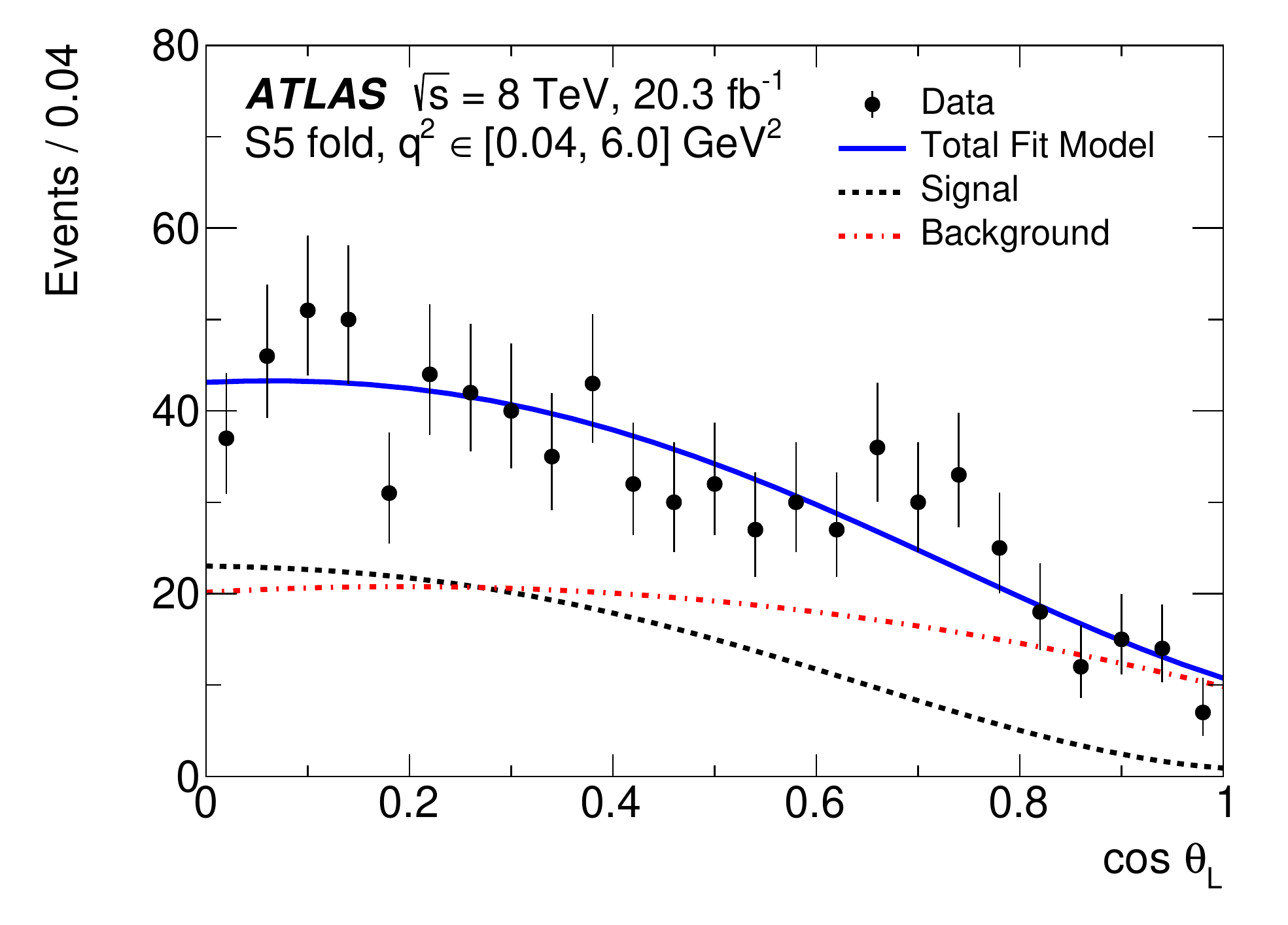}
 }
}
\end{picture}
\end{center}
\caption{
The distributions of (top left) $m_{K\pi\mu\mu}$,  (top right) $\phi$, (bottom left) \ctk, and (bottom right) \ctl obtained for $\qsqr\in\qsqrbinF$ \qsqrunits.
The (blue) solid line is a projection of the total pdf, the (red) dot-dashed line represents the background, and the (black) dashed line represents the signal component. These plots are obtained from a fit using the \Sfive folding scheme.
\label{fig:resultqsqrF}}
\end{figure}

The event yields obtained from the fits are summarised in Table~\ref{tbl:yields} where only 
statistical uncertainties are reported.
Figures~\ref{fig:resultqsqrA} through~\ref{fig:resultqsqrF} show for the different \qsqr bins the distributions of the variables used in the 
fit for the \Sfive folding scheme (corresponding to the transformation of Equation~(\ref{eq:s5transformation})) with the total, signal and background fitted pdfs superimposed. 
Similar sets of distributions are obtained for the three other folding schemes: \Sfour, \Sseven and \Seight.
The results of the angular fits to the data in terms of the $S_i$ and $P_j^{(\prime)}$ can be found in Tables~\ref{tbl:angularone} 
and \ref{tbl:angulartwo}. 
Statistical and systematic uncertainties are quoted in the tables.
The distributions of \fl and the $S_i$ parameters as a function of \qsqr are shown in Figure~\ref{fig:results:s} and those for 
$P_j^{(\prime)}$ are shown in Figure~\ref{fig:results}.  The correlations between \fl and the $S_i$ parameters and 
between \fl and the $P_j^{(\prime)}$ are given in Appendix~\ref{sec:correlations}.

\begin{table*} [!ht]
\renewcommand{\arraystretch}{1.2}
\caption{The values of fitted signal, $n_\mathrm{signal}$, and background, $n_\mathrm{background}$,
yields obtained for different bins in \qsq. 
The uncertainties indicated are statistical.} \label{tbl:yields}
{
\small
\begin{center}
\begin{tabular}{r|cc}\hline\hline
\qsq [$\qsqrunits$] & $n_\mathrm{signal}$      & $n_\mathrm{background}$  \\ \hline
\qsqrbinA           & $128 \pm 22$ & $122 \pm 22$  \\
\qsqrbinB           & $106 \pm 23$ & $113\pm 23$\\
\qsqrbinC           & $114\pm 24$ & $204\pm 26$\\ \hline
\qsqrbinD           & $236\pm 31$ & $233\pm 32$\\ 
\qsqrbinE           & $275\pm 35$ & $363\pm 36$\\
\qsqrbinF           & $342\pm 39$ & $445\pm 40$\\[0.2pt] 
\hline\hline
\end{tabular}
\end{center}
}
\end{table*}


\begin{sidewaystable} [!ht]
\renewcommand{\arraystretch}{1.2}
\caption{The values of
    \fl, and \Sthree, \Sfour, \Sfive, \Sseven and \Seight parameters obtained 
for different bins in \qsq.  The uncertainties indicated are statistical and systematic, respectively.
} \label{tbl:angularone}
{
\small
\begin{center}
\begin{tabular}{r|rrrrrr}\hline\hline
\qsq [$\qsqrunits$]& \multicolumn{1}{c}{\fl} & \multicolumn{1}{c}{\Sthree} & \multicolumn{1}{c}{\Sfour} & \multicolumn{1}{c}{\Sfive} & \multicolumn{1}{c}{\Sseven} & \multicolumn{1}{c}{\Seight} \\ \hline
\qsqrbinA        & $0.44\pm 0.08\pm 0.07$ & $-0.02\pm 0.09\pm 0.02$ & $0.15\pm 0.20 \pm 0.10$ & $0.33\pm  0.13\pm 0.08$ & $-0.09\pm 0.10\pm 0.02$ & $-0.14\pm 0.24\pm 0.09$\\
\qsqrbinB        & $0.64\pm 0.11\pm 0.05$ & $-0.15\pm 0.10\pm 0.07$ & $-0.37\pm 0.15\pm 0.10$ & $-0.16\pm 0.15\pm 0.06$ & $0.15\pm  0.14\pm 0.09$ & $0.52\pm  0.20\pm 0.19$\\
\qsqrbinC        & $0.42\pm 0.13\pm 0.12$ & $0.00\pm 0.12 \pm 0.07$ & $0.32\pm 0.16 \pm 0.09$ & $0.13\pm  0.18\pm 0.09$ & $0.03\pm  0.13\pm 0.07$ & $-0.12\pm 0.21\pm 0.05$\\ \hline
\qsqrbinD        & $0.52\pm 0.07\pm 0.06$ & $-0.05\pm 0.06\pm 0.04$ & $-0.15\pm 0.12\pm 0.09$ & $0.16\pm  0.10\pm 0.05$ & $0.01\pm  0.08\pm 0.05$ & $0.19\pm  0.16\pm 0.12$\\ 
\qsqrbinE        & $0.56\pm 0.07\pm 0.06$ & $-0.04\pm 0.07\pm 0.03$ & $0.03\pm 0.11 \pm 0.07$ & $0.00\pm  0.10\pm 0.04$ & $0.02\pm  0.08\pm 0.06$ & $0.11\pm  0.14\pm 0.10$\\
\qsqrbinF        & $0.50\pm 0.06\pm 0.04$ & $-0.04\pm 0.06\pm 0.03$ & $0.03\pm 0.10 \pm 0.07$ & $0.14\pm  0.09\pm 0.03$ & $0.02\pm  0.07\pm 0.05$ & $0.07\pm  0.13\pm 0.09$\\ 
\hline\hline
\end{tabular}
\end{center}
}
\vspace{3.0cm}

\renewcommand{\arraystretch}{1.2}
\caption{The values of \Pone, \Pfour, \Pfive, \Psix and \Peight parameters obtained 
for different bins in \qsq.  
The uncertainties indicated are statistical and systematic, respectively.
} \label{tbl:angulartwo}
{
\small
\begin{center}
\begin{tabular}{r|rrrrrr}\hline\hline
\qsq [$\qsqrunits$]& \multicolumn{1}{c}{\Pone} & \multicolumn{1}{c}{\Pfour} & \multicolumn{1}{c}{\Pfive} & \multicolumn{1}{c}{\Psix} & \multicolumn{1}{c}{\Peight} \\ \hline
\qsqrbinA        &  $-0.05\pm 0.30\pm 0.08$ & $0.31\pm 0.40 \pm 0.20$ & $0.67\pm 0.26 \pm 0.16$ & $-0.18\pm 0.21\pm 0.04$ & $-0.29\pm 0.48\pm 0.18$ \\
\qsqrbinB        &  $-0.78\pm 0.51\pm 0.34$ & $-0.76\pm 0.31\pm 0.21$ & $-0.33\pm 0.31\pm 0.13$ & $0.31\pm 0.28 \pm 0.19$ & $1.07\pm 0.41 \pm 0.39$ \\
\qsqrbinC        &  $0.14\pm 0.43 \pm 0.26$ & $0.64\pm 0.33 \pm 0.18$ & $0.26\pm 0.35 \pm 0.18$ & $0.06\pm 0.27 \pm 0.13$ & $-0.24\pm 0.42\pm 0.09$ \\ \hline
\qsqrbinD        &  $-0.22\pm 0.26\pm 0.16$ & $-0.30\pm 0.24\pm 0.17$ & $0.32\pm 0.21 \pm 0.11$ & $0.01\pm 0.17 \pm 0.10$ & $0.38\pm 0.33 \pm 0.24$ \\ 
\qsqrbinE        &  $-0.17\pm 0.31\pm 0.13$ & $0.05\pm 0.22 \pm 0.14$ & $0.01\pm 0.21 \pm 0.08$ & $0.03\pm 0.17 \pm 0.12$ & $0.23\pm 0.28 \pm 0.20$ \\
\qsqrbinF        &  $-0.15\pm 0.23\pm 0.10$ & $0.05\pm 0.20 \pm 0.14$ & $0.27\pm 0.19 \pm 0.06$ & $0.03\pm 0.15 \pm 0.10$ & $0.14\pm 0.27 \pm 0.17$ \\ 
\hline\hline
\end{tabular}
\end{center}
}
\end{sidewaystable}

\clearpage

\section{Systematic uncertainties}
\label{sec:systematic}
Systematic uncertainties in the parameter values obtained from the angular analysis come from several sources.
The methods for determining these uncertainties are based either on a comparison of nominal and modified fit results, or on
observed fit biases in modified pseudo-experiments.
The systematic uncertainties are symmetrised.
The most significant ones are described in the following, in decreasing order of
importance.
\begin{itemize}
%
%
\item
A systematic uncertainty is assigned for the combinatorial $K\pi$ (fake \kstar) background 
peaking at \ctk values around $1.0$ obtained by comparing results of the nominal fit to 
that where data above $\ctk=0.9$ are excluded from the fit.
\item
A systematic uncertainty is derived to account for background arising from partially reconstructed
$B \to D^0/D^+/D^+_s X $ decays, that manifest in an accumulation of events at $|\ctl|$ values 
around $0.7$.
Two-track or three-track combinations are formed from the signal candidate tracks, and are reconstructed assuming the 
pion or kaon mass hypothesis. A veto is then applied for events in which a track combination has a mass in a 
window of 30~\mev around the $D^0$, $D^+$ or $D^+_s$ meson mass.
Similarly, a veto is implemented to reject $\Bu\to K^+ \mumu $ and $\Bu\to \pi^+ \mumu $ events that pass the 
event selection. Here \Bp candidates are reconstructed from one of the hadrons from the \kstar candidate 
and the muons in the signal candidate. Signal candidates that have a three-track mass within 50~\mev
of the \Bp mass are excluded from the fit.
A few percent of signal events are removed when applying these vetoes,
with a corresponding effect on the acceptance distributions.
The fit results obtained from the data samples with vetoes applied are compared to those
obtained from the nominal fit and the change in each result is taken as the systematic uncertainty
from these backgrounds. 
This systematic uncertainty dominates the measurement of \fl at higher values of \qsqr.
\item 
The combinatorial background pdf shape has an uncertainty arising from the choice of the
model.  For the mass distribution it is assumed that an exponential function
model is adequate; however, for the angular variables the data are 
re-fitted using third-order Chebychev polynomials.  The change 
from the nominal result is taken as the uncertainty from this source.
\item
The acceptance function is assumed to factorise into three separate components, 
for \ctk, \ctl and $\phi$. To validate this assumption, the signal simulated events are 
fitted with the acceptance function obtained from that same MC sample. Differences in the fit results from expectation are small
and taken as the uncertainty resulting from this assumption.
\item
A systematic uncertainty is assigned for the angular pdf model for the background by comparing the nominal result to that with a reduced fit range of 
$m_{K\pi\mu\mu} \in [5200, 5700]\ \mev$, in particular to account for possible residues of the partially reconstructed $B$-decays.
\item
A correction is applied to the data by shifting the track $\pT$ according to
the uncertainties arising from biases in rapidity and momentum scale. 
The change in results obtained is ascribed 
to the uncertainty in the ID alignment and knowledge of the magnetic field.
\item
The maximum-likelihood estimator used is intrinsically
biased.  Ensembles of MC simulated events are used in order to 
ascertain the bias in the extracted values of the parameters of interest.
The bias is assigned as a systematic uncertainty.
\item 
The $\pT$ spectrum of \Bd candidates observed in data is not accurately reproduced by the MC simulation. 
This difference in the kinematics results in a slight modification of the acceptance functions. This
is accounted for by reweighting signal MC simulated events to resemble the $\pT$ spectrum found in data.
The change in fitted parameter values
obtained due to the reweighting is taken as the systematic uncertainty resulting from
this difference.
\item
The signal decay mode is resonant $\kstar\to K\pi$ decay, but 
scalar contributions from non-resonant $K\pi$ transitions may also exist.  
The LHCb Collaboration reported an $S$-wave
contribution at the level of 5\% of the signal~\cite{Aaij:2015oid,Aaij:2016flj}.
Ensembles of MC simulated events are fitted with 5\% of the signal being drawn
from an $S$-wave sample of events and the remaining 95\% from signal.
The observed change in fit bias is assigned as the systematic uncertainty from this 
source.  Any variation in $S$-wave content as a function of \qsqr would not
significantly affect the results reported here.
\item 
The values of the nuisance parameters of the fit model obtained from MC control samples and fits to the data mass distribution
have associated uncertainties.  These parameters include $m_0$, $\xi$, 
the signal and background yields, the shape parameter of the combinatorial background mass distribution,
and the parameters of the signal acceptance functions.
The uncertainty in the value of each of these parameters
is varied independently in order to assess the 
effect on parameters of interest.  This source of uncertainty has a small effect
on the measurements reported here. 
\item
Background from exclusive modes peaking in $m_{K\pi\mu\mu}$ is neglected in the nominal
fit. This may affect the fitted results and is accounted for by computing 
the fit bias obtained when embedding MC simulated samples
of $\Lambda_b\to \Lambda(1520) \mu^+\mu^-$, $\Lambda_b\to p K^- \mu^+\mu^-$,
$B^+\to K^{(*)+} \mu^+\mu^-$ and $B_s^0 \to \phi \mu^+\mu^-$ into 
ensembles of pseudo-data generated from the fit model containing 
only combinatorial background and signal components. The
change in fit bias observed when adding exclusive backgrounds is taken as the systematic
error arising from neglecting those modes in the fit.
\item
The difference from nominal results obtained when 
fitting the \Bd signal MC events with the acceptance function for \Bdbar is taken 
as an upper limit of the systematic error resulting from event migration due to mistagging the \Bd flavour. 
\item
The parameters $S_5$ and $S_8$, as well as the respective $P_j^{(\prime)}$ parameters are affected by
dilution and thus have a multiplicative scaling applied to them. 
This dilution factor depends on the kinematics of the \kstar decay and 
has a systematic uncertainty associated with it.
The effect of data/MC differences in the $\pT$ spectrum of \Bd candidates
on the mistag probability was studied and found to be negligible.
The uncertainty due to the limited number of MC events
is used to compute the statistical uncertainty of $\omega$ and $\overline\omega$.
Studies of MC simulated events indicate that there is no significant difference between the 
mistag probability for \Bd and \Bdbar events and the analysis assumes that the average
mistag probability provides an adequate description of this effect.  
The magnitude of the mistag probability difference, $|\omega -\overline \omega|$, is included as a systematic uncertainty
resulting from this assumption.
\end{itemize}
The total systematic uncertainties of the fitted $S_i$ and $P_j^{(\prime)}$ parameter values
are presented in Tables~\ref{tbl:angularone} and \ref{tbl:angulartwo}, where the dominant 
contributions for \fl come from the modelling of the angular distributions of the combinatorial background
and the partially reconstructed decays peaking in \ctk and \ctl.  These contributions
and in addition also ID alignment and magnetic field calibration affect \Sthree (\Pone).
The largest systematic uncertainty contribution to \Sthree (\Pone) comes from 
partially reconstructed decays entering the signal region.
This also affects the measurement of \Sfive (\Pfive) and \Sseven (\Psix).  
The partially reconstructed decays peaking in \ctl affect the
measurement of \Sfour (\Pfour) and \Seight (\Peight), whereas the fake \kstar background in
\ctk affects \Sfour (\Pfour), \Sfive (\Pfive), and \Seight (\Peight). The parameterization
of the signal acceptance is another significant systematic uncertainty source for 
\Sfour (\Pfour).  The systematic uncertainties are smaller than the statistical uncertainties
for all parameters measured.

\section{Comparison with theoretical computations}
\label{sec:comparison}

\begin{figure}[!ht]
\begin{center}
\resizebox{\columnwidth}{!}{
  \includegraphics[width=0.45\columnwidth]{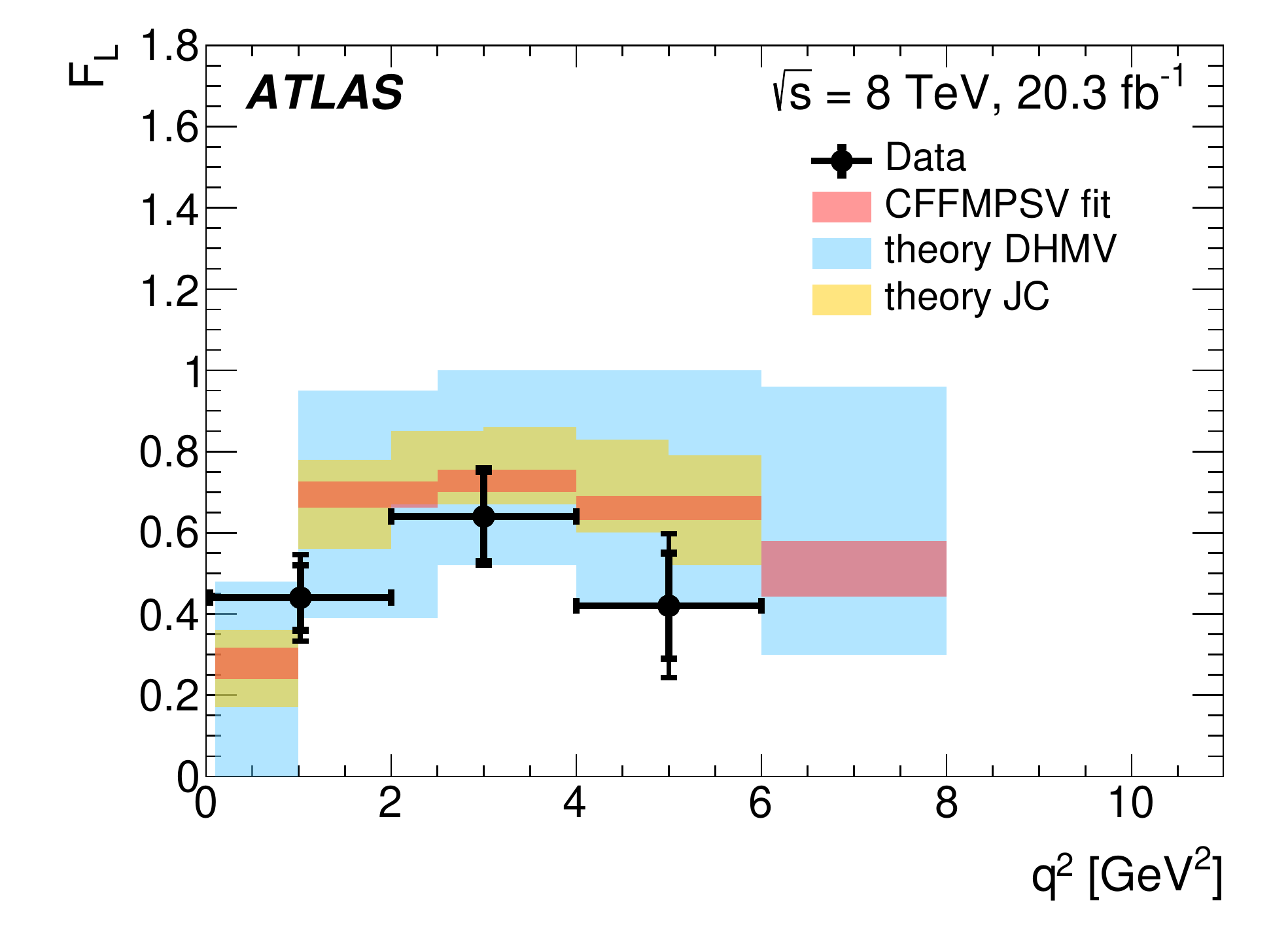}
  \includegraphics[width=0.45\columnwidth]{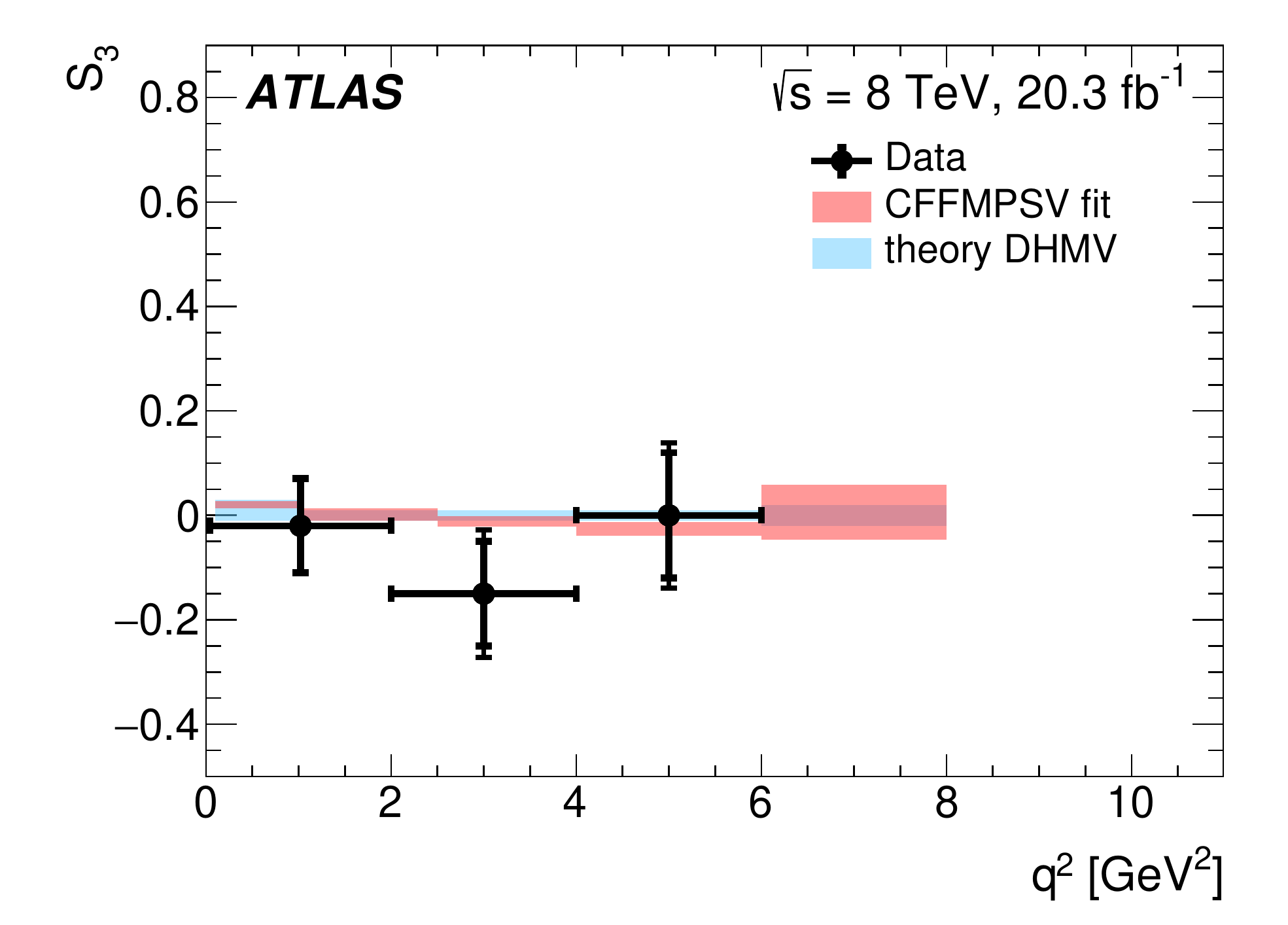}
}
\resizebox{\columnwidth}{!}{
  \includegraphics[width=0.45\columnwidth]{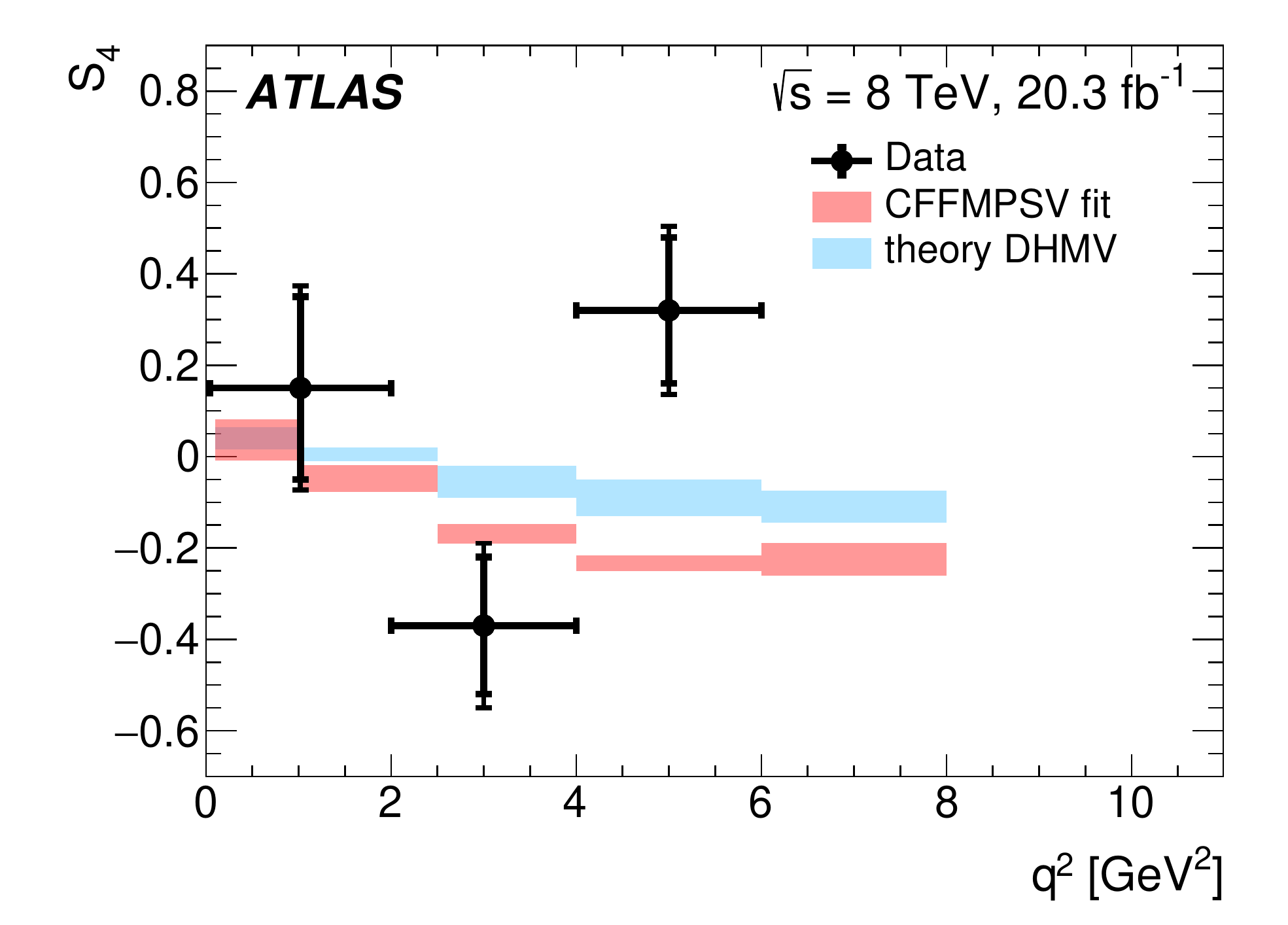}
  \includegraphics[width=0.45\columnwidth]{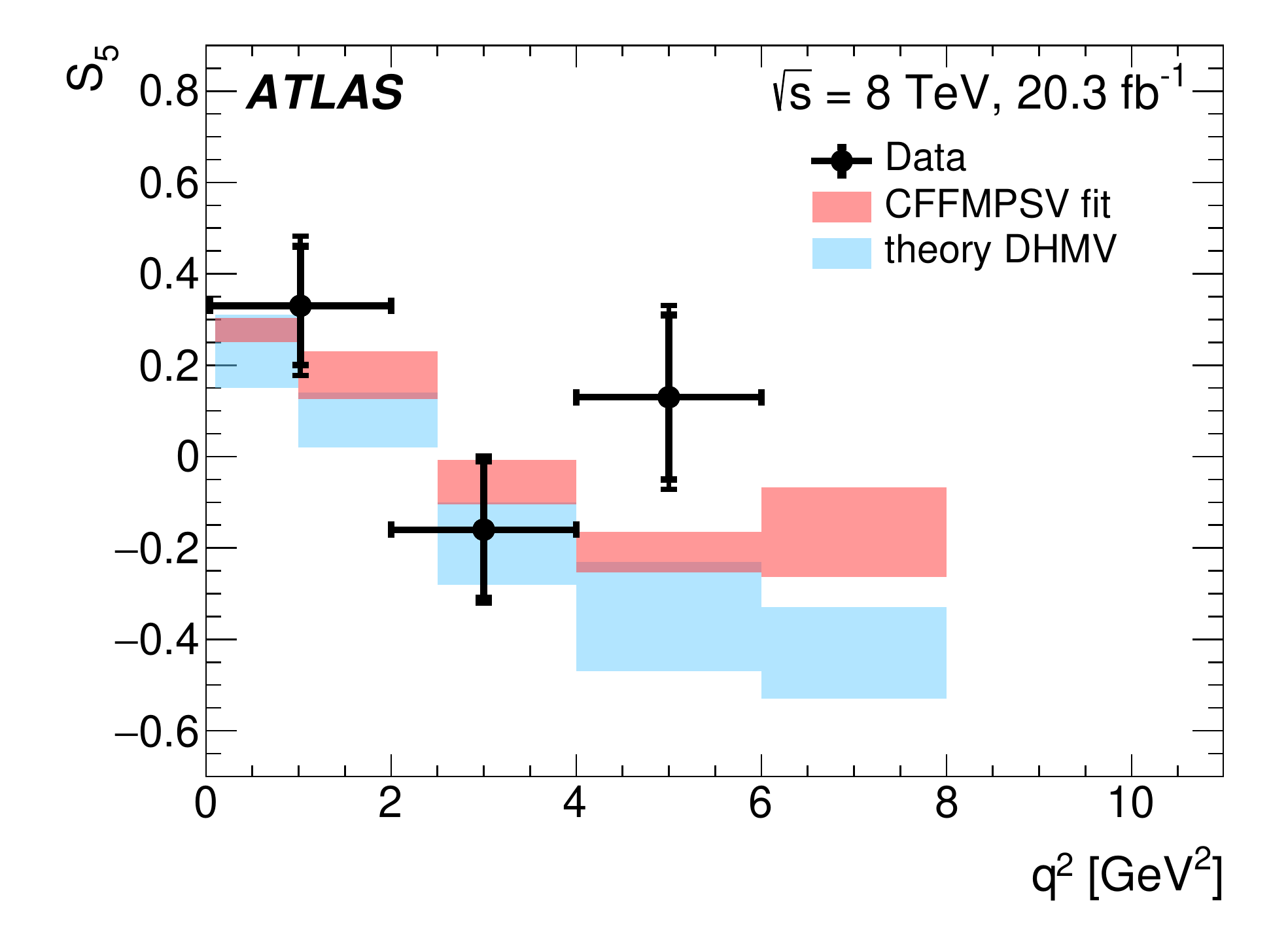}
}
\resizebox{\columnwidth}{!}{
  \includegraphics[width=0.45\columnwidth]{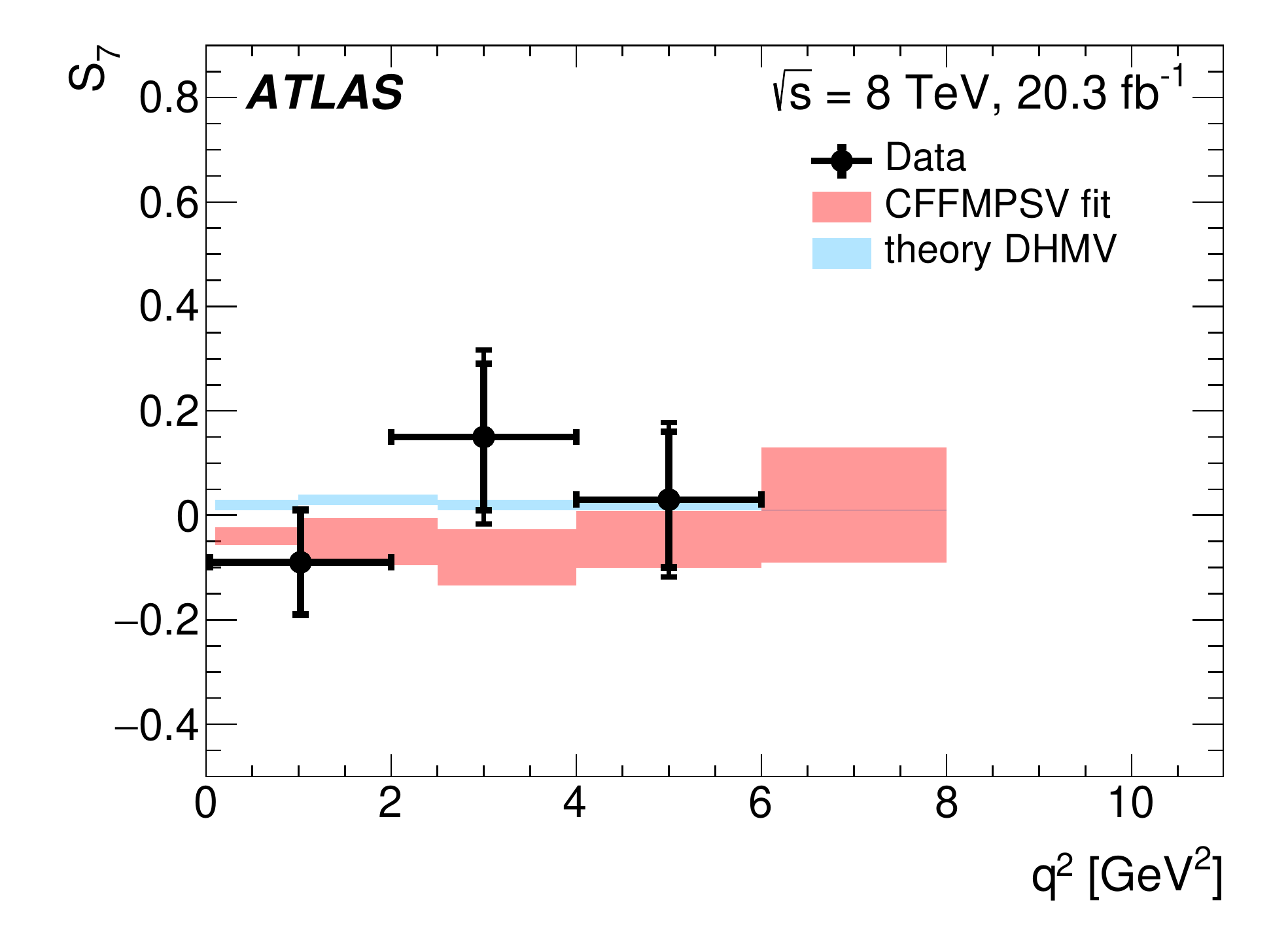}
  \includegraphics[width=0.45\columnwidth]{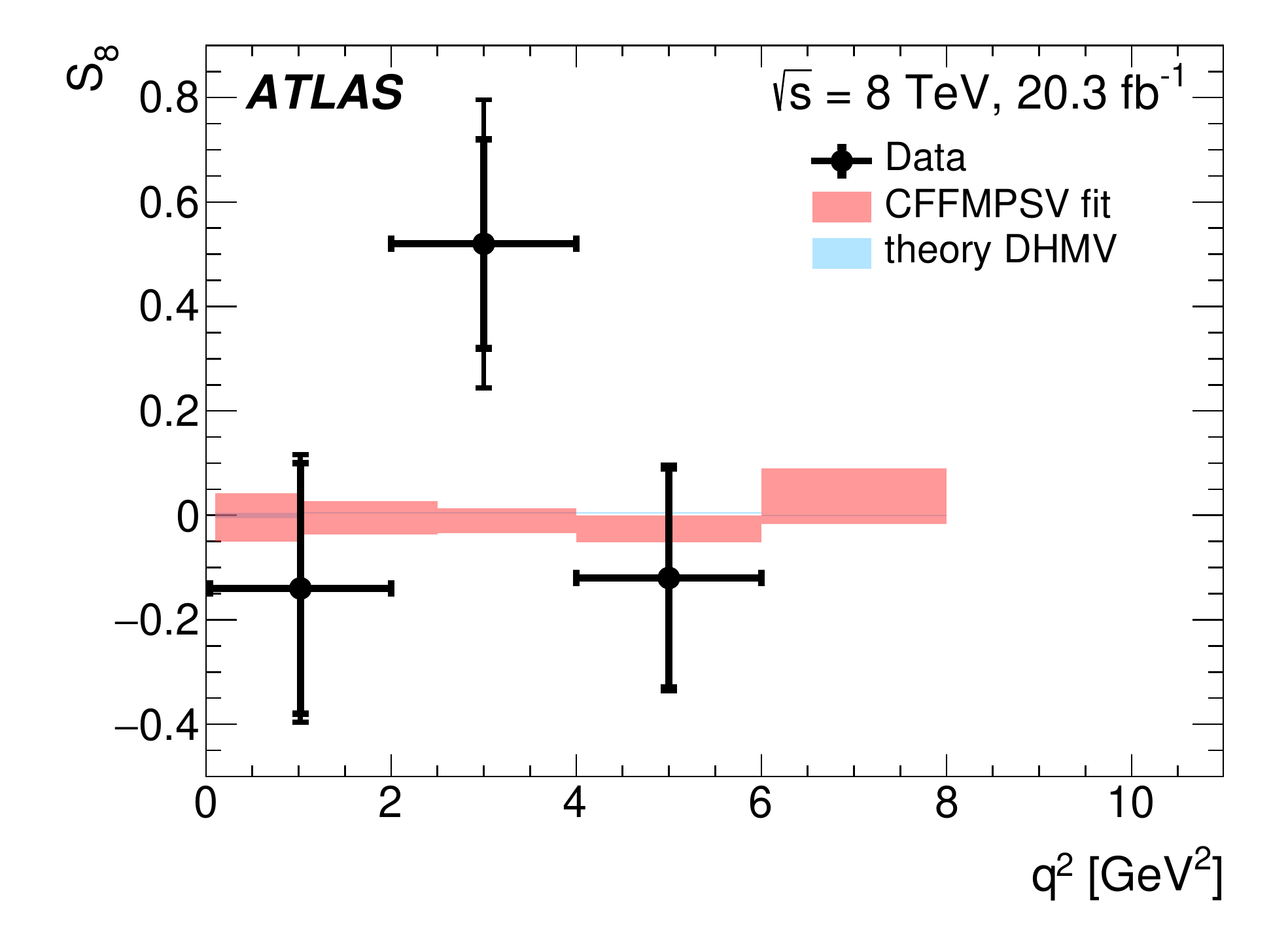}
}
\end{center}
\caption{The measured values of \fl, \Sthree, \Sfour, \Sfive, \Sseven, \Seight compared with predictions from
 the theoretical calculations discussed in the text (Section~\ref{sec:comparison}).
 Statistical and total uncertainties are shown for the data, i.e.\ the inner mark 
 indicates the statistical uncertainty and the total error bar the total uncertainty.
\label{fig:results:s}}
\end{figure}

\begin{figure}[!ht]
\begin{center}
\resizebox{0.5\columnwidth}{!}{
  \includegraphics[width=0.45\columnwidth]{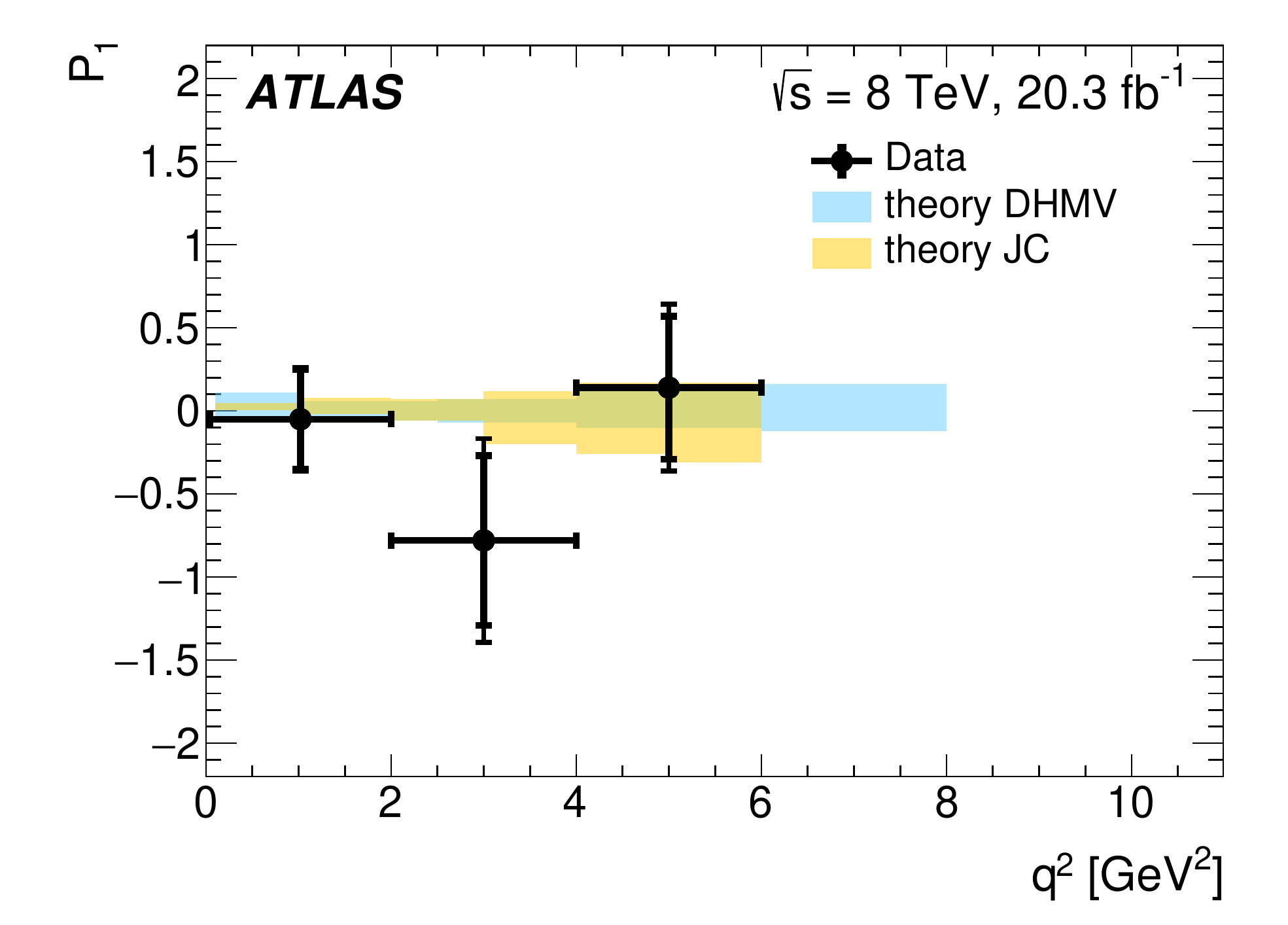}
}
\resizebox{\columnwidth}{!}{
  \includegraphics[width=0.45\columnwidth]{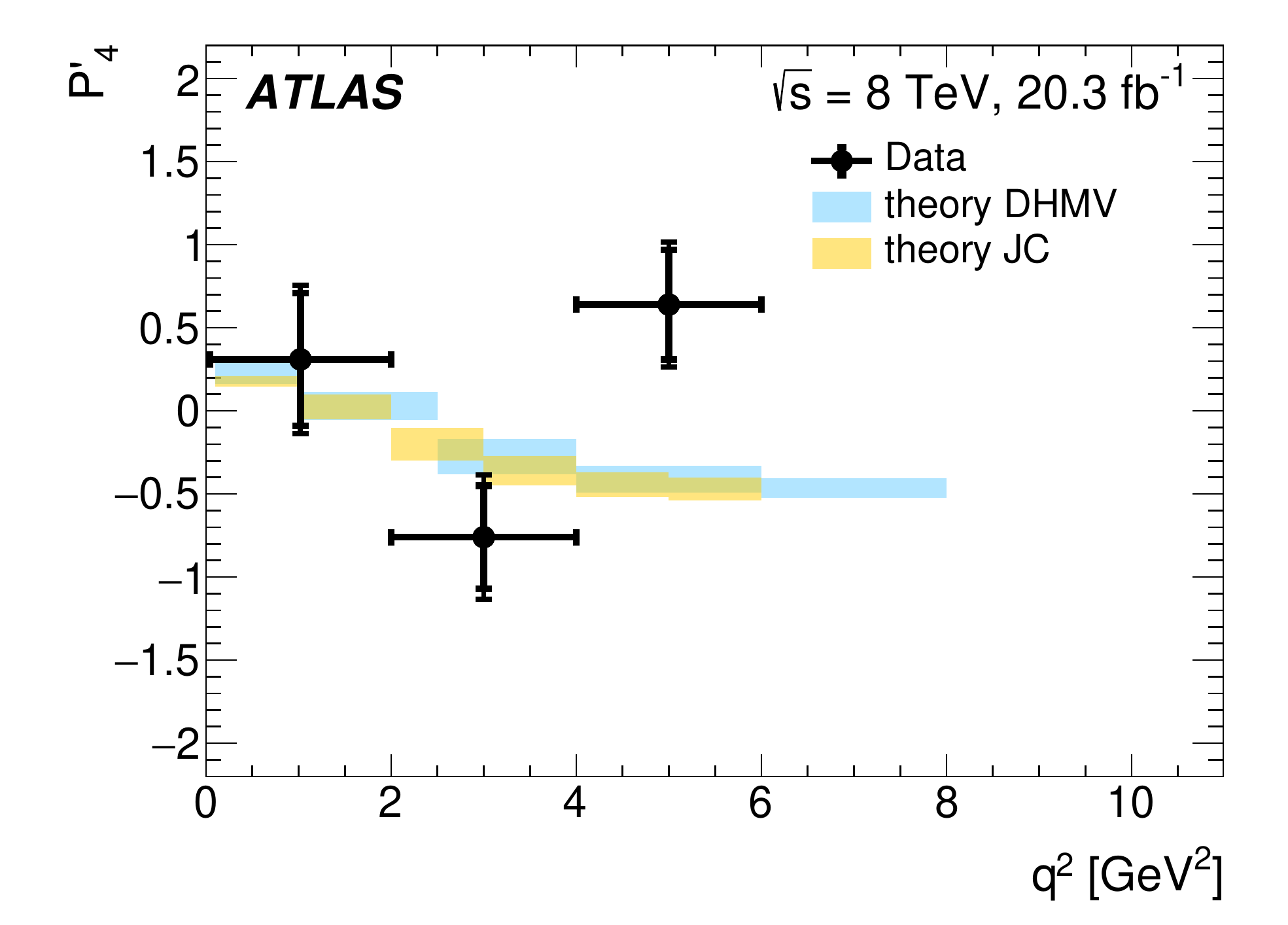}
  \includegraphics[width=0.45\columnwidth]{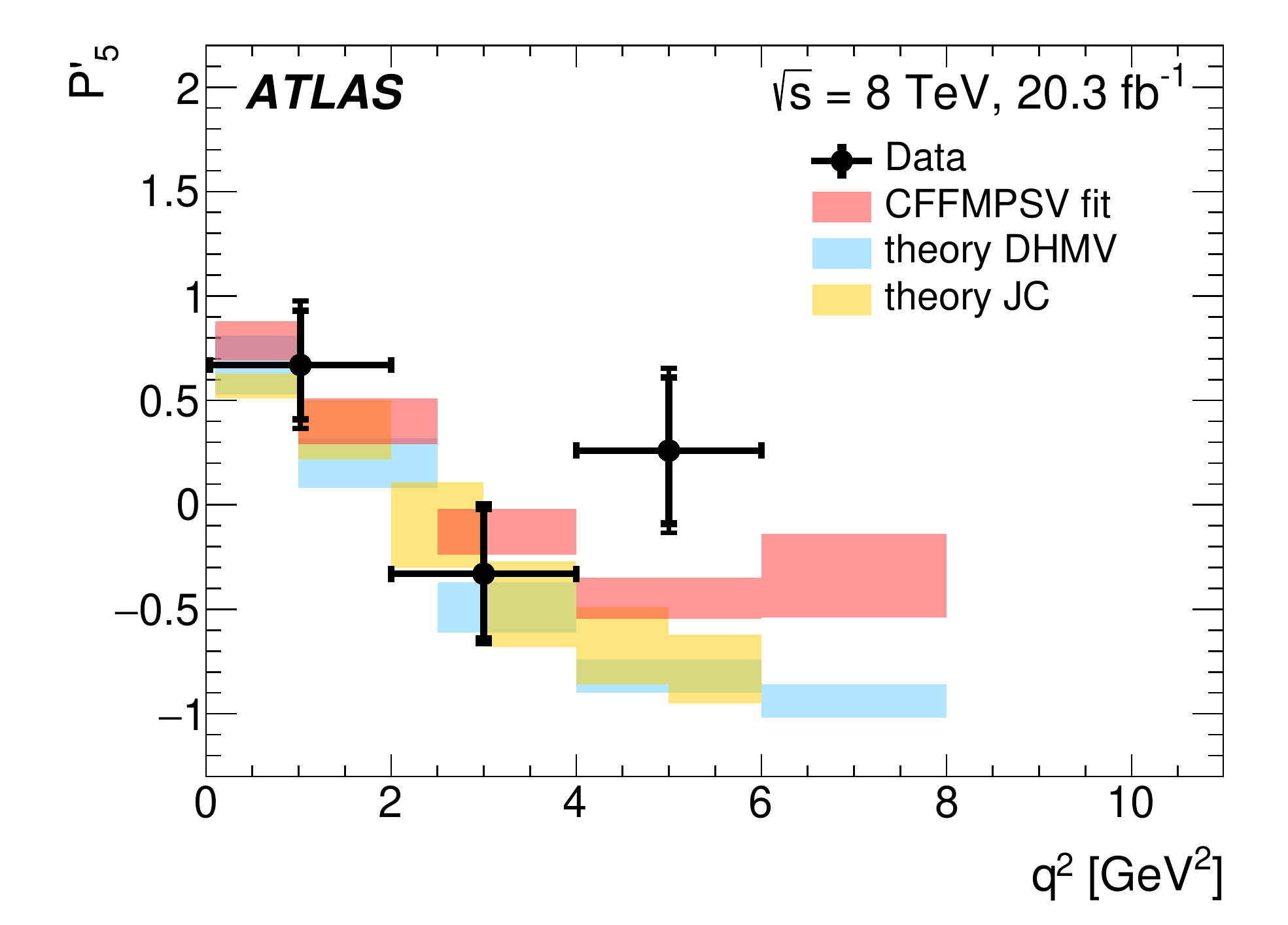}
}
\resizebox{\columnwidth}{!}{
  \includegraphics[width=0.45\columnwidth]{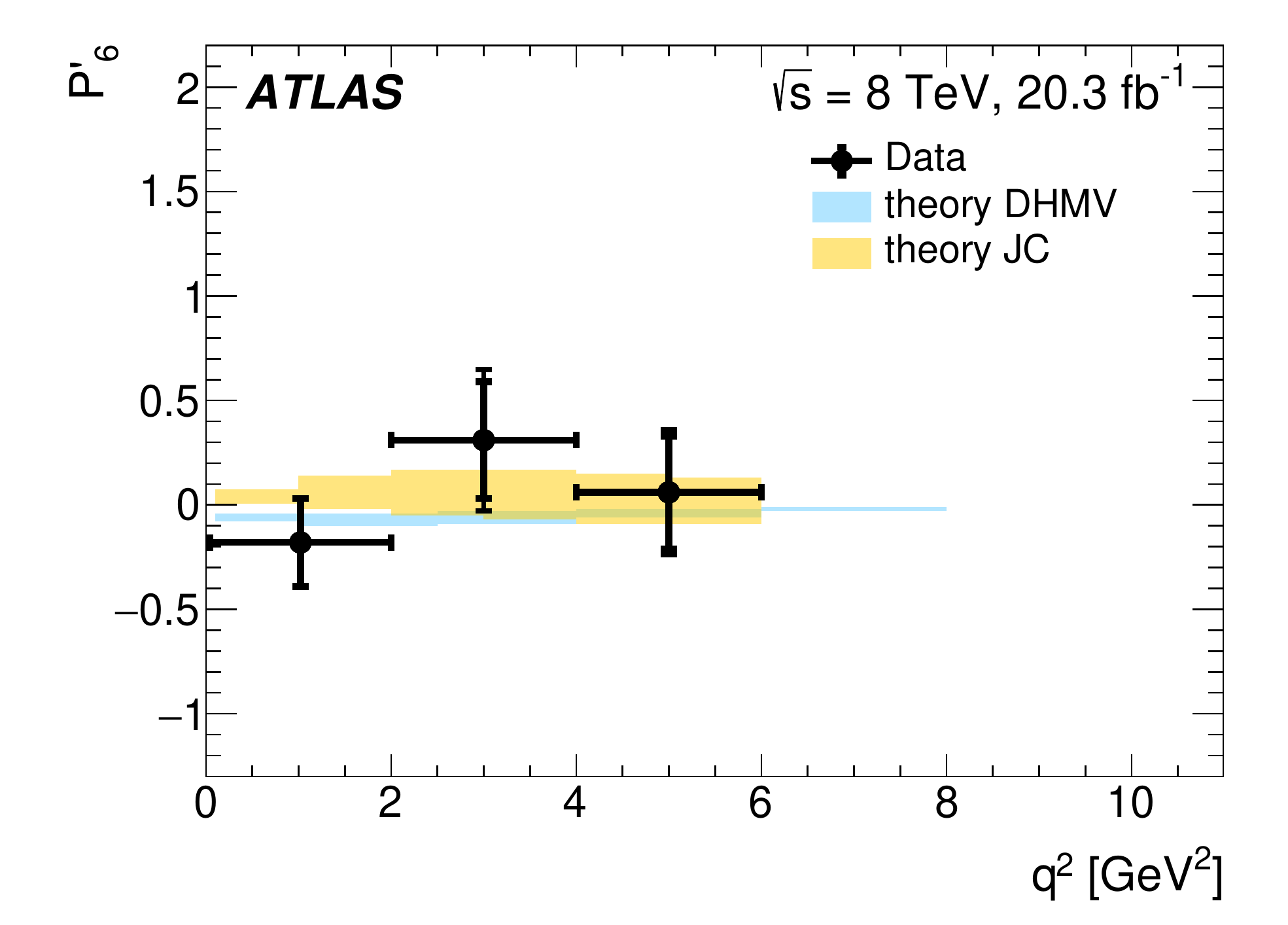}
  \includegraphics[width=0.45\columnwidth]{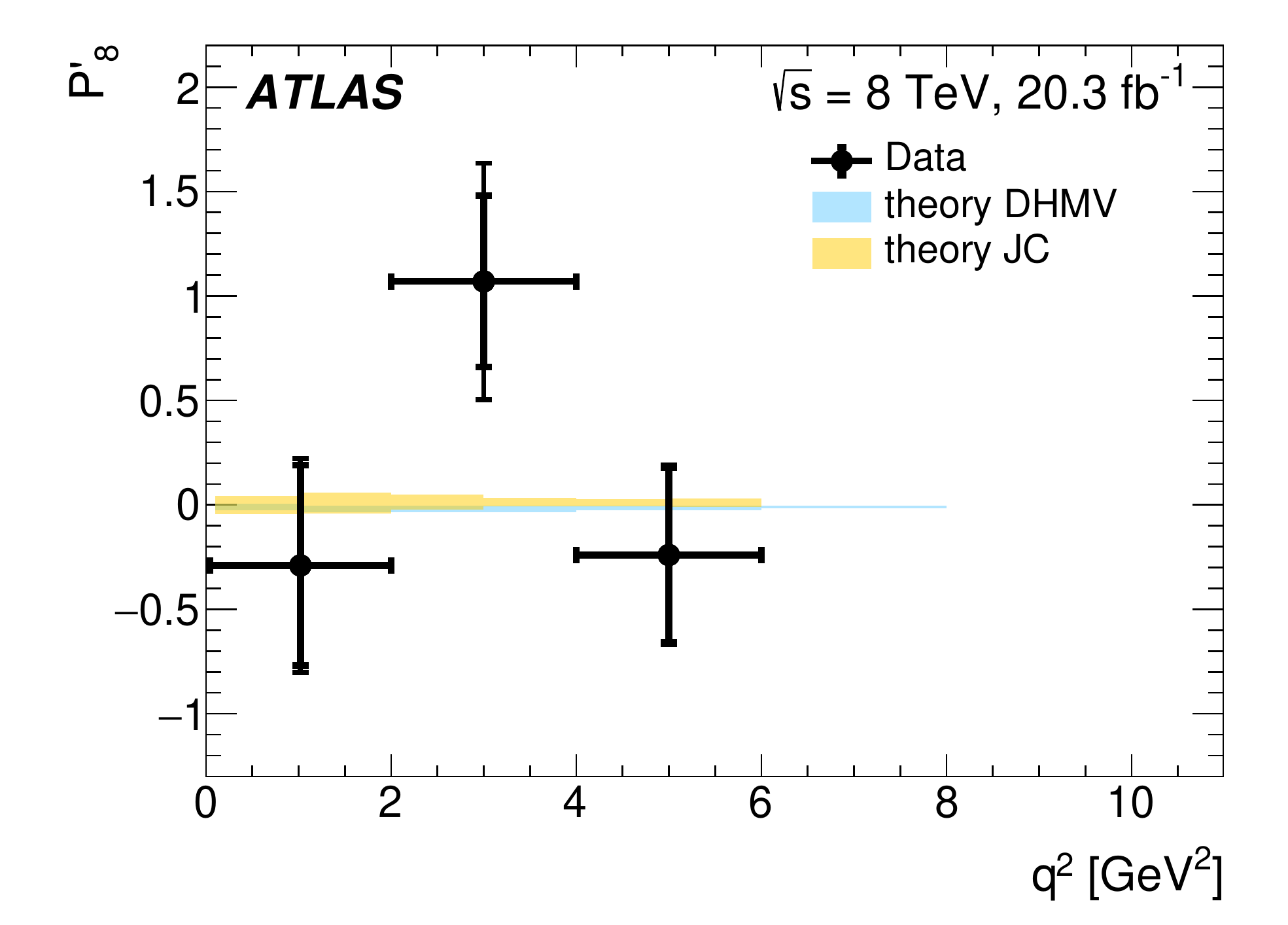}
}
\end{center}
\caption{The measured values of \Pone, \Pfour, \Pfive, \Psix, \Peight compared with predictions from
 the theoretical calculations discussed in the text (Section~\ref{sec:comparison}).
 Statistical and total uncertainties are shown for the data, i.e.\ the inner mark 
 indicates the statistical uncertainty and the total error bar the total uncertainty.
\label{fig:results}}
\end{figure}

The results of theoretical approaches of Ciuchini et al.\
(CFFMPSV)~\cite{Ciuchini:2015qxb}, Descotes-Genon et al.\ \linebreak (DHMV)~\cite{Descotes-Genon:2014uoa}, and J\"ager and Camalich (JC)~\cite{Jager:2012uw,Jager:2014rwa}
are shown in Figure~\ref{fig:results:s} for the $S$ parameters, and in Figure~\ref{fig:results}
for the $P^{(\prime)}$ parameters, along with the results presented
here.\footnote{This result uses the experimental convention of Equations~(\ref{eq:pone})--(\ref{eq:pprime}) following the LHCb
Collaboration's notation in Ref.~\cite{Aaij:2013qta}. \\In the DHMV calculation, a different convention is used as explained by Equation~(16) in Ref.~\cite{Descotes-Genon:2015uva}.
}

QCD factorisation is used by DHMV and JC, where the latter focus on the impact of long-distance corrections
using a helicity amplitude approach.
The CFFMPSV group takes a different approach, using the QCD factorisation framework to perform compatibility checks 
of the LHCb data with theoretical predictions.  This approach also allows information from a given experimentally measured 
parameter of interest to be excluded in order to make a fit-based prediction of the expected value of that parameter
from the rest of the data.

With the exception of the \Pfour and \Pfive measurements in $\qsqr\in\qsqrbinC$ \qsqrunits
 and \Peight in $\qsqr\in\qsqrbinB$ \qsqrunits there is good agreement between theory and measurement. 
The \Pfour and \Pfive parameters have statistical correlation of 0.37 in the $\qsqr\in\qsqrbinC$ \qsqrunits bin.
The observed deviation from the SM prediction of \Pfour and \Pfive
is for both parameters approximately 2.7 standard deviations (local) away from the calculation of DHMV
for this bin.
The deviations are less significant for the other calculation and the fit approach.
All measurements are found to be within three standard 
deviations of the range covered by the different predictions. Hence, including 
experimental and theoretical uncertainties, the measurements presented here are 
found to agree with the predicted SM contributions to this decay.

\FloatBarrier

\section{Conclusion}
\label{sec:conclusion}

The results of an angular analysis of the rare decay \bdkstmumu are presented.
This flavour-changing neutral current process is sensitive to potential new-physics contributions.
The \bdkstmumu analysis presented here uses a total of \lumi~\invfb 
of $\sqrt{s}=8$ \tev\ $pp$ collision data collected by the ATLAS experiment at the LHC in 2012.
An extended unbinned maximum-likelihood fit of the angular distribution of the signal decay
is performed in order to extract the parameters \fl, $S_i$ and $P_j^{(\prime)}$ in six bins of \qsqr.
Three of these bins overlap in order to report results in ranges compatible with
other experiments and phenomenology studies.
All measurements are found to be within three standard
deviation of the range covered by the different predictions.
The results are also compatible with the results of the LHCb, CMS and Belle collaborations.

%% file: acknowledgements/Acknowledgements.tex

We thank CERN for the very successful operation of the LHC, as well as the
support staff from our institutions without whom ATLAS could not be
operated efficiently.

We acknowledge the support of ANPCyT, Argentina; YerPhI, Armenia; ARC, Australia; BMWFW and FWF, Austria; ANAS, Azerbaijan; SSTC, Belarus; CNPq and FAPESP, Brazil; NSERC, NRC and CFI, Canada; CERN; CONICYT, Chile; CAS, MOST and NSFC, China; COLCIENCIAS, Colombia; MSMT CR, MPO CR and VSC CR, Czech Republic; DNRF and DNSRC, Denmark; IN2P3-CNRS, CEA-DRF/IRFU, France; SRNSFG, Georgia; BMBF, HGF, and MPG, Germany; GSRT, Greece; RGC, Hong Kong SAR, China; ISF, I-CORE and Benoziyo Center, Israel; INFN, Italy; MEXT and JSPS, Japan; CNRST, Morocco; NWO, Netherlands; RCN, Norway; MNiSW and NCN, Poland; FCT, Portugal; MNE/IFA, Romania; MES of Russia and NRC KI, Russian Federation; JINR; MESTD, Serbia; MSSR, Slovakia; ARRS and MIZ\v{S}, Slovenia; DST/NRF, South Africa; MINECO, Spain; SRC and Wallenberg Foundation, Sweden; SERI, SNSF and Cantons of Bern and Geneva, Switzerland; MOST, Taiwan; TAEK, Turkey; STFC, United Kingdom; DOE and NSF, United States of America. In addition, individual groups and members have received support from BCKDF, the Canada Council, CANARIE, CRC, Compute Canada, FQRNT, and the Ontario Innovation Trust, Canada; EPLANET, ERC, ERDF, FP7, Horizon 2020 and Marie Sk{\l}odowska-Curie Actions, European Union; Investissements d'Avenir Labex and Idex, ANR, R{\'e}gion Auvergne and Fondation Partager le Savoir, France; DFG and AvH Foundation, Germany; Herakleitos, Thales and Aristeia programmes co-financed by EU-ESF and the Greek NSRF; BSF, GIF and Minerva, Israel; BRF, Norway; CERCA Programme Generalitat de Catalunya, Generalitat Valenciana, Spain; the Royal Society and Leverhulme Trust, United Kingdom.

The crucial computing support from all WLCG partners is acknowledged gratefully, in particular from CERN, the ATLAS Tier-1 facilities at TRIUMF (Canada), NDGF (Denmark, Norway, Sweden), CC-IN2P3 (France), KIT/GridKA (Germany), INFN-CNAF (Italy), NL-T1 (Netherlands), PIC (Spain), ASGC (Taiwan), RAL (UK) and BNL (USA), the Tier-2 facilities worldwide and large non-WLCG resource providers. Major contributors of computing resources are listed in Ref.~\cite{ATL-GEN-PUB-2016-002}.

%% file: texts/appendix.tex
\appendix
\section{Correlation Matrices}
\label{sec:correlations}

Four folding schemes are applied to the data in order to extract \fl, \Sthree, \Sfour, \Sfive, \Sseven and \Seight
from four separate fits.  The $P^{(\prime)}$ parameters are subsequently derived from the fit results
using Equations~(\ref{eq:pone})--(\ref{eq:pprime}).
It is not possible to extract a full correlation matrix between fitted parameters obtained from 
different fits.  In order to reconstruct the correlation matrix, ensembles of pseudo-experiments 
are simulated using the pdf corresponding to the nominal angular distributions. Each simulated
ensemble has the four folding schemes applied to it and four fits are performed on the resulting
samples.  The distributions obtained for pairs of parameters obtained from fits to these ensembles
are used to compute Pearson correlation coefficients for those pairs. Correlation matrices
for \fl and the $S$ parameters are reconstructed from all possible pairings for a given
\qsqr bin.  A similar method is used to extract the correlation matrices for the $P^{(\prime)}$ 
parameters.  This procedure is repeated for each \qsqr bin studied in order to obtain 
correlation matrices given in the remainder of this appendix. The correlation matrices are statistical only.
Contributions from systematic uncertainties are not included, since the measurement precision is statistically limited.
\begin{itemize}
  \item Table~\ref{tbl:binone} (\ref{tbl:binonep}) shows the statistical correlation matrix for $\fl$ and $S$ ($P^{(\prime)}$) parameters for the \qsqr bin $[0.04, 2.0]$ \qsqrunits.
  \item Table~\ref{tbl:bintwo} (\ref{tbl:bintwop}) shows the statistical correlation matrix for $\fl$ and $S$ ($P^{(\prime)}$) parameters for the \qsqr bin $[2.0, 4.0]$ \qsqrunits.
  \item Table~\ref{tbl:binthree} (\ref{tbl:binthreep}) shows the statistical correlation matrix for $\fl$ and $S$ ($P^{(\prime)}$) parameters for the \qsqr bin $[4.0, 6.0]$ \qsqrunits.
  \item Table~\ref{tbl:binfour} (\ref{tbl:binfourp}) shows the statistical correlation matrix for $\fl$ and $S$ ($P^{(\prime)}$) parameters for the \qsqr bin $[0.04, 4.0]$ \qsqrunits.
  \item Table~\ref{tbl:binfive} (\ref{tbl:binfivep}) shows the statistical correlation matrix for $\fl$ and $S$ ($P^{(\prime)}$) parameters for the \qsqr bin $[1.1, 6.0]$ \qsqrunits.
  \item Table~\ref{tbl:binsix} (\ref{tbl:binsixp}) shows the statistical correlation matrix for $\fl$ and $S$ ($P^{(\prime)}$) parameters for the \qsqr bin $[0.04, 6.0]$ \qsqrunits.
\end{itemize}

\begin{table}[!ht]
\begin{center}
\caption{Statistical correlation matrix for the $\fl$ and $S$ parameters obtained for $\qsqr\in[0.04, 2.0]$ \qsqrunits.}\label{tbl:binone}
\begin{tabular}{l|rrrrrr} \hline\hline
         & \fl  & \Sthree & \Sfour & \Sfive & \Sseven & \Seight \\ \hline
 \fl     & 1.00 & 0.11    & $-0.13$& 0.03   & 0.16    & 0.24\\
 \Sthree &      & 1.00    & 0.31   & 0.28   & 0.73    & 0.45\\
 \Sfour  &      &         & 1.00   & 0.58   & 0.19    & 0.22\\
 \Sfive  &      &         &        & 1.00   & 0.14    & 0.28\\
 \Sseven &      &         &        &        & 1.00    & 0.59\\
 \Seight &      &         &        &        &         & 1.00\\
\hline\hline
\end{tabular}
\end{center}
\end{table}

\begin{table}[!ht]
\begin{center}
\caption{Statistical correlation matrix for the $P^{(\prime)}$ parameters obtained for $\qsqr\in[0.04, 2.0]$ \qsqrunits.}\label{tbl:binonep}
\begin{tabular}{l|rrrrrr} \hline\hline
          & \Pone & \Pfour & \Pfive & \Psix & \Peight \\ \hline
 \Pone    & 1.00  & 0.04    & 0.05    & 0.62  & 0.32\\
 \Pfour   &       & 1.00     & 0.53   & $-$0.08  & $-$0.06\\
 \Pfive   &       &          & 1.00   & 0.00     & 0.22\\
 \Psix    &       &          &        & 1.00     & 0.55\\
 \Peight  &       &          &        &          & 1.00\\
\hline\hline
\end{tabular}
\end{center}
\end{table}

\begin{table}[!ht]
\begin{center}
\caption{Statistical correlation matrix for the $\fl$ and $S$ parameters obtained for $\qsqr\in[2.0, 4.0]$ \qsqrunits.}\label{tbl:bintwo}
\begin{tabular}{l|rrrrrr} \hline\hline
         & \fl  & \Sthree & \Sfour & \Sfive  & \Sseven & \Seight \\ \hline
 \fl     & 1.00 & 0.27    & 0.35   & $-0.04$ & $-0.15$  & $-0.37$\\
 \Sthree &      & 1.00    & $-0.08$& $-0.44$ & $-0.09$  & $-0.20$\\
 \Sfour  &      &         & 1.00   & 0.60    & $-0.02$  & $-0.12$\\
 \Sfive  &      &         &        & 1.00    & $-0.11$  & $-0.20$\\
 \Sseven &      &         &        &         & 1.00     & 0.63\\
 \Seight &      &         &        &         &          & 1.00\\
\hline\hline
\end{tabular}
\end{center}
\end{table}

\begin{table}[!ht]
\begin{center}
\caption{Statistical correlation matrix for the $P^{(\prime)}$ parameters obtained for $\qsqr\in[2.0, 4.0]$ \qsqrunits.}\label{tbl:bintwop}
\begin{tabular}{l|rrrrrr} \hline\hline
           & \Pone & \Pfour & \Pfive & \Psix & \Peight \\ \hline
 \Pone     & 1.00    & $-$0.12  & $-$0.21  & 0.05  & 0.05\\
 \Pfour    &         & 1.00     & 0.51     & 0.08  & 0.03\\
 \Pfive    &         &          & 1.00     & $-$0.23  & 0.22\\
 \Psix     &         &          &          & 1.00     &  0.66\\
 \Peight   &         &          &          &          & 1.00\\
\hline\hline
\end{tabular}
\end{center}
\end{table}

\begin{table}[!ht]
\begin{center}
\caption{Statistical correlation matrix for the $\fl$ and $S$ parameters obtained for $\qsqr\in[4.0, 6.0]$ \qsqrunits.}\label{tbl:binthree}
\begin{tabular}{l|rrrrrr} \hline\hline
         & \fl  & \Sthree & \Sfour  & \Sfive  & \Sseven & \Seight \\ \hline
 \fl     & 1.00 & 0.33    & $-0.18$ & 0.04    & 0.22    & 0.28\\
 \Sthree &      & 1.00    & 0.15    & 0.23    & 0.60    & 0.05\\
 \Sfour  &      &         & 1.00    & 0.52    & 0.03    & 0.01\\
 \Sfive  &      &         &         & 1.00    & 0.28    & 0.27\\
 \Sseven &      &         &         &         & 1.00    & 0.60\\
 \Seight &      &         &         &         &         & 1.00\\
\hline\hline
\end{tabular}
\end{center}
\end{table}

\begin{table}[!ht]
\begin{center}
\caption{Statistical correlation matrix for the $P^{(\prime)}$ parameters obtained for $\qsqr\in[4.0, 6.0]$ \qsqrunits.}\label{tbl:binthreep}
\begin{tabular}{l|rrrrrr} \hline\hline
           & \Pone & \Pfour & \Pfive & \Psix & \Peight \\ \hline
\Pone      & 1.00    & 0.11  & 0.34   & 0.41  & 0.16\\
 \Pfour    &         & 1.00     & 0.37   & 0.06  & 0.04\\
 \Pfive    &         &          & 1.00     & 0.39   & 0.33\\
 \Psix     &         &          &          & 1.00     & 0.62\\
 \Peight   &         &          &          &          & 1.00\\
\hline\hline
\end{tabular}
\end{center}
\end{table}

\begin{table}[!ht]
\begin{center}
\caption{Statistical correlation matrix for the $\fl$ and $S$ parameters obtained for $\qsqr\in[0.04, 4.0]$ \qsqrunits.}\label{tbl:binfour}
\begin{tabular}{l|rrrrrr} \hline\hline
         & \fl  & \Sthree & \Sfour  & \Sfive  & \Sseven & \Seight \\ \hline
 \fl     & 1.00 & 0.08    & 0.05    & 0.01    & 0.18    & 0.14\\
 \Sthree &      & 1.00    & $-0.04$ & 0.03    & 0.29    & $-0.16$\\
 \Sfour  &      &         & 1.00    & 0.79    & 0.08    & 0.03\\
 \Sfive  &      &         &         & 1.00    & 0.03    & $-0.02$\\
 \Sseven &      &         &         &         & 1.00    & 0.60\\
 \Seight &      &         &         &         &         & 1.00\\
\hline\hline
\end{tabular}
\end{center}
\end{table}

\begin{table}[!ht]
\begin{center}
\caption{Statistical correlation matrix for the $P^{(\prime)}$ parameters obtained for $\qsqr\in[0.04, 4.0]$ \qsqrunits.}\label{tbl:binfourp}
\begin{tabular}{l|rrrrrr} \hline\hline
           & \Pone & \Pfour & \Pfive & \Psix & \Peight \\ \hline
 \Pone     & 1.00    & $-$0.07  & 0.00     & 0.21     & 0.12\\
 \Pfour    &         & 1.00     & 0.78     & 0.08     & 0.02\\
 \Pfive    &         &          & 1.00     & 0.03     & $-$0.04\\
 \Psix     &         &          &          & 1.00     & 0.59\\
 \Peight   &         &          &          &          & 1.00\\
\hline\hline
\end{tabular}
\end{center}
\end{table}

\begin{table}[!ht]
\begin{center}
\caption{Statistical correlation matrix for the $\fl$ and $S$ parameters obtained for $\qsqr\in[1.1, 6.0]$ \qsqrunits.}\label{tbl:binfive}
\begin{tabular}{l|rrrrrr} \hline\hline
         & \fl  & \Sthree & \Sfour & \Sfive & \Sseven & \Seight \\ \hline
 \fl     & 1.00 & 0.05    & 0.00    & 0.07    & 0.18    & $-0.03$\\
 \Sthree &      & 1.00    & 0.41    & 0.46    & 0.32    & $-0.01$\\
 \Sfour  &      &         & 1.00    & 0.60    & 0.09    & 0.03\\
 \Sfive  &      &         &         & 1.00    & 0.17    & 0.24\\
 \Sseven &      &         &         &         & 1.00    & 0.67\\
 \Seight &      &         &         &         &         & 1.00\\
\hline\hline
\end{tabular}
\end{center}
\end{table}

\begin{table}[!ht]
\begin{center}
\caption{Statistical correlation matrix for the $P^{(\prime)}$ parameters obtained for $\qsqr\in[1.1, 6.0]$ \qsqrunits.}\label{tbl:binfivep}
\begin{tabular}{l|rrrrrr} \hline\hline
           & \Pone & \Pfour & \Pfive & \Psix & \Peight \\ \hline
 \Pone     & 1.00    & 0.23     & $-$0.09  & 0.08     & $-$0.07\\
 \Pfour    &         & 1.00     & 0.53     & 0.15     & 0.08\\
 \Pfive    &         &          & 1.00     & 0.28     & 0.24\\
 \Psix     &         &          &          & 1.00     & 0.67\\
 \Peight   &         &          &          &          & 1.00\\
\hline\hline
\end{tabular}
\end{center}
\end{table}

\begin{table}[!ht]
\begin{center}
\caption{Statistical correlation matrix for the $\fl$ and $S$ parameters obtained for $\qsqr\in[0.04, 6.0]$ \qsqrunits.}\label{tbl:binsix}
\begin{tabular}{l|rrrrrr} \hline\hline
         & \fl  & \Sthree & \Sfour & \Sfive & \Sseven & \Seight \\ \hline
 \fl     & 1.00 & 0.03    & 0.01   & $-0.10$& 0.13    & 0.06\\
 \Sthree &      & 1.00    & $-0.02$& $-0.09$& 0.32    & $-0.01$\\
 \Sfour  &      &         & 1.00   & 0.68   & 0.00    & 0.04\\
 \Sfive  &      &         &        & 1.00   & $-0.05$ & 0.03\\
 \Sseven &      &         &        &        & 1.00    & 0.65\\
 \Seight &      &         &        &        &         & 1.00\\
\hline\hline
\end{tabular}
\end{center}
\end{table}

\begin{table}[!ht]
\begin{center}
\caption{Statistical correlation matrix for the $P^{(\prime)}$ parameters obtained for $\qsqr\in[0.04, 6.0]$ \qsqrunits.}\label{tbl:binsixp}
\begin{tabular}{l|rrrrrr} \hline\hline
           & \Pone & \Pfour & \Pfive & \Psix & \Peight \\ \hline
 \Pone     & 1.00    & $-$0.02  & $-$0.14  & 0.17     & 0.04\\
 \Pfour    &         & 1.00     & 0.65     & 0.00     & 0.05\\
 \Pfive    &         &          & 1.00     & $-$0.06  & 0.09\\
 \Psix     &         &          &          & 1.00     & 0.61\\
 \Peight   &         &          &          &          & 1.00\\
\hline\hline
\end{tabular}
\end{center}
\end{table}

\clearpage

%% file: atlas_authlist.tex
 
\begin{flushleft}
{\Large The ATLAS Collaboration}

\bigskip

M.~Aaboud$^\textrm{\scriptsize 34d}$,    
G.~Aad$^\textrm{\scriptsize 99}$,    
B.~Abbott$^\textrm{\scriptsize 124}$,    
O.~Abdinov$^\textrm{\scriptsize 13,*}$,    
B.~Abeloos$^\textrm{\scriptsize 128}$,    
S.H.~Abidi$^\textrm{\scriptsize 165}$,    
O.S.~AbouZeid$^\textrm{\scriptsize 143}$,    
N.L.~Abraham$^\textrm{\scriptsize 153}$,    
H.~Abramowicz$^\textrm{\scriptsize 159}$,    
H.~Abreu$^\textrm{\scriptsize 158}$
% ...
\bigskip
\\

% ...
$^{13}$Institute of Physics, Azerbaijan Academy of Sciences, Baku; Azerbaijan.\\
% ...
$^{34}$$^{(a)}$Facult\'e des Sciences Ain Chock, R\'eseau Universitaire de Physique des Hautes Energies - Universit\'e Hassan II, Casablanca;$^{(b)}$Centre National de l'Energie des Sciences Techniques Nucleaires (CNESTEN), Rabat;$^{(c)}$Facult\'e des Sciences Semlalia, Universit\'e Cadi Ayyad, LPHEA-Marrakech;$^{(d)}$Facult\'e des Sciences, Universit\'e Mohamed Premier and LPTPM, Oujda;$^{(e)}$Facult\'e des sciences, Universit\'e Mohammed V, Rabat; Morocco.\\
% ...
$^{99}$CPPM, Aix-Marseille Universit\'e, CNRS/IN2P3, Marseille; France.\\
% ...
$^{124}$Homer L. Dodge Department of Physics and Astronomy, University of Oklahoma, Norman OK; United States of America.\\
% ...
$^{128}$LAL, Universit\'e Paris-Sud, CNRS/IN2P3, Universit\'e Paris-Saclay, Orsay; France.\\
% ...
$^{143}$Santa Cruz Institute for Particle Physics, University of California Santa Cruz, Santa Cruz CA; United States of America.\\
% ...
$^{153}$Department of Physics and Astronomy, University of Sussex, Brighton; United Kingdom.\\
% ...
$^{158}$Department of Physics, Technion, Israel Institute of Technology, Haifa; Israel.\\
$^{159}$Raymond and Beverly Sackler School of Physics and Astronomy, Tel Aviv University, Tel Aviv; Israel.\\
% ...
$^{165}$Department of Physics, University of Toronto, Toronto ON; Canada.\\
% ...
$^{a}$ Also at  Department of Physics, University of Malaya, Kuala Lumpur; Malaysia.\\
$^{b}$ Also at Borough of Manhattan Community College, City University of New York, NY; United States of America.\\
$^{c}$ Also at Centre for High Performance Computing, CSIR Campus, Rosebank, Cape Town; South Africa.\\
$^{d}$ Also at CERN, Geneva; Switzerland.\\
$^{e}$ Also at CPPM, Aix-Marseille Universit\'e, CNRS/IN2P3, Marseille; France.\\
% ...
$^{*}$ Deceased

\end{flushleft}
